\documentclass[11pt,a4paper]{article}
\pdfoutput=1
\usepackage{jstyle}
\usepackage{amsmath, amsfonts, amssymb}
\usepackage{graphicx, hyperref, color, verbatim}
\usepackage[dvipsnames]{xcolor}
\usepackage{float}
\usepackage{caption}
\usepackage{subcaption}

\DeclareGraphicsExtensions{.pdf,.png,.jpg,.mps}
\usepackage{booktabs}
\usepackage{tikz}
\usetikzlibrary{calc,matrix,decorations.pathmorphing,decorations.markings,arrows,positioning,intersections,mindmap,backgrounds}
\usepackage{cancel}

\usepackage{physics}
\usepackage{diagbox}
\usepackage{pifont}

\newcommand{\zb}{\bar{z}}
\newcommand{\hsv}{G^\text{SV}}
\newcommand{\dst}{\mathtt{d}_{st}}
\newcommand{\dsu}{\mathtt{d}_{su}}
\newcommand{\dtu}{\mathtt{d}_{tu}}

\title{AdS super gluon scattering up to two loops: A position space approach}

\author[]{Zhongjie Huang$^{a,b}$,}
\author[]{Bo Wang$^{a,b}$,}
\author[]{Ellis Ye Yuan$^{a,b}$,}
\author[]{Xinan Zhou$^{c}$}
\affiliation[]{$^{a}$Zhejiang Institute of Modern Physics, School of Physics, Zhejiang University, \\Hangzhou, Zhejiang 310058, China }
\affiliation[]{$^{b}$Joint Center for Quanta-to-Cosmos Physics, Zhejiang University,
\\Hangzhou, Zhejiang 310058, China}
\affiliation[]{$^{c}$Kavli Institute for Theoretical Sciences,
University of Chinese Academy of Sciences, \\Beijing 100190, China. }

\abstract{We carry out a bootstrap study of four-point correlators in 4d $\mathcal{N}=2$ SCFTs which are dual to super Yang-Mills on $AdS_5\times S^3$. We focus on the simplest $\frac{1}{2}$-BPS operators which correspond to the  super gluons in the massless current multiplet. Our computation is based on an ansatz in position space which is inspired by a hidden symmetry structure manifest in the leading terms of the Lorentzian singularities of the correlators. By using other consistency conditions, we completely fix the super gluon correlators at one and two loops in the bulk genus expansion, up to possible counterterms. Our results reveal a number of interesting properties enriched by the color structures. In particular, the  implication of hidden conformal symmetry on the full super gluon reduced correlator exhibits an analogous pattern as in the $AdS_5\times S^5$ supergravity correlators recently computed up to two loops.}

\emailAdd{zjhuang@zju.edu.cn}
\emailAdd{b\_w@zju.edu.cn}
\emailAdd{eyyuan@zju.edu.cn}
\emailAdd{xinan.zhou@ucas.ac.cn}

\begin{document}
\maketitle
\tableofcontents

\newpage

\section{Introduction}

The AdS/CFT correspondence maps correlation functions of local operators in the CFT to on-shell scattering amplitudes in AdS. In the holographic limit, these observables are expanded in powers of $1/c$ with respect to the large central charge. At the leading order, the holographic correlators are just given by the generalized free field theory due to the large $N$ factorization and they can be computed simply by Wick contractions. However, to extract nontrivial dynamical information one needs to go to higher orders in $1/c$ . Computing these subleading contributions is in general intractable from the CFT side alone as the theory is strongly coupled. The weakly coupled dual description makes it possible, at least in principle, as holographic correlators can be computed as amplitudes at various loop orders by using the AdS generalization of the standard Feynman diagram expansion. However, it should be noted that such a recipe is rather impractical to use beyond the few simplest cases \cite{Freedman:1998tz,DHoker:1999pj,Arutyunov:2000py,Arutyunov:2002fh,Arutyunov:2003ae}, due to the proliferation of diagrams and complicated AdS vertices \cite{Arutyunov:1999fb}. In fact, just at the tree level, {\it i.e.}, at order $1/c$, the computation of general four-point functions remained an unsolved problem for almost two decades. 

A much better strategy, initiated in \cite{Rastelli:2016nze,Rastelli:2017udc}, is the bootstrap approach, which led to the complete tree-level four-point functions of $\frac{1}{2}$-BPS operators with arbitrary Kaluza-Klein (KK) levels for IIB supergravity in $AdS_5\times S^5$. The bootstrap approach exploits both the amplitude intuition from the bulk and the superconformal  constraints from the boundary, and is currently the most efficient method for computing holographic correlators. At the moment, there is already a wealth of results at tree level. For example, general four-point functions of arbitrary $\frac{1}{2}$-BPS operators have been computed in closed forms in all maximally superconformal theories \cite{Alday:2020lbp,Alday:2020dtb}, as well as in theories with half the amount of maximal superconformal symmetry \cite{Rastelli:2019gtj,Giusto:2020neo,Alday:2021odx}.\footnote{See \cite{Bissi:2022mrs} for a recent review.} By contrast, our understanding for loop level correlators is much more limited, even in the paradigmatic example of IIB supergravity on $AdS_5\times S^5$. The first one-loop correlator was computed in \cite{Aprile:2017bgs,Alday:2017xua} for the stress tensor multiplet in position space and later in Mellin space \cite{Alday:2018kkw}. The calculation was generalized to four-point functions with higher KK levels in \cite{Aprile:2017qoy,Aprile:2019rep,Alday:2019nin}. However, explicit one-loop results are still case-by-case with the exception for the $\langle 22pp\rangle$ family in \cite{Alday:2019nin}. At two loops and higher, the situation is more difficult. The strategy at one loop, which is based on the AdS unitarity method \cite{Aharony:2016dwx}, now requires the additional input of multi-trace operators. Such information is not yet available in the literature.\footnote{For example, at two loops there are exchange contributions from triple-trace operators. These can be in principle extracted from tree-level five-point functions. However, only five-point functions of the form $\langle pp222\rangle$ have been computed \cite{Goncalves:2019znr,Goncalves:2023oyx} while extracting the data requires all $\langle pqr22\rangle$ five-point functions.} Therefore, one can in principle only compute a part of the correlator that corresponds to the iterated s-channel cuts in flat space \cite{Bissi:2020wtv,Bissi:2020woe}. However, it turns out that this difficulty can be overcome at two loops by formulating an ansatz that is structured by an observed extra hidden symmetry in the leading Lorentzian singularities, together with additional physical constraints such as the behavior in the flat-space limit \cite{Huang:2021xws}. In this way, the four-point two-loop correlator of stress tensor multiplets has also been bootstrapped \cite{Huang:2021xws,Drummond:2022dxw}.

In this paper, we continue to explore the loop-level calculation of holographic correlators. However, instead of considering correlators of super gravitons, we will focus on super gluons of SYM in AdS. More precisely, we consider a decoupling sector of certain 4d $\mathcal{N}=2$ SCFTs in the holographic limit. These SCFTs can be engineered by using either a stack of $N$ D3-branes probing F-theory singularities \cite{Fayyazuddin:1998fb,Aharony:1998xz} or  D3-branes with probe D7-branes \cite{Karch:2002sh}. The near horizon geometries in both cases include an $AdS_5\times S^3$ subspace which hosts localized degrees of freedom corresponding to the gluons. In the limit of $N\to\infty$, the gluon degrees of freedom effectively decouple from the graviton degrees of freedom living in the full 10d bulk via $1/N$ suppressions in the vertices \cite{Alday:2021odx}. The resulting physics in 8d is the same regardless of the model we choose. Strictly speaking, the decoupling happens only at the leading order and correlators at subleading orders include gravity contributions as well. However, in this paper we will choose to turn off gravity to all orders in $1/N$ and our goal is to compute the super gluon four-point correlators in this SYM theory in $AdS_5\times S^3$ to two loops. 

The motivations for considering super gluon correlators in such a setup are two fold. First, as we already mentioned, holographic correlators are on-shell scattering amplitudes in AdS. It is natural to wonder if various remarkable properties of flat-space amplitudes admit generalizations in curved backgrounds. In particular, does the double copy relation \cite{Bern:2010ue}, which famously states gravity is the ``square'' of YM, still holds in AdS? To this end, it makes sense to decouple gravity and study the amplitudes of just SYM in AdS. In fact, analysis of this model at tree level already showed evidence for such a generalization at four points \cite{Zhou:2021gnu}. Here we will compute the loop corrections of the super gluon four-point functions which will serve as the starting point for exploring further generalizations of double copy at higher genus. Second, the super gluon case also provides a useful playground for acquiring deeper understandings of various results from the supergravity setup. The position space method for computing loop-level correlators so far have only been tested in $AdS_5\times S^5$ and it is {\it a priori} unclear whether it can be applied to other backgrounds. In this paper, we will show that such a method can be successfully applied to $AdS_5\times S^3$ and leads to similar results to the supergravity case. In the process, we also provide a nontrivial consistency check of the one-loop result which was previously obtained in \cite{Alday:2021ajh} using Mellin space techniques. Moreover, the various different color structures allow us to have a more refined understanding of the dynamical structures of the correlators which are similar in the two cases, whereas in the supergravity case all structures are mixed up due to the absence of colors. 

Let us briefly outline our strategy and the key results of the paper. Our approach is similar to that of \cite{Aprile:2017bgs,Aprile:2019rep,Huang:2021xws}. We first make an ansatz in position space which requires a set of building block functions. Due to the similarity with the supergravity case at tree level, we assume that  single-valued multiple polylogarithms (SVMPLs) continue to be a good basis for the super gluon correlators at one and two loops. In other words, the correlators are assumed to be linear combinations of SVMPLs with rational functions of the cross ratios as coefficients. However, this turns out to be a bit too general. In the supergravity case, the existence of a tree-level 10d superconformal symmetry \cite{Caron-Huot:2018kta} highlights a special eighth-order differential operator $\Delta^{(8)}$ which relates the correlators of the top and bottom components of the super graviton multiplet. By unitarity this symmetry extends to the leading part of the Lorentzian singularities at arbitrary loops. Using this operator at loop levels, the supergravity correlators can be more succinctly written in terms of the pre-correlators  $\mathcal{L}$ \cite{Aprile:2019rep,Huang:2021xws,Drummond:2022dxw}
\begin{subequations}
\begin{align}
\mathcal{H}_{\rm sugra}^{\text{1-loop}}={}&\Delta^{(8)}\mathcal{L}_{\rm sugra}^{\text{1-loop}}+\frac{1}{4}\mathcal{H}_{\rm sugra}^{\text{tree}}\;,\\
\mathcal{H}_{\rm sugra}^{\text{2-loop}}={}&\left[\Delta^{(8)}\right]^2\mathcal{L}_{\rm sugra}^{\text{2-loop}}+\frac{5}{4}\mathcal{H}_{\rm sugra}^{\text{1-loop}}-\frac{1}{16}\mathcal{H}_{\rm sugra}^{\text{tree}}\;,
\end{align}
\end{subequations}
together with additional lower-order correlators. A similar 8d superconformal symmetry also appears in the tree-level super gluon correlators \cite{Alday:2021odx} and the role of $\Delta^{(8)}$ is replaced by a fourth-order operator $\Delta^{(4)}$. In analogy with the supergravity case, we assume that similar pre-correlators can also be defined for super gluons
\begin{subequations}
\begin{align}
\mathcal{H}_{\rm SYM}^{\text{1-loop}}={}&\Delta^{(4)}\mathcal{L}_{\rm SYM}^{\text{1-loop}}+\bar{\mathcal{H}}_{\rm SYM}^{\text{tree}}\;,\\
\mathcal{H}_{\rm SYM}^{\text{2-loop}}={}&\left[\Delta^{(4)}\right]^2\mathcal{L}_{\rm SYM}^{\text{2-loop}}+\widetilde{\mathcal{H}}_{\rm SYM}^{\text{1-loop}}+\widetilde{\mathcal{H}}_{\rm SYM}^{\text{tree}}\;,
\end{align}
\end{subequations}
where $\bar{\mathcal{H}}_{\rm SYM}^{\text{tree}}$ and $\widetilde{\mathcal{H}}_{\rm SYM}^{\text{tree}}$ are ``tree-like'' correlators and $\widetilde{\mathcal{H}}_{\rm SYM}^{\text{1-loop}}$ is a ``one-loop-like'' correlator. We will be more precise about the meaning of ``tree-like'' and ``one-loop-like''. But for the moment it suffices to say they are characterized by the transcendental degrees of SVMPLs expected at each loop order. Then the position space ansatz in terms of SVMPLs is formulated in terms of the pre-correlators $\mathcal{L}$ and the lower-order objects $\mathcal{H}_i$, in parallel with  the supergravity story. Note that, unlike supergravity, super gluon correlators have different color structures. Therefore, we make such an ansatz for each independent color structure and assume the correlator to be a linear combination of all these structures. To perform the bootstrap, we impose a number of consistency conditions. These are 
\begin{itemize}
\item Leading logarithmic singularities
\item Crossing symmetry
\end{itemize}
together with a few other constraints. Here the leading logarithmic singularities rely only on the tree-level data and can be computed at any loop order. At two loops, the additional constraints further include comparison with the scattering amplitude in a proper flat-space limit that can be computed independently using flat-space techniques, and the data of twist-4 operators which can be extracted from the tree and one-loop correlators. Imposing these constraints, we find that all parameters in the ansatz are fixed except for those corresponding to the counterterms needed for the UV divergences. Moreover, the tree-like and one-loop-like terms turn out to be exactly the tree-level and one-loop correlators except for simple replacements for the color structures. 

The rest of the paper is organized as follows. We review in Section \ref{Sec:preliminaries} some preliminaries of super gluon four-point functions which include the superconformal kinematics, color structure and superconformal block decomposition. In Section  \ref{Sec:unitarityrecursion} we review how the leading logarithmic singularities can be constructed from the tree-level data and compute them in closed forms using hidden conformal symmetry. In Section  \ref{sec:oneloop} we introduce the position space method and demonstrate it by bootstrapping the one-loop correlator. In Section  \ref{sec:twoloops} we apply the method to the two-loop correlator and obtain the full answer by imposing constraints. In Section  \ref{Sec:discussions} we outline a few future directions. The paper also has several appendices where we include further technical details. In Appendix \ref{sec:mpls} we give a brief review of the properties of SVMPLs. Appendix \ref{sec:H2analytic} contains the complete analytic result for the reduced correlator at one loop. The details of the flat-space two-loop amplitude are presented in Appendix \ref{sec:flatspacelimit}. In Appendix \ref{sec:recursion} we discuss the computations related to the twist-4 data.

\section{Preliminaries}\label{Sec:preliminaries}

In this paper, we consider holographic correlators corresponding to super gluon scattering in AdS. To be concrete, we consider SYM in $AdS_5\times S^3$ which arises as a decoupling sector of certain 4d $\mathcal{N}=2$ SCFTs. One can construct these SCFTs from a stack of $N$ D3-branes, by either using them to probe F-theory singularities \cite{Fayyazuddin:1998fb,Aharony:1998xz} or  by adding a few probe D7-branes \cite{Karch:2002sh}. In either case, in the near horizon limit there is an $AdS_5\times S^3$ subspace in the total ten dimensional spacetime which is locally $AdS_5\times S^5$. On this subspace there are localized degrees of freedom transforming as an $\mathcal{N}=1$ vector multiplet and in the adjoint representation of certain flavor group $G_F$ of the boundary CFT. Here $G_F$ depends on the theory and is a gauge group from the bulk perspective.\footnote{Therefore, in the following we will use ``flavor'', ``color'' and ``gauge'' interchangeably.} Since the $\mathcal{N}=1$ vector multiplet contains fields with Lorentz spin at most 1, its KK reduction with respect to $S^3$ also leads to fields with the same maximal spin. These are the massless and massive AdS gluons and their super partners. Because of the bound on spins all the KK modes have to reside in $\frac{1}{2}$-BPS multiplets by the 4d $\mathcal{N}=2$ representation theory. Their conformal dimensions are fully fixed by R-symmetry and therefore are independent of the bulk theory. More precisely, the superconformal primaries of these $\frac{1}{2}$-BPS are scalar operators $\mathcal{O}_k$ labelled by an integer $k=2,3,\ldots$. They have conformal dimension $\Delta=k$ and transform in the spin-$\frac{k}{2}$ representation of the $SU(2)_R$ R-symmetry group. Moreover, they transform in the adjoint representaiton of the flavor group $G_F$. We will call the fields dual to these superprimaries the super gluons. The real spinning gluon fields are superconformal descendants in the multiplets. 

By contrast, the super gravitons and their super partners live in the full ten dimensional spacetime. Unlike the supergluons, their KK spectrum depends on the specific theory. However, an interesting fact of all these 4d $\mathcal{N}=2$ SCFTs is that there is a hierarchy in the couplings at large $N$. For example, the cubic coupling of three super gluons (or their superconformal descendants) is of order $1/\sqrt{N}$, while the coupling involving two super gluons and one super graviton is of order $1/N$. Therefore, for large $N$ the leading contribution to the super gluon correlators comes only from the 8d SYM. Subleading corrections in $1/N$ will in general contain graviton contributions as well. 

As mentioned in the introduction, we will continue to study loop corrections of the four-point correlator of the super gluon operator $\mathcal{O}_2$ in the $AdS_5\times S^3$ SYM. Although this does not give the full answer for this correlator in $\mathcal{N}=2$ SCFTs, it makes sense from the perspective of exploring curved space generalizations of gauge theory amplitudes. This section serves to provide some preliminary features of this correlator, which will be used in our bootstrap computation.

\subsection{Four-point correlators}

With all indices restored, the super gluon operator $\mathcal{O}_2$ has the form
\begin{equation}
\mathcal{O}^{I;a_1a_2}_2(x)\;,
\end{equation}
where $I=1,\ldots,\operatorname{dim}(G_F)$ is the flavor symmetry index and $a_i=1,2$ are the $SU(2)_R$ R-symmetry indices. It is convenient to contract the R-symmetry indices with two-dimensional polarization spinors 
\begin{equation}
\mathcal{O}^I_2(x;v)=\mathcal{O}^{I;a_1a_2}_2 v_{a_1}v_{a_2}\;.
\end{equation}
The four-point function
\begin{equation}
    G^{I_1 I_2 I_3 I_4}(x_i;v_i)=\langle \mathcal{O}^{I_1}_2(x_1;v_1) \mathcal{O}^{I_2}_2(x_2;v_2) \mathcal{O}^{I_3}_2(x_3;v_3) \mathcal{O}^{I_4}_2(x_4;v_4) \rangle
\end{equation}
is therefore a function of both the spacetime coordinates $x_i$ and the internal space spinors $v_i$. Exploiting the bosonic symmetries, i.e.~conformal symmetry and R-symmetry, we can write the correlator as a function of the cross ratios
\begin{equation}\label{4pointcorrelator}
G^{I_1 I_2 I_3 I_4}=\frac{\left(v_1 \cdot v_2\right)^2\left(v_3 \cdot v_4\right)^2}{x_{12}^{4} x_{34}^{4}} \mathcal{G}^{I_1 I_2 I_3 I_4}(u, v ; \alpha)
\end{equation}
where $x_{ij}=x_i-x_j$, $v_i\cdot v_j=v^a_i v^b_j \epsilon_{ab}$ ($\epsilon_{ab}$ being the 2d Levi--Civita symbol), and the cross ratios are 
\begin{equation}
u=\frac{x_{12}^2 x_{34}^2}{x_{13}^2 x_{24}^2}=z \bar  z\;,\quad v=\frac{x_{23}^2 x_{14}^2}{x_{13}^2 x_{24}^2}=(1-z)(1- \bar  z)\;,\quad \alpha=\frac{\left(v_1 \cdot v_3\right)\left(v_2 \cdot v_4\right)}{\left(v_1 \cdot v_2\right)\left(v_3 \cdot v_4\right)}\;.
\end{equation}
In addition, the ferminonic generators in the superconformal algebra impose further constraints known as the superconformal Ward identities \cite{Nirschl:2004pa}
\begin{equation}
\left.\left(x \partial_x-\alpha \partial_\alpha\right) \mathcal{G}^{I_1 I_2 I_3 I_4}(z, \bar{z} ; \alpha)\right|_{\alpha=1 / x}=0\;,\quad x=z\text{ or }\zb.
\end{equation}
Solving these identities, we can decompose the correlator into two parts
\begin{equation}\label{solSCFWI}
\mathcal{G}^{I_1 I_2 I_3 I_4}(z,\bar{z};\alpha)=\mathcal{G}_{0}^{I_1 I_2 I_3 I_4}(z,\bar{z};\alpha)+\mathcal{R} \mathcal{H}^{I_1 I_2 I_3 I_4}(z,\bar{z})\;,
\end{equation}
where 
\begin{equation}
\mathcal{R} =\frac{(1-z\alpha)(1-\bar z  \alpha)}{z \zb}\;.
\end{equation} 
Note that our definiton of $\mathcal{R}$ is different from that of \cite{Alday:2021odx} by a $z\zb$ in the denominator. The first term $\mathcal{G}_{0}^{I_1 I_2 I_3 I_4}$ is protected, while dynamical information of the correlator is encoded in the reduced correlator $\mathcal{H}^{I_1 I_2 I_3 I_4}$. In our case of $\mathcal{O}_2$ correlator, $\mathcal{H}^{I_1 I_2 I_3 I_4}$ is simply a function of the spacetime cross ratios $\{z,\bar{z}\}$ (or equivalently $\{u,v\}$) and is free of the R-symmetry cross ratio $\alpha$.

We study the expansion of the correlator with respect to the large flavor central charge $C_{\mathcal{J}}$\footnote{The flavor central charge $C_{\mathcal{J}}$ appears in the flavor current two-point functions as $\langle \mathcal{J}_\mu^I(x)\mathcal{J}_\nu^J(0)\rangle=\frac{C_{\mathcal{J}}}{2\pi^2}\frac{\delta^{IJ}(\delta_{\mu\nu}-2\frac{x_\mu x_\nu}{x^2})}{x^6}$. Moreover, via supersymmetry, it is related to the three-point function coefficient $C_{2,2,2}^2$ of $\langle \mathcal{O}_{2}\mathcal{O}_{2}\mathcal{O}_{2}\rangle$ by $\mathcal{C}_{\mathcal{J}}=\frac{1}{6}C_{2,2,2}^2$.}. For convenience, we use the small parameter $a_F=6/C_{\mathcal{J}}$, with respect to which the expansion reads
\begin{equation}\label{eq:aFexpansion}
\mathcal{H}_{2222}^{I_1 I_2 I_3 I_4}\equiv\mathcal{H}= \mathcal{H}^{(0)}+a_F\, \mathcal{H}^{(1)}+a^{2}_F\, \mathcal{H}^{(2)}+ a^{3}_F\, \mathcal{H}^{(3)}+\cdots\;.
\end{equation}
This expansion has a nice interpretation from the bulk point of view. The leading contribution $\mathcal{H}^{(0)}$ is associated with the disconnected part of scattering in AdS and can be evaluated by generalized free field theory. The first correction $\mathcal{H}^{(1)}$ is the tree-level scattering of the super gluons, which has been obtained in \cite{Alday:2021odx}. The higher-order correction $\mathcal{H}^{(L+1)}$ corresponds to scattering at $L$ loops, where the one-loop case has been computed in \cite{Alday:2021ajh} using Mellin space techniques.

\subsection{Projectors and color decomposition}\label{projection method}

Since we are studying gluon scattering, as usual the correlator $\mathcal{H}^{(L+1)}$ at each perturbative order splits into various color factors and their corresponding dynamical factors \footnote{In the context of scattering amplitudes these coefficients of color factors are more frequently called kinematic factors (when referring to numerators in Feynman diagrams). In this paper we call them dynamical factors to remind the readers that they contains the dynamical information of the theory.}
\begin{equation}\label{eq:CIIIIexpansion}
    \qty(\mathcal{H}^{(L+1)})^{I_1I_2I_3I_4}=\sum_CC^{I_1I_2I_3I_4}\mathcal{H}_C^{(L+1)}.
\end{equation}
A color factor $C^{I_1I_2I_3I_4}$ is constructed out of the structure constants $f^{IJK}$ of the gauge group according to the topology of a diagram that may arise at the given loop order according to Feynman rules (or Witten rules in AdS), and so the summation above carries over all possible topologies at $L$ loops. The dynamical factors $\mathcal{H}_C^{(L+1)}$ are functions of kinematic variables $\{z,\bar{z}\}$, and with the above decomposition they only rely on diagram topologies as well, regardless of any specific choice of the gauge group $G_F$.

The decomposition \eqref{eq:CIIIIexpansion} is not the most convenient for practical computations as the color factors $C^{I_1I_2I_3I_4}$ are highly redundant. So instead one often seeks for other types of color decompositions. Because our computation requires the input from CFT data of the spectrum and the coefficients arising in OPEs, it is preferable to decompose the color factors in a way that resembles the conformal block expansion. This can be fulfilled by specifying a particular channel (say the s-channel) and introduce an operation $P_{\mathbf{a}}^{I_1I_2\vert I_3I_4}$ that picks out irreducible represetation $\mathbf{a}$ of the flavor group from the tensor products of two adjoints $\mathbf{adj}\otimes\mathbf{adj}$. This is called an s-channel projector, and by definition it satisfies the symmetry properties
\begin{equation}\label{eq:Pparity}
    P^{I_1 I_2 \vert I_3 I_4}_{\mathbf{a}} = (-1)^{\abs{R_{\mathbf{a}}}} P^{I_2 I_1 \vert I_3 I_4}_{\mathbf{a}},\qquad P^{I_1 I_2 \vert I_3 I_4}_{\mathbf{a}} = P^{I_3 I_4 \vert I_1 I_2}_{\mathbf{a}},
\end{equation}
where $\abs{R_{\mathbf{a}}}$ stands for the parity of representation $\mathbf{a}$, and the idempotency condition
\begin{equation}\label{eq:Pidempotency}
 P^{I_1 I_2 \vert I_5 I_6}_{\mathbf{a}} P^{I_6 I_5 \vert I_3 I_4}_{\mathbf{b}} = \delta_{{\mathbf{a}}{\mathbf{b}}}P^{I_1 I_2 \vert I_3 I_4}_{\mathbf{a}}.
\end{equation}
In particular from \eqref{eq:Pidempotency} we also have $P^{I_1 I_2 \vert I_3 I_4}_{\mathbf{a}} P^{I_1 I_2 \vert I_3 I_4}_{\mathbf{b}} = \delta_{{\mathbf{a}}{\mathbf{b}}}{\rm dim}(R_{\mathbf{a}})$. Therefore every color factor appearing in \eqref{eq:CIIIIexpansion} receives a \emph{unique} decomposition onto the s-channel projectors
\begin{equation}\label{eq:cons}
    C^{I_1 I_2 I_3 I_4} = \sum_{\mathbf{a}\in \mathbf{adj}\otimes\mathbf{adj}} P^{I_1 I_2 \vert I_3 I_4}_{\mathbf{a}} C_{\mathbf{a}}\;,
\end{equation}
with coefficients $C_{\mathbf{a}}$, or equivalently
\begin{equation}
   C_{\mathbf{a}} = \text{dim}(R_{\mathbf{a}})^{-1} P^{I_1 I_2 \vert I_3 I_4}_{\mathbf{a}} C^{I_1 I_2 I_3 I_4}\;.
\end{equation}

The efficiency of these projectors comes from the fact that the set of irreducible representations arising in $\mathbf{adj}\otimes\mathbf{adj}$ depends only on the gauge group $G_F$ but not on the perturbative order. As a result, the color decomposition of the reduced correlator $\mathcal{H}$ as well as any term $\mathcal{H}^{(L+1)}$ in the expansion \eqref{eq:aFexpansion} can be carried out in a uniform manner. Generically, we have
\begin{equation}\label{eq:PHdecomposition}
    \mathcal{H}^{I_1I_2I_3I_4}=\sum_{\mathbf{a}\in\mathbf{adj}\otimes\mathbf{adj}}P_{\mathbf{a}}^{I_1I_2\vert I_3I_4}\mathcal{H}_{\mathbf{a}}\;,
\end{equation}
and $\mathcal{H}^{(L+1)}$ follows similarly. Furthermore, the idempotency condition \eqref{eq:Pidempotency} also makes the recursive relation between different loop levels very simple, as will be further illustrated in the next section.

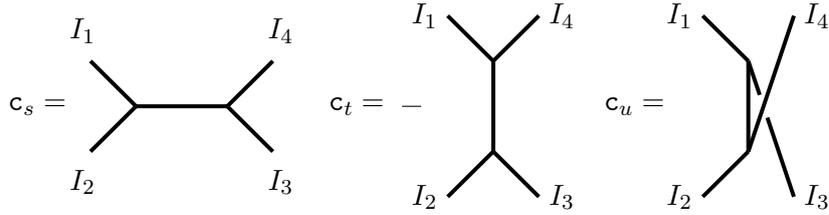
\begin{figure}[h]
    \centering
   \begin{tikzpicture}
        \centering
        \draw [line width=1.5pt] (-4.7,0) -- (-5.3,0.6);
        \draw [line width=1.5pt](-4.7,0) -- (-5.3,-0.6);
        \draw [line width=1.5pt](-4.7,0) -- (-3.5,0);
        \draw [line width=1.5pt](-3.5,0) -- (-2.9,-0.6);
        \draw [line width=1.5pt](-3.5,0) -- (-2.9,0.6);

        \node [anchor=east] at (-5.1,1) {$I_1$};
        \node [anchor=west] at (-3.1,1) {$I_4$};
        \node [anchor=west] at (-3.1,-1) {$I_3$};
        \node [anchor=east] at (-5.1,-1) {$I_2$};
        \node at (-6,0) {${\mathtt{c}}_{s} = $};

        \draw [line width=1.5pt] (0,0.6) -- (-0.6,1.2);
        \draw [line width=1.5pt](0,0.6) -- (0.6,1.2);
        \draw [line width=1.5pt](0,0.6) -- (0,-0.6);
        \draw [line width=1.5pt] (0,-0.6) -- (-0.6,-1.2);
        \draw [line width=1.5pt] (0,-0.6) -- (0.6,-1.2);
        \draw   (-1.2,0) -- (-0.95,0);
        \node at (-1.8,0) {${\mathtt{c}}_{t} =$};
        \node [anchor=east] at (-0.6,1.2) {$I_1$};
        \node [anchor=west] at (0.6,1.2) {$I_4$};
        \node [anchor=west] at (0.6,-1.2) {$I_3$};
        \node [anchor=east] at (-0.6,-1.2) {$I_2$};
    \end{tikzpicture}
    \begin{tikzpicture}
        \draw [line width=1.5pt] (0,0.6) -- (-0.6,1.2);
        \draw [line width=1.5pt](0,0.6) -- (0.6,-1.2);
        \draw [line width=1.5pt](0,0.6) -- (0,-0.6);
        \draw [line width=1.5pt] (0,-0.6) -- (-0.6,-1.2);
        \draw [draw=white,line width=2pt] (0.15,0.15) -- (0.25,-0.15);
        \draw [line width=1.5pt] (0,-0.6) -- (0.6,1.2);
        \node at (-1.5,0) {${\mathtt{c}}_{u} =$};
        \node [anchor=east] at (-0.6,1.2) {$I_1$};
        \node [anchor=west] at (0.6,1.2) {$I_4$};
        \node [anchor=west] at (0.6,-1.2) {$I_3$};
        \node [anchor=east] at (-0.6,-1.2) {$I_2$};
    \end{tikzpicture}
    \caption{Tree color structures $\mathtt{c}_s$, $\mathtt{c}_t$ and $\mathtt{c}_u$.}
    \label{fig:fourpttree}
\end{figure}

As a simple example for the use of projectors, let us quickly review the tree-level correlator $\mathcal{H}^{(1)}$, which was computed in \cite{Alday:2021odx}. Its takes the following form 
\begin{equation}\label{eq:tree1}
    \mathcal{H}^{(1)} = \mathtt{c}_s\mathcal{H}^{(1)}_{s}  +\mathtt{c}_t\mathcal{H}^{(1)}_{t}  +\mathtt{c}_u\mathcal{H}^{(1)}_{u} ,
\end{equation}
with
\begin{subequations}\label{eq:tree1stu}
    \begin{align}
        \mathcal{H}^{(1)}_{s} &=  \frac{u^3}{3} \left(2\partial_u + (1+v)\partial_u \partial_v + u\partial_u^2 \right) \bar{D}_{1111},\\
        \mathcal{H}^{(1)}_{t} &= -\frac{u^3}{3} \left(2\partial_v + v\partial_v^2 + (1+u)\partial_u\partial_v \right) \bar{D}_{1111},\\
        \mathcal{H}^{(1)}_{u} &= \frac{u^3}{3} \left(2\partial_v + v\partial_v^2 -2\partial_u + (u-v)\partial_u\partial_v - u\partial_u^2 \right) \bar{D}_{1111}\;.
    \end{align}
\end{subequations}
Here $\bar{D}_{1111}$ is an example of  the $\bar{D}$-functions which are contact Witten diagrams in AdS \footnote{For a review of the precise definition and general properties of $\bar{D}$-functions, see Appendix C of \cite{Bissi:2022mrs}.}
\begin{equation}\label{eq:barD1111}
    \bar{D}_{1111}(z,\bar{z})=\frac{1}{z-\bar{z}}\qty(2\mathrm{Li}_2(z)-2\mathrm{Li}_2(\bar{z})+\log(z\bar{z})\log\qty(\frac{1-z}{1-\bar{z}}))\;.
 \end{equation}
$\mathtt{c}_{s/t/u}$ are color factors built from structure constants
\begin{equation}
    (\mathtt{c}_s)^{I_1 I_2 I_3 I_4} = f^{I_1 I_2 J} f^{J\, I_3 I_4},\quad (\mathtt{c}_t)^{I_1 I_2 I_3 I_4} = f^{I_1 I_4 J} f^{J\, I_2 I_3},\quad (\mathtt{c}_u)^{I_1 I_2 I_3 I_4} = f^{I_1 I_3 J} f^{J\, I_4 I_2},
\end{equation}
which are diagrammatically depicted in  Figure \ref{fig:fourpttree}. Note again in this decomposition the kinematic factors $\mathcal{H}_{s/t/u}^{(1)}$ are independent of the gauge group $G_F$.

When decomposing using the projectors, let us assume that we are working with the gauge group $E_8$. In this case $\mathbf{adj}\otimes\mathbf{adj}$ includes altogether five irreducible representations $\bf{1}$, $\bf{3875}$, $\bf{27000}$, $\bf{248}\,({\mathbf{adj}})$, and $\bf{30380}$. The former three representations are parity even and the latter two are paritty odd. 

Note that $(\mathtt{c}_s)^{I_1 I_2 I_3 I_4}$ already represents the exchange of the adjoint representation ${\bf 248}$ in the s-channel, it is therefore proportional to the projector $P_{\mathbf{248}}^{I_1I_2\vert I_3I_4}$, and we have 
\begin{equation}\label{eq:csdef}
    (\mathtt{c}_s)_{\mathbf{a}}\equiv P_{\mathbf{248}}^{I_1I_2\vert I_3I_4}(\mathtt{c}_s)^{I_1 I_2 I_3 I_4} = \psi^2 h^\vee (\underset{\substack{\scriptsize\uparrow\\\mathbf{1}}}{0},\underset{\substack{\scriptsize\uparrow\\\mathbf{3875}}}{0},\underset{\substack{\scriptsize\uparrow\\\mathbf{27000}}}{0},\underset{\substack{\scriptsize\uparrow\\\mathbf{248}}}{1},\underset{\substack{\scriptsize\uparrow\\\mathbf{30380}}}{0})^T,
\end{equation}
where $h^\vee$ is the dual Coxeter number, $\psi^2$ is the length squared of the longest root. By contrast the decomposition of $\mathtt{c}_t$, $\mathtt{c}_u$ involves a mixture of different s-channel projectors
\begin{subequations}\label{eq:ctcu}
\begin{align}
    (\mathtt{c}_t)_{\mathbf{a}} &= -\psi^2 h^\vee \left(1,\frac{1}{5},-\frac{1}{30},\frac{1}{2},0 \right)^T,\quad 
    (\mathtt{c}_u)_{\mathbf{a}} = \psi^2 h^\vee \left(1,\frac{1}{5},-\frac{1}{30},-\frac{1}{2},0 \right)^T.
\end{align}
\end{subequations}
One easily sees that the Jacobi identity $(\mathtt{c}_s)_{\mathbf{a}}+ (\mathtt{c}_t)_{\mathbf{a}}+ (\mathtt{c}_u)_{\mathbf{a}} = 0$ is satisfied. Consequently the coefficients in the projector decomposition of the whole tree-level correlator $\mathcal{H}^{(1)}$ are
\begin{subequations}\label{eq:tree2}
    \begin{align}
        \mathcal{H}^{(1)}_{\bf{1}} =& \psi^2 h^\vee \left(-\mathcal{H}^{(1)}_{t} + \mathcal{H}^{(1)}_{u}\right),\\
        \mathcal{H}^{(1)}_{\bf{3875}} =&  \frac{\psi^2 h^\vee}{5}\left(-\mathcal{H}^{(1)}_{t} + \mathcal{H}^{(1)}_{u}\right),\\
        \mathcal{H}^{(1)}_{\bf{27000}} =& -\frac{\psi^2 h^\vee}{30} \left(-\mathcal{H}^{(1)}_{t} + \mathcal{H}^{(1)}_{u}\right),\\
        \mathcal{H}^{(1)}_{\bf{248}} =& \frac{\psi^2 h^\vee}{2} \left(2\mathcal{H}^{(1)}_{s} - \mathcal{H}^{(1)}_{t} - \mathcal{H}^{(1)}_{u}\right),\\
        \mathcal{H}^{(1)}_{\bf{30380}} =& 0.
    \end{align}
\end{subequations}
Quite remarkably, at this specific level the coefficients of projectors with equal parity are in fact the same up to some overall constant factors, as was observed in a more general setup in \cite{Alday:2021odx}. 

\subsection{Spectrum and conformal block decomposition}

As mentioned before our computation partly relies on the existing data of operators obtained from lower loops, so it is helpful to have a quick look at the structure of OPE and the related block expansion. Thanks to the 4d $\mathcal{N}=2$ superconformal symmetry, the correlator $\mathcal{G}^{I_1I_2I_3I_4}$ admits a decomposition into superconformal blocks in correspondence to the exchanges of different superconformal multiplets in the four-point function. The relevant sueprmultiplets are listed in Table \ref{multiplet} and a complete classification can be found in \cite{Nirschl:2004pa}. The OPE of two $\frac{1}{2}$-BPS multipelts $\mathcal{B}_1$ contains the following supermultiplets
\begin{equation}
{\mathcal{B}}_{1} \otimes {\mathcal{B}}_{1} \simeq \bigoplus_{p=0}^{2} {\mathcal{B}}_p \oplus  \bigoplus_{\ell \geq 0}\left(\bigoplus_{p=0}^{1} {\mathcal{C}}_{p,\left(\frac{\ell}{2}, \frac{\ell}{2}\right)} \bigoplus_{\Delta} \mathcal{A}_{0,\left(\frac{\ell}{2}, \frac{\ell}{2}\right)}^{\Delta}\right)\;.
\end{equation}
Here $\mathcal{B}_p$ and $\mathcal{C}_{p,(\frac{\ell}{2},\frac{\ell}{2})}$ are protected multiplets and their twists $\tau=\Delta-\ell$ are bounded from above by the allowed R-symmetry charges. In contrast, there is no upper bound on the twists of the long multiplets $\mathcal{A}^\Delta_{0,(\frac{\ell}{2},\frac{\ell}{2})}$ and their dimensions are not protected. Instead, they have a lower bound in the holographic limit as they are double-trace (and more generally multi-trace) operators formed by single-trace operators.\footnote{This bound is stronger than the unitarity bound in Table \ref{multiplet}.} Let us also note that the superprimaries of the long multiplets are only allowed to be R-symmetry singlets in order for the representations of the entire multiplet to fit into the four-point function. The long multiplets play a key role in the paper as the loop corrections correspond to precisely the contribution of these multipelts.
\begin{table}[h]
\centering
\begin{tabular}{c|c|c|c}
\hline  Multiplet & Label & SU(2)$_R$ & Dimension $\Delta$ and spin $\ell$ \\
\hline Half-BPS &$\mathcal{B}_{R}$ & {$R$} & $\Delta=2 R $, $\ell=0$ \\
Semi-short &$\mathcal{C}_{R,(\ell/2,\ell/2)}$ & {$R$} & $\Delta=2+2R+\ell$ \\
Long &  $\mathcal{A}^{\Delta}_{R,(\ell/2,\ell/2)}$ & ${ {R} }$ & $\Delta\geq2+2R+\ell$ \\
\hline
\end{tabular}
\caption{Supermultiplets that appear in the fusion rules of two $\mathcal{B}$'s for $\mathcal{N}=2$ SCFTs.}
\label{multiplet}
\end{table}

We will focus on the reduced correlator $\mathcal{H}$ which has already taken superconformal symmetry into account. In this way the superconformal block decomposition simply reduces to just the ordinary conformal block decomposition. As superconformal symmetry and gauge symmetry commute, this directly passes through the color projector decomposition, and in terms of each component in \eqref{eq:PHdecomposition} this reads \cite{Nirschl:2004pa} 
\begin{equation}\label{eq:blockexpand}
\mathcal{H}_{\mathbf{a}}(z, \bar{z})=\sum_{\tau_{\mathbf{a}}, \ell} a_{\mathbf{a}} g_{\tau_{\mathbf{a}}+2, \ell}(z, \bar{z})\;,
\end{equation}
where $\tau_{\mathbf{a}}$ and $\ell$ sum over the spectrum of the supermultiplets. Note that the shift in $\tau$ by 2 in the ordinary conformal block $g_{\tau+2,\ell}$ is a consequence of the superconformal symmetry. The detailed expression of these blocks is \cite{Dolan:2000ut}
\begin{equation}\label{eq:defblock}
g_{\tau, \ell}=\frac{z \bar{z}}{\bar{z}-z}\left(k_{\frac{\tau-2}{2}}(z) k_{\frac{\tau+2 \ell}{2}}(\zb)-k_{\frac{\tau-2}{2}}(\zb) k_{\frac{\tau+2 \ell}{2}}(z)\right) ,\quad k_h(z)=z^h\  _2 F_1(h, h, 2 h, z) .
\end{equation}

Since the long multiplets are not protected, in the limit of $N\to\infty$ their twists as well as OPE coefficients receive perturbative corrections with respect to small $a_F$ 
\begin{subequations}
    \begin{align}\label{tauandaexpansion}
        \tau_{\mathbf{a}} =& \tau_0 + a_F \gamma^{(1)}_{\mathbf{a}} + a_F^2 \gamma^{(2)}_{\mathbf{a}} + \dots\;,\\
        a_{\mathbf{a}}(\tau,\ell) =& a^{(0)}_{\mathbf{a}} + a_F\, a^{(1)}_{\mathbf{a}} + a_F^2\, a^{(2)}_{\mathbf{a}} + \dots\;.
    \end{align}
\end{subequations}
Substituting the above expansion into \eqref{eq:blockexpand} gives the following series expansion for $\mathcal{H}_{\mathbf{a}}$
\begin{equation}\label{eq:blockexp}
    \begin{split}
        \mathcal{H}_{\mathbf{a}}=&\underbrace{\sum_{\tau_0,\ell} a^{(0)}_{\mathbf{a}}  g_{\tau_0+2,\ell}(z,\zb)  }_{\mathcal{H}_{\mathbf{a}}^{(0)}}
        + a_F\underbrace{\sum_{\tau_0,\ell}\left(  a^{(0)}_{\mathbf{a}}\gamma^{(1)}_{\mathbf{a}}  \partial_{\tau_0} +  a^{(1)}_{\mathbf{a}}  \right)   g_{\tau_0 + 2,\ell}(z,\zb)}_{\mathcal{H}_{\mathbf{a}}^{(1)}}\\
        &+a_F^2 \underbrace{\sum_{\tau_0,\ell}\left( \frac{1}{2} a^{(0)}_{\mathbf{a}}(\gamma^{(1)}_{\mathbf{a}}) ^2  \partial_{\tau_0}^2  + (a^{(1)}_{\mathbf{a}}\gamma^{(1)}_{\mathbf{a}} + a^{(0)}_{\mathbf{a}}\gamma^{(2)}_{\mathbf{a}})  \partial_{\tau_0} +  a^{(2)}_{\mathbf{a}}  \right)g_{\tau_0+2,\ell}(z,\zb)}_{\mathcal{H}_{\mathbf{a}}^{(2)}} +\dots  .
    \end{split}
\end{equation}
The first term $\mathcal{H}_{\mathbf{a}}^{(0)}$ receives contributions only from long operators whose $a_{\mathbf{a}}^{(0)}$ are non-vanishing. From large $N$ factorization, $\mathcal{H}_{\mathbf{a}}^{(0)}$ is given by the disconnected correlator and these contributing operators can only be double-trace operators. However, these operators are degenerate at the classical level. For instance, among the double-trace operators
\begin{equation}\label{eq:degeneracy2n}
    : \mathcal{O}_2 \square^{n-2} \partial^\ell \mathcal{O}_2 :\ ,\ : \mathcal{O}_3 \square^{n-3} \partial^\ell \mathcal{O}_3 :\ , \dots, \  : \mathcal{O}_{n}  \partial^\ell \mathcal{O}_{n}:\ 
\end{equation}
all have classical twist $\tau^{(0)}=2n$ and spin $\ell$. Consequently, each term in \eqref{eq:blockexp} should not be literally understood as the contribution from a single operator, but rather in an averaged sense. Moreover, at higher orders in $a_F$ there are also higher-trace operators appearing in the OPE \footnote{For example, triple-trace operators first appear at two loops. That higher-trace operators can only be seen at higher orders is because their coefficients in the OPE are suppressed by powers of $a_F$.}, which can have the same twist as the double-trace operators and will enter the mixing as well. Therefore, in a precise description it is necessary to use an extra label $i$ to distinguish different operators in the degeneracy. Then the coefficient $a_{\mathbf{a}}^{(0)}\gamma^{(1)}_{\mathbf{a}}$ should in fact be understood as $\langle  a^{(0)}_{\mathbf{a}}\gamma^{(1)}_{\mathbf{a}} \rangle \equiv \sum_i a^{(0)}_{i,\mathbf{a}}\gamma^{(1)}_{i,\mathbf{a}}$, and $a_{\mathbf{a}}^{(0)}\qty(\gamma^{(1)}_{\mathbf{a}})^2$ as $\langle  a^{(0)}_{\mathbf{a}}\qty(\gamma^{(1)}_{\mathbf{a}})^2 \rangle \equiv \sum_i a^{(0)}_{i,\mathbf{a}}\qty(\gamma^{(1)}_{i,\mathbf{a}})^2$, and so on.

\section{Leading logarithmic singularities}\label{Sec:unitarityrecursion}

As an analytic function of the kinematic variables $z$ and $\bar{z}$, a conformal correlator can in principle be constructed out of its singularities by dispersion-type relations, in a similar way as the dispersion relation that generates a four-point scattering amplitude from its physical channel discontinuities. For generic CFTs such relations were formulated in \cite{Carmi:2019cub}. This means that the defining data for a correlator is necessarily encoded in its singularities. While our computation does not rely on the dispersion relations, these data still provide a vital input in determining the loop-level corrections to the reduced correlator $\mathcal{H}$.

When viewed in the perturbative expansion \eqref{eq:blockexp} these singularities are sourced at small $u$ by the $\log(u)$ factors arising from the derivatives acting on the conformal block. Recall in the definition \eqref{eq:blockexp} that $g_{\tau,\ell}(z,\bar{z})\propto u^{\tau/2}$, so at each order $a_F^p$ the reduced correlator can be organized in terms of powers of $\log(u)$
\begin{equation}
    \mathcal{H}_{\mathbf{a}}^{(p)}(z,\bar{z})=\frac{1}{2^pp!}\log^p(u)\sum_{\tau_0,\ell}\langle a_{\mathbf{a}}^{(0)}(\gamma^{(1)}_{\mathbf{a}})^p\rangle\,g_{\tau_0,\ell}(z,\bar{z})+\qty(\text{terms with }\log^{k<p}(u))\;.
\end{equation}
In the above expression we explicitly write out the terms with the maximal power of $\log(u)$, which are named the leading logarithmic singularities. While they are not the only source of singularities in general, they make up the simplest and in some sense the most important contribution to the correlator. On the one hand, the apparent proportionality to $a_{\mathbf{a}}^{(0)}$ and $\gamma^{(1)}_{\mathbf{a}}$ suggests that they can be computed once the data up to tree-level $\mathcal{H}^{(1)}$ are available, which involve only double-trace operators. On the other hand, it was observed in \cite{Caron-Huot:2018kta} that the leading log singularities turn out to enjoy a well-organized structure, ruled by a conjectural hidden conformal symmetry in higher dimensions. We discuss these two points in detail in the following two subsections. In particular, the latter point provides a crucial hint to the ansatz that we are going to use in the computation.

\subsection{Recursion by unitarity}\label{sec:recursionunitarity}

By the appearance of the leading log coefficients $\langle a_{\mathbf{a}}^{(0)}(\gamma^{(1)}_{\mathbf{a}})^k\rangle$ it is very tempting to write down the following recursion relation
\begin{equation}\label{eq:naiverecursion}
    \langle a_{\mathbf{a}}^{(0)}(\gamma^{(1)}_{\mathbf{a}})^k\rangle\overset{?}{=}\frac{\langle a_{\mathbf{a}}^{(0)}(\gamma^{(1)}_{\mathbf{a}})^{k-1}\rangle\,\langle a_{\mathbf{a}}^{(0)}\gamma^{(1)}_{\mathbf{a}}\rangle}{\langle a_{\mathbf{a}}^{(0)}\rangle}\;,
\end{equation}
so that once the coefficients at the two lowest orders of this class are known, {\it i.e.}, $\langle a_{\mathbf{a}}^{(0)}\rangle$ and $\langle a_{\mathbf{a}}^{(0)}\gamma^{(1)}_{\mathbf{a}}\rangle$, the coefficients at arbitrary higher orders can be recursively determined. This is almost correct, but the validity of \eqref{eq:naiverecursion} is polluted by the existence of degeneracy described around \eqref{eq:degeneracy2n}. The solution to this problem is to unmix the degeneracy by considering a larger set of correlators $\langle ppqq\rangle\equiv\langle \mathcal{O}_p\mathcal{O}_p\mathcal{O}_q\mathcal{O}_q\rangle$ involving higher KK modes. All the operators with the same classical twist $\tau^{0}=2n$ in \eqref{eq:degeneracy2n} may enter the decomposition of every $\langle ppqq\rangle$ where $2\leq p,q\leq n$. We select all such correlators and label the corresponding data in each by $\langle a_{\mathbf{a}}^{(0)}\rangle_{ppqq}$, $\langle a_{\mathbf{a}}^{(0)}\gamma^{(1)}_{\mathbf{a}}\rangle_{ppqq}$, etc. Note that $\langle a_{\mathbf{a}}^{(0)}\rangle_{ppqq}=0$ for $p\neq q$, while $\langle a_{\mathbf{a}}^{(0)}\gamma^{(1)}_{\mathbf{a}}\rangle_{ppqq}$ is generically non-vanishing for any choice of $p,q$, which hints at the mixed contribution from these degenerate operators. With these additional input, the correct recursion for data at the classical twist $\tau^{(0)}=2n$ and spin $\ell$ is
\begin{eqnarray}\label{eq:llkrecursion}
    \langle a_{\mathbf{a}}^{(0)}(\gamma^{(1)}_{\mathbf{a}})^k\rangle_{ppqq}=\sum_{r=2}^n\langle a_{\mathbf{a}}^{(0)}(\gamma^{(1)}_{\mathbf{a}})^{k-1}\rangle_{pprr}\,\qty(\langle a_{\mathbf{a}}^{(0)}\rangle_{rrrr})^{-1}\,\langle a_{\mathbf{a}}^{(0)}\gamma^{(1)}_{\mathbf{a}}\rangle_{rrqq}\;.
\end{eqnarray}
We will not present the derivation of this formula in this paper since in each color channel it is the same as in the supergravity case \cite{Alday:2017xua,Aprile:2017bgs,Aprile:2017xsp}. We  refer interested readers to these references for details. For our problem, the desired coefficients $\langle a_{\mathbf{a}}^{(0)}(\gamma^{(1)}_{\mathbf{a}})^k\rangle$ in $\mathcal{H}_{\mathbf{a}}^{(k)}$ are obtained by setting $p=q=2$ in \eqref{eq:llkrecursion}.  The explicit computation at one loop for $\langle a_{\mathbf{a}}^{(0)}(\gamma^{(1)}_{\mathbf{a}})^2 \rangle$ in $\mathcal{H}_{\mathbf{a}}^{(2)}$ can be found in \cite{Alday:2021ajh}.

The recursion formula interprets the leading log singularity of $\mathcal{H}^{(k)}$ as gluing (along the s-channel) the leading log singularity of the correlator at order $a_F^{k-1}$ and that at the tree level. Such a relation is the CFT counterpart of the unitarity cut (or Cutkosky cut) relations commonly used in the flat-space scattering amplitudes \cite{Cutkosky:1960sp,Bern:1994zx,Bern:1994cg}, and was first utilized in the AdS perturbative computation in \cite{Aharony:2016dwx}. For the super gluons under our study, this gluing operation involves summing over both the intermediate modes mentioned above and the irreducible representations $\mathbf{a}\in\mathbf{adj}\otimes\mathbf{adj}$ of $G_F$. The fact that there is no mixing between different color representations in \eqref{eq:llkrecursion} is guaranteed by the idempotency condition \eqref{eq:Pidempotency} of the projectors, as can be easily seen by the color projector decomposition of $\mathcal{H}$ \eqref{eq:PHdecomposition}.

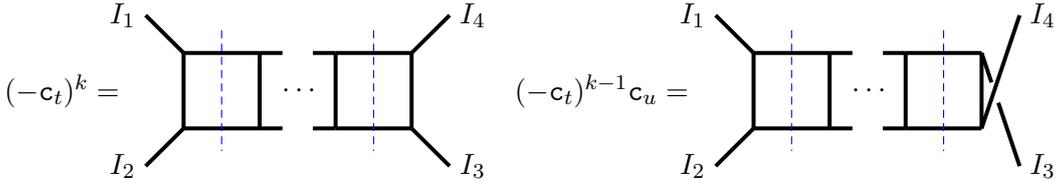
\begin{figure}[h]
    \centering
    \begin{tikzpicture}
        \draw [line width=1.5pt] (-0.5,0.5) -- (-0.5,-0.5);
        \draw [line width=1.5pt](-0.5,-0.5) -- (-1.5,-0.5);
        \draw [line width=1.5pt](-1.5,-0.5) -- (-1.5,0.5);
        \draw [line width=1.5pt](-1.5,0.5) -- (-0.5,0.5);
        \draw [line width=1.5pt](-0.5,0.5) -- (-0.2,0.5);
        \node  at (0.05, 0) {$\cdots$};
        \draw [line width=1.5pt](-0.5,-0.5) -- (-0.2,-0.5);
        \draw [line width=1.5pt](0.2,0.5) -- (0.5,0.5);
        \draw [line width=1.5pt](0.2,-0.5) -- (0.5,-0.5);
        \draw [line width=1.5pt](0.5,0.5) -- (0.5,-0.5);
        \draw [line width=1.5pt](0.5,0.5) -- (1.5,0.5);
        \draw [line width=1.5pt](0.5,-0.5) -- (1.5,-0.5);
        \draw [line width=1.5pt](1.5,0.5) -- (1.5,-0.5);
        \draw [line width=1.5pt] (-1.5,-0.5) -- (-2.0,-1.0);
        \draw [line width=1.5pt] (-1.5,0.5) -- (-2.0,1.0);
        \draw [line width=1.5pt] (1.5,0.5) -- (2.0,1.0);
        \draw [line width=1.5pt] (1.5,-0.5) -- (2.0,-1.0);
        
        \draw [draw=blue, densely dashed ](-1,0.8) -- (-1,-0.8);
        \draw [draw=blue, densely dashed ](1,0.8) -- (1,-0.8);
        
        \node [anchor=east] at (-2.0,1.0) {$I_1$};
        \node [anchor=west] at (2,1.0) {$I_4$};
        \node [anchor=west] at (2,-1.0) {$I_3$};
        \node [anchor=east] at (-2.0,-1.0) {$I_2$};
    \node at (-3.1,0) {$(-{\mathtt{c}}_{t})^{k} = $};
        
    \end{tikzpicture}
    \begin{tikzpicture}
        \draw [line width=1.5pt] (-0.5,0.5) -- (-0.5,-0.5);
        \draw [line width=1.5pt](-0.5,-0.5) -- (-1.5,-0.5);
        \draw [line width=1.5pt](-1.5,-0.5) -- (-1.5,0.5);
        \draw [line width=1.5pt](-1.5,0.5) -- (-0.5,0.5);
        \draw [line width=1.5pt](-0.5,0.5) -- (-0.2,0.5);
        \node  at (0.05, 0) {$\cdots$};
        \draw [line width=1.5pt](-0.5,-0.5) -- (-0.2,-0.5);
        \draw [line width=1.5pt](0.2,0.5) -- (0.5,0.5);
        \draw [line width=1.5pt](0.2,-0.5) -- (0.5,-0.5);
        \draw [line width=1.5pt](0.5,0.5) -- (0.5,-0.5);
        \draw [line width=1.5pt](0.5,0.5) -- (1.5,0.5);
        \draw [line width=1.5pt](0.5,-0.5) -- (1.5,-0.5);
        \draw [line width=1.5pt](1.5,0.5) -- (1.5,-0.5);
        \draw [line width=1.5pt] (-1.5,-0.5) -- (-2.0,-1.0);
        \draw [line width=1.5pt] (-1.5,0.5) -- (-2.0,1.0);
        \draw [line width=1.5pt] (1.5,0.5) -- (2,-1.0);
        \draw [draw=white,line width=2pt] (1.63,0.11) -- (1.72,-0.16);
        \draw [line width=1.5pt] (1.5,-0.5) -- (2,1.0);
        \draw [draw=blue, densely dashed ](-1,0.8) -- (-1,-0.8);
        \draw [draw=blue, densely dashed ](1,0.8) -- (1,-0.8);

        \node at (-3.5,0) {$(-{\mathtt{c}}_{t})^{k-1}{\mathtt{c}}_{u} = $};

        \node [anchor=east] at (-2.0,1.0) {$I_1$};
        \node [anchor=west] at (2,1.0) {$I_4$};
        \node [anchor=west] at (2,-1.0) {$I_3$};
        \node [anchor=east] at (-2.0,-1.0) {$I_2$};
    \end{tikzpicture}    
    \caption{The color structures of planar ladder diagrams at $(k-1)$ loops, generated by gluing $k$ exchange diagrams together.}
    \label{fig:leadinglog}
\end{figure}

Although the recursion \eqref{eq:llkrecursion} applies to each color representation individually, the coefficients directly obtained in this way depend on the choice of gauge group $G_F$. Nevertheless the full leading log singularity admits a $G_F$-independent form once its color factors are turned back into structure constants. In order to see this we modify the definition of the OPE coefficients by splitting out numerical factors that arise from the projector decomposition of color factors. At tree level we have
\begin{equation}\label{eq:a0g1}
    \begin{pmatrix}
        \langle a^{(0)}_{\bf 1}\gamma^{(1)}_{\bf 1} \rangle \\ \langle a^{(0)}_{\bf 3875}\gamma^{(1)}_{\bf 3875} \rangle \\ \langle a^{(0)}_{\bf 27000}\gamma^{(1)}_{\bf 27000} \rangle \\ \langle a^{(0)}_{\bf 248}\gamma^{(1)}_{\bf 248} \rangle \\ \langle a^{(0)}_{\bf 30380}\gamma^{(1)}_{\bf 30380} \rangle
    \end{pmatrix} =\qty( \underbrace{\psi^2 h^\vee \begin{pmatrix}
        1 \\ \frac{1}{5} \\ -\frac{1}{30} \\ \frac{1}{2} \\ 0
    \end{pmatrix}}_{-(\texttt{c}_t)_{\mathbf{a}}}  + \underbrace{\psi^2 h^\vee \begin{pmatrix}
        1 \\ \frac{1}{5} \\ -\frac{1}{30} \\ -\frac{1}{2} \\ 0
    \end{pmatrix}}_{(\texttt{c}_u)_{\mathbf{a}}} (-1)^\ell )\langle a^{(0)}\gamma^{(1)} \rangle_{\rm dyn.},
\end{equation}
where
\begin{equation}
    \langle a^{(0)}\gamma^{(1)} \rangle_{\rm dyn.} = \frac{2 \Gamma ( \ell+3)^2 }{\Gamma (2  \ell+5)}
\end{equation}
encodes the dynamical information, which is the same for every $\mathbf{a}$ and is $G_F$-independent. Similarly we have
\begin{equation}\label{eq:a0}
    \begin{pmatrix}
        \langle a^{(0)}_{\bf 1} \rangle \\ \langle a^{(0)}_{\bf 3875} \rangle \\ \langle a^{(0)}_{\bf 27000} \rangle \\ \langle a^{(0)}_{\bf 248} \rangle \\ \langle a^{(0)}_{\bf 30380} \rangle
    \end{pmatrix} = \qty( \underbrace{\begin{pmatrix}
        1 \\ 1 \\ 1 \\ 1 \\ 1
    \end{pmatrix}}_{(\delta^{I_1I_4}\delta^{I_2I_3})_{\mathbf{a}}} +  \underbrace{\begin{pmatrix}
        1 \\ 1 \\ 1 \\ -1 \\ -1
    \end{pmatrix}}_{(\delta^{I_1I_3}\delta^{I_2I_4})_{\mathbf{a}}}(-1)^\ell )\langle a^{(0)} \rangle_{\rm dyn.},
\end{equation}
where
\begin{equation}
    \langle a^{(0)} \rangle_{\rm dyn.} = (\ell+1)(\ell+4)\frac{ \Gamma ( \ell+3)^2 }{2\Gamma (2  \ell+5)}.
\end{equation}
Turning to other correlators $\langle ppqq \rangle$ will change the data $\langle a^{(0)} \rangle_{\rm dyn.}$ and $\langle a^{(0)}\gamma^{(1)} \rangle_{\rm dyn.}$, but the color factors in front remain the same as those in \eqref{eq:a0g1} and \eqref{eq:a0}. As a consequence, in doing the recursion it suffices to replace $\langle a_{\mathbf{a}}^{(0)}\cdots\rangle$ in \eqref{eq:llkrecursion} by $\langle a^{(0)}\cdots\rangle_{\rm dyn.}$
\begin{eqnarray}\label{eq:llkrecursiondyn}
    \langle a^{(0)}(\gamma^{(1)})^k\rangle_{\substack{\rm dyn.\\ppqq}}=\sum_{r=2}^n\langle a^{(0)}(\gamma^{(1)})^{k-1}\rangle_{\substack{\rm dyn.\\pprr}}\,\qty(\langle a^{(0)}\rangle_{\substack{\rm dyn.\\rrrr}})^{-1}\,\langle a^{(0)}\gamma^{(1)}\rangle_{\substack{\rm dyn.\\rrqq}},
\end{eqnarray}
and treat this as a recursive definition for $\langle a^{(0)}(\gamma^{(1)})^k\rangle_{\rm dyn.}$ at higher $k$. Then the original recursion \eqref{eq:llkrecursion} can be expressed as
\begin{equation}\label{eq:llk}
\begin{split}
        \begin{pmatrix}
        \langle a^{(0)}_{\bf 1}(\gamma^{(1)}_{\bf 1})^k \rangle \\ \langle a^{(0)}_{\bf 3875}(\gamma^{(1)}_{\bf 3875})^k \rangle \\ \langle a^{(0)}_{\bf 27000}(\gamma^{(1)}_{\bf 27000})^k \rangle \\ \langle a^{(0)}_{\bf 248}(\gamma^{(1)}_{\bf 248})^k \rangle \\ \langle a^{(0)}_{\bf 30380}(\gamma^{(1)}_{\bf 30380})^k \rangle
    \end{pmatrix} =& \qty( \begin{pmatrix}
        (\psi^2 h^\vee)^k \\ \left(\frac{1}{5}\psi^2 h^\vee\right)^k \\ \left(-\frac{1}{30}\psi^2 h^\vee\right)^k \\ \left(\frac{1}{2}\psi^2 h^\vee\right)^k \\ 0
    \end{pmatrix} + \begin{pmatrix}
        (\psi^2 h^\vee)^k \\ \left(\frac{1}{5}\psi^2 h^\vee\right)^k \\ \left(-\frac{1}{30}\psi^2 h^\vee\right)^k \\ -\left(\frac{1}{2}\psi^2 h^\vee\right)^k \\ 0
    \end{pmatrix} (-1)^\ell )\langle a^{(0)}(\gamma^{(1)})^k \rangle_{\rm dyn.} \\
    =& \qty( ((-\texttt{c}_\mathtt{t})_{\mathbf{a}})^k + ((-\texttt{c}_\mathtt{t})_{\mathbf{a}})^{k-1} (\texttt{c}_\mathtt{u})_{\mathbf{a}}\, (-1)^\ell )\langle a^{(0)}(\gamma^{(1)})^k \rangle_{\rm dyn.}\\
    =& \qty( (-\texttt{c}_\mathtt{t})^k + (-\texttt{c}_\mathtt{t})^{k-1} \texttt{c}_\mathtt{u}\, (-1)^\ell )_{\mathbf{a}}\,\langle a^{(0)}(\gamma^{(1)})^k \rangle_{\rm dyn.}.
\end{split}
\end{equation}
Here the last equality utilizes the idempotency condition \eqref{eq:Pidempotency}, and hence in the last line the color factors are multiplied according to $(C_1C_2)^{I_1I_2I_3I_4}=C_1^{I_1I_2I_5I_6}C_2^{I_6I_5I_3I_4}$. This gluing rule immediately implies that the color structures $(-\texttt{c}_\mathtt{t})^k$ and $(-\texttt{c}_\mathtt{t})^{k-1} \texttt{c}_\mathtt{u}$ are in correspondence to planar ladder diagrams at $(k\!-\!1)$ loops, as illustrated in Figure \ref{fig:leadinglog}. The dynamical part of the leading log singularities receives a similar diagrammatic interpretation, which was first analyzed in the context of ${AdS}_5\times S^5$ supergravity in \cite{Bissi:2020woe}. 

It is worth emphasizing again that, although we have worked with the special gauge group $E_8$ in this section, the final result \eqref{eq:llk} is independent of gauge group $G_F$. In fact, the whole computation can be done in a $G_F$-independent manner by manifesting the color structures $\texttt{c}_\mathtt{t}$ and $\texttt{c}_\mathtt{u}$, instead of going through the decomposition into different color representations using explicit projectors. 

\subsection{Hidden conformal symmetry}\label{Hidden conformal symmetry}

As we discussed in the previous subsection, $\langle a_{\mathbf{a}}^{(0)}(\gamma^{(1)})^k\rangle_{\rm dyn.}$  determines the leading log singularities. However, obtaining this data by solving the mixing among degenerate operators is usually a bit cumbersome in practice. A much more convenient strategy makes use of the so-called hidden conformal symmetry, which was first discovered in ${AdS}_5 \times S^5$ \cite{Caron-Huot:2018kta}, and later in ${AdS}_3 \times S^3$ \cite{Rastelli:2019gtj} and ${AdS}_5 \times S^3$ \cite{Alday:2021odx}. The hidden conformal symmetry organizes all the tree-level correlators into a single generating function. Moreover, it allows us to  generate the leading log singularities by differentiation. 

More precisely, the use of the hidden conformal symmetry automatically diagonalizes the mixing matrices, and gives the tree-level anomalous dimensions $\gamma^{(1)}$ for all double-trace operators are a simple rational function of the 8d conformal spin $\ell_{\text{8d}}$ \cite{Drummond:2022dxd}. Using the results of $\gamma^{(1)}$, the leading log for four-point holographic correlators can be determined to all loop order as
\begin{equation}\label{eq:leadinglog}
    \left. \mathcal{H}^{(k)}\right\vert_{\log^k{u}} = \left[ \Delta^{(4)} \right]^{k-1} \mathcal{D}_{(2)}\,  h^{(k)}(z).
\end{equation}
Here \footnote{The operator $\mathcal{D}_{(2)}$ here is to build the seed functions $-\frac{12}{(\ell+1) (\ell+2)} z^{\ell+1} \, _2F_1(\ell+1,\ell+3,2 \ell+6,z)$ to 8d conformal blocks on unitarity bound $G^{\rm 8d}_{6+\ell,\ell}(z,\zb)$. See Sections 4 and 5 of \cite{Dolan:2011dv}.}
\begin{equation}
    \begin{split}
      \mathcal{D}_{(2)}f(z) = &\left[ \left(\frac{z \zb}{\zb - z}\right)^5 f(z) + \left(\frac{z \zb}{\zb - z}\right)^4 \frac{z^2}{2}\partial_z f(z)+ \left(\frac{z \zb}{\zb - z}\right)^3 \frac{z^3}{12}\partial^2_z \left(z f(z)\right)  \right] \\&+ (z\leftrightarrow\zb),
    \end{split}
\end{equation}
and $\Delta^{(4)}$ is a forth-order differential operator
\begin{equation}
    \Delta^{(4)} = \frac{z\zb}{\zb-z} D_z D_{\zb} \frac{\zb-z}{z\zb}\;,
\end{equation}
where $D_z=z^2\partial_z (1-z)\partial_z$ the Casimir of $\text{SL}(2,\mathbb{R})$. The function $h^{(k)}(z)$ is defined by an infinite sum
\begin{equation}\label{eq:hkdef}
    h^{(k)}(z) = \frac{1}{2^k k!} \sum_{\ell=0}^\infty  \frac{-24\Gamma(\ell+1)\Gamma(\ell+3)}{\Gamma (2 \ell+5)} \frac{\left(-\mathtt{c}_t+(-1)^\ell \mathtt{c}_u\right)^k}{\left[(\ell+1)_4\right]^{k-1}} {}_2F_1(\ell+1,\ell+3,2 \ell+6,z).
\end{equation}
For fixed $k$, this sum can be written in a closed form using multiple polylogarithms (MPLs). We list the first few examples of $h^{(k)}(z)$ here
\begin{subequations}\label{eq:h}
    \begin{align}
    h^{(1)}(z) =& -\mathtt{c}_t\left( \frac{2 \left(z^2+3 z-6\right)}{z^2}+\frac{12 (z-1) }{z^3}G_{1}(z) \right) \nonumber\\
    &\quad + \mathtt{c}_u \left( \frac{2 \left(2 z^2-9 z+6\right)}{z^2}+\frac{12 (z-1)^2 }{z^3}G_{1}(z) \right),\label{eq:h:1}\\
    h^{(2)}(z) =& \ \mathtt{d}_{st} \left(\frac{61 z^2-135 z+72}{36 z^2}+\frac{(z-1)^2 }{z^3}G_{1}(z)+\frac{(z-1)^3}{z^3}\left( G_{1,1}(z)-G_{0,1}(z)\right) \right)\nonumber \\
    &+ \mathtt{d}_{su} \left(\frac{2 z^2+9 z-72}{36 z^2}+\frac{(z-1)}{z^3}G_{1}(z)- \frac{1}{z^3}G_{0,1}(z) \right) ,\label{eq:h:2}\\
    h^{(3)}(z) =&\ \mathtt{e}_{s_1} \left(\frac{31 z^2-243 z-828}{1944 z^2}+ \frac{(z-1) (z+23) }{108 z^3}G_{1}(z) \right.\nonumber\\
    & \left.+\frac{(z-1) \left(z^2-8 z-17\right)}{108 z^3}\left( G_{1,1}(z)-G_{0,1}(z)\right)+\frac{(3 z+1)}{18
   z^3}\left(G_{0,1,1}(z)- G_{0,0,1}(z)\right) \right) \nonumber\\
    &+ \mathtt{e}_{s_2} \left(\frac{1040 z^2-1899 z+828}{1944 z^2} +\frac{(24 z-23) (z-1)
   }{108 z^3}G_{1}(z) \right. \nonumber\\
   &+ \left. \frac{\left(24 z^2-42 z+17\right) }{108 z^3}G_{0,1}(z) + \frac{(4 z-1) (z-1)^2 }{18 z^3}\left(G_{1,0,1}(z)-G_{0,0,1}(z)\right) \right) \label{eq:h:3},
\end{align}
\end{subequations}
where $G_{\vec{a}}(z)$ are multiple polylogarithms, whose definition is reviewed in Appendix \ref{sec:mpls}. The same appendix also shows how to write the $G_{\vec{a}}(z)$ appearing above in terms of classical polylogarithms $\text{Li}_n(z)$, from which we can observe that each $h^{(k)}$ (and hence the corresponding leading log singularity) has a maximal transcendentality $k$. The $\mathtt{d}_{st},\mathtt{d}_{su},\mathtt{e}_{s_1},\mathtt{e}_{s_2}$ are color structures of box and planar double-box diagrams coming from expanding $\left(-\mathtt{c}_t+(-1)^\ell \mathtt{c}_u\right)^k$, and will be discussed in Section \ref{sec:oneloop} and Section \ref{sec:twoloops}. As can be expected from the definition \eqref{eq:hkdef} the two terms with different color factors in each $h^{(k)}(z)$ are in fact related by exchanging $1\leftrightarrow2$.

\section{One-loop correlator}\label{sec:oneloop}
The super gluon one-loop amplitude has already been computed in \cite{Alday:2021ajh} in Mellin space using techniques developed in \cite{Alday:2018kkw,Alday:2019nin}. Here we compute the same quantity directly in position space. A simple idea is to use the CFT dispersion relation \cite{Carmi:2019cub}, through which one can reconstruct the whole correlator from its double discontinuity. For the one-loop correlator, the double discontinuity purely comes from the leading log singularity which has been computed in \eqref{eq:h:2} using the tree-level data. Therefore, the one-loop correlator $\mathcal{H}^{(2)}$ in principle can be obtained by substituting the leading log into the CFT dispersion relation. However, this approach requires one to work out a highly complicated integral, which is beyond our current technical capability. A more practical way is  to follow the position space method for $AdS_5\times S^5$ supergravity correlators \cite{Aprile:2017bgs,Aprile:2019rep}, which starts with a position space ansatz for the correlator and then bootstrap it by using the leading log data together with physical constraints such as crossing symmetry. In particular it was noted in \cite{Aprile:2019rep} that imposing an educated ansatz inspired by the hidden-symmetry structure of the leading log singularities \eqref{eq:leadinglog} greatly improves the efficiency in the one-loop computation of super graviton scattering. This idea was later further extended to make the two-loop computation accessible \cite{Huang:2021xws,Drummond:2022dxw}.  In this section we apply the same strategy to the case of $AdS_5\times S^3$ super gluons at one loop. As discussed previously this theory enjoys similar hidden symmetries in its tree-level scattering and its leading log data, hence it is very interesting to check whether the structure observed in the super gravitons holds in the super gluons as well. This is indeed the case as we will see. We will also compare this result against the Mellin space result and find an exact match. This one-loop computation also serves as a careful preparation for a more elaborate bootstrap computation for the two-loop scattering of super gluons, to be presented in the next section.

\subsection{Ansatz}\label{sec:oneloopansatz}

To give a precise description of our ansatz, let us first introduce some relevant ingredients, which include details on the one-loop color structures, a basis of functions for decomposing $\mathcal{H}^{(2)}$ and an observation on the structure of one-loop scattering related to hidden conformal symmetries.
\begin{itemize}
    \item {\bf Color structures}
    \begin{figure}[h]
    \centering
   \begin{tikzpicture}
        \draw [line width=1.5pt] (0.5,0.5) -- (0.5,-0.5);
        \draw [line width=1.5pt](0.5,-0.5) -- (-0.5,-0.5);
        \draw [line width=1.5pt](-0.5,-0.5) -- (-0.5,0.5);
        \draw [line width=1.5pt](-0.5,0.5) -- (0.5,0.5);
        \draw [line width=1.5pt](0.5,0.5) -- (1.0,1.0);
        \draw [line width=1.5pt](0.5,-0.5) -- (1.0,-1.0);
        \draw [line width=1.5pt] (-0.5,-0.5) -- (-1.0,-1.0);
        \draw [line width=1.5pt] (-0.5,0.5) -- (-1.0,1.0);
        \node at (-2,0) {${\mathtt{d}}_{st} = $};
        \node [anchor=east] at (-1.0,1.0) {$I_1$};
        \node [anchor=west] at (1.0,1.0) {$I_4$};
        \node [anchor=west] at (1.0,-1.0) {$I_3$};
        \node [anchor=east] at (-1.0,-1.0) {$I_2$};
    \end{tikzpicture}
    \begin{tikzpicture}
        \draw [line width=1.5pt] (0.5,0.5) -- (0.5,-0.5);
        \draw [line width=1.5pt](0.5,-0.5) -- (-0.5,-0.5);
        \draw [line width=1.5pt](-0.5,-0.5) -- (-0.5,0.5);
        \draw [line width=1.5pt](-0.5,0.5) -- (0.5,0.5);
        \draw [line width=1.5pt](0.5,0.5) -- (1.0,1.0);
        \draw [line width=1.5pt](0.5,-0.5) -- (1.0,-1.0);
        \draw [line width=1.5pt] (-0.5,-0.5) -- (-1.0,-1.0);
        \draw [line width=1.5pt] (-0.5,0.5) -- (-1.0,1.0);
        \node at (-2,0) {${\mathtt{d}}_{su} =$};
        \node [anchor=east] at (-1.0,1.0) {$I_1$};
        \node [anchor=west] at (1.0,1.0) {$I_3$};
        \node [anchor=west] at (1.0,-1.0) {$I_4$};
        \node [anchor=east] at (-1.0,-1.0) {$I_2$};
    \end{tikzpicture}
    \begin{tikzpicture}
        \draw [line width=1.5pt] (0.5,0.5) -- (0.5,-0.5);
        \draw [line width=1.5pt](0.5,-0.5) -- (-0.5,-0.5);
        \draw [line width=1.5pt](-0.5,-0.5) -- (-0.5,0.5);
        \draw [line width=1.5pt](-0.5,0.5) -- (0.5,0.5);
        \draw [line width=1.5pt](0.5,0.5) -- (1.0,1.0);
        \draw [line width=1.5pt](0.5,-0.5) -- (1.0,-1.0);
        \draw [line width=1.5pt] (-0.5,-0.5) -- (-1.0,-1.0);
        \draw [line width=1.5pt] (-0.5,0.5) -- (-1.0,1.0);
        \node at (-2,0) {${\mathtt{d}}_{tu} =$};
        \node [anchor=east] at (-1.0,1.0) {$I_1$};
        \node [anchor=west] at (1.0,1.0) {$I_4$};
        \node [anchor=west] at (1.0,-1.0) {$I_2$};
        \node [anchor=east] at (-1.0,-1.0) {$I_3$};
    \end{tikzpicture}
    \caption{1-loop color structures $\mathtt{d}_{st}$, $\mathtt{d}_{su}$ and $\mathtt{d}_{tu}$.}
    \label{fig:fourpt1loop}
\end{figure}
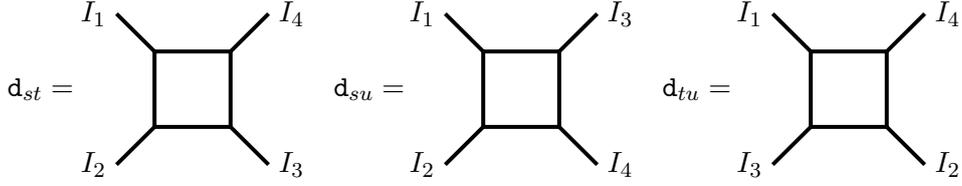

The Mellin amplitude result of \cite{Alday:2021ajh} (see \eqref{eq:mellinresult} in Section \ref{Subsec:compareMellin}) has already shown that the color structures at one-loop are just three box diagrams drawn in Figure \ref{fig:fourpt1loop}. Among these, $\texttt{d}_{st}$ and $\texttt{d}_{su}$ can already be generated from the color gluing operation in the recursion of leading log singularities as discussed in \eqref{eq:llk}. Working with the gauge group $E_8$ for example, s-channel projections of these color factors explicitly are
\begin{align}
    (\mathtt{d}_{st})_{\mathbf{a}} &= ((-\mathtt{c}_t)^2)_{\mathbf{a}} = \left( \psi^2 h^\vee \right)^2 \left(1,\frac{1}{25},\frac{1}{900},\frac{1}{4},0 \right),\\
    (\mathtt{d}_{su})_{\mathbf{a}} &= (-\mathtt{c}_t \mathtt{c}_u)_{\mathbf{a}} = \left( \psi^2 h^\vee \right)^2 \left(1,\frac{1}{25},\frac{1}{900},-\frac{1}{4},0 \right).
\end{align}
Expression for the remaining color factor $\mathtt{d}_{tu}$ can be obtained by crossing. The method of projectors provides an efficient way to perform crossing operations, by means of color crossing matrices. Imagine that we cross into the t-channel by exchanging the operators $\mathcal{O}(x_1)$ and $\mathcal{O}(x_3)$, which transforms a generic color factor $C_s \equiv C^{I_1 I_2 I_3 I_4}$ to $C_t \equiv C^{I_3 I_2 I_1 I_4}$. Note the old and new factors receive the same projection coefficients in s- and in t-channel, using projectors $P_{\mathbf{a}}^{I_1I_2\vert I_3I_4}$ and $P_{\mathbf{a}}^{I_3I_2\vert I_1I_4}$ respectively. The connection between $C_s$ and $C_t$ is then encoded in the overlap of these two types of projectors, hence we construct the so-called t-channel color crossing matrix
\begin{equation}
    {({\rm F}_t)_{\mathbf{a}}}^{{\mathbf{a}}'} \equiv \sum_{I_1,I_2,I_3,I_4} \text{dim}(R_{\mathbf{a}})^{-1} P^{I_3 I_2 \vert I_1 I_4}_{\mathbf{a}} P^{I_1 I_2 \vert I_3 I_4}_{{\mathbf{a}}'}\;.
\end{equation}
From this definition, we obviously have the following relation between the s-channel projections before and after crossing
\begin{equation}
    (C_t)_{\mathbf{a}} = ({\rm F}_t)_{\mathbf{a}}^{\ {\mathbf{a}}'} (C_s)_{{\mathbf{a}}'}\;.
\end{equation}
In the same spirit we can define the u-channel color crossing matrix, in correspondence to the crossing $\mathcal{O}(x_1)\leftrightarrow\mathcal{O}(x_4)$
\begin{equation}
    ({\rm F}_u)_{\mathbf{a}}^{\ {\mathbf{a}}'} \equiv \sum_{I_1,I_2,I_3,I_4} \text{dim}(R_{\mathbf{a}})^{-1} P^{I_4 I_2 \vert I_3 I_1}_{\mathbf{a}} P^{I_1 I_2 \vert I_3 I_4}_{{\mathbf{a}}'}\;.
\end{equation}
Still in the gauge group $E_8$, these matrices explicitly read \cite{Chang:2017xmr}
\begin{equation}\label{eq:E8FtFu}
{\rm F}_t=\left(
\begin{array}{ccccc}
 \frac{1}{248} & \frac{125}{8} & \frac{3375}{31} & 1 & \frac{245}{2} \\
 \frac{1}{248} & -\frac{3}{8} & \frac{27}{31} & \frac{1}{5} & -\frac{7}{10} \\
 \frac{1}{248} & \frac{1}{8} & \frac{23}{62} & -\frac{1}{30} & -\frac{7}{15} \\
 \frac{1}{248} & \frac{25}{8} & -\frac{225}{62} & \frac{1}{2} & 0 \\
 \frac{1}{248} & -\frac{5}{56} & -\frac{90}{217} & 0 & \frac{1}{2} \\
\end{array}
\right)\;,\quad {\rm F}_u=\left(
\begin{array}{ccccc}
 \frac{1}{248} & \frac{125}{8} & \frac{3375}{31} & -1 & -\frac{245}{2} \\
 \frac{1}{248} & -\frac{3}{8} & \frac{27}{31} & -\frac{1}{5} & \frac{7}{10} \\
 \frac{1}{248} & \frac{1}{8} & \frac{23}{62} & \frac{1}{30} & \frac{7}{15} \\
 -\frac{1}{248} & -\frac{25}{8} & \frac{225}{62} & \frac{1}{2} & 0 \\
 -\frac{1}{248} & \frac{5}{56} & \frac{90}{217} & 0 & \frac{1}{2} \\
\end{array}
\right),\;
\end{equation}
With this tool $\texttt{d}_{tu}$ can be generated by
\begin{align}
    (\mathtt{d}_{tu})_{\mathbf{a}} = {({\rm F}_u)_{\mathbf{a}}}^{\mathbf{a}'} (\mathtt{d}_{st})_{\mathbf{a}'} &= \left( \psi^2 h^\vee \right)^2 \left(\frac{1}{2},-\frac{3}{50},\frac{4}{225},0,0 \right).
\end{align}
In principle, we should write down five different ansatz for $E_8$, since there are five components under projection and we do not know what color structures will appear a priori. However, noting that $\mathcal{H}^{(2)}$ can be obtained by the CFT dispersion relation, which utilizes its leading log singularities in two different channels, only color factors in \eqref{eq:h:2} and their crossing can appear in $\mathcal{H}^{(2)}$. Therefore only $\mathtt{d}_{st}$, $\mathtt{d}_{su}$ and $\mathtt{d}_{tu}$ are required. 

\item{\bf Basis functions.}

In the previous sections we observed that the tree-level correlator $\mathcal{H}^{(1)}$ \eqref{eq:tree1stu} and the leading log singularities at loop levels \eqref{eq:h} are some linear combinations of MPL functions which are generalizations of the familiar classical polylogarithms $\text{Li}_n(x)$, and the coefficients are rational functions. Ideally one can expect that the full correlator belongs to the same class of functions as well, at least in the first several orders in the perturbative expansion. This has been extensively confirmed in a large class of four-point correlators in $AdS_5\times S^5$ IIB supergravity, including those of $\frac{1}{2}$-BPS operators with various KK levels at both tree and one-loop level, and that of stress-tensor operators at two loops. Therefore it is natural to assume that this continues to hold for the super gluon correlator in $AdS_5\times S^3$.

While the space of MPLs in general has a rich and complicated structure, under proper conditions one can restrict to a finite dimensional linear subspace. Specific for the need of our current investigation, the first condition is that the maximal transcendentality of $\mathcal{H}^{(2)}$ is limited by its leading log singularities, which (including the factor $\log^2{u}$) is $4$. MPLs with different transcendental weights do not mix under rational transformations of its variables, and so the MPLs in need can be classified into a finite number of subspaces according to the weight. 

The second condition has to do with the singularities in the MPLs, which are constrained by the physically allowed singularities of the correlator. The obvious singularities are located at $z=0$ and $\bar{z}=0$ in the Lorentzian region, as explicitly shown by the $\log{u}$ expansion, and by crossing, at $z=1$, $\bar{z}=1$, $z=\infty$ and $\bar{z}=\infty$ as well. In practice one may also encounter singularities at $z=\bar{z}$, which have to do with the so-called bulk-point limit of perturbative scattering in AdS. Locations of these singularities further restricts the set of MPLs that can show up.

The third condition is that the full reduced correlator $\mathcal{H}^{(2)}$ has to be single-valued on the Euclidean slice $\bar{z}=z^*$. Although not absolutely necessary, it is quite convenient to already impose this condition on the set of MPLs in use. This is because the single-valued multiple polylogarithms (SVMPLs) by themselves can form a linear subspace of MPLs, which is much smaller than the latter.

The above conditions characterize a finite-dimensional linear space of SVMPLs to be conveniently used for our position space bootstrap computation. To construct the ansatz we select a complete basis for this space of functions and assume a linear decomposition of $\mathcal{H}^{(2)}$ on this basis with rational function coefficients. The necessary reviews on MPLs and SVMPLs and the selection of such basis are described in Appendix \ref{sec:mpls}. Here we just set up notation for the basis elements for the clarity of later discussions. We require that each basis element has a uniform transcendental weight $w$, which ranges from $0$ to $4$ for $\mathcal{H}^{(2)}$, and denote it as $G_{w,i}^{\rm SV}$. The extra index $i$ is required to distinguish  independent basis elements with the same weight. The  range of $i$ depends on the value of $w$.

\item {\bf Hidden symmetry structures.} 

As was already pointed out in Section \ref{Hidden conformal symmetry}, the  hidden conformal symmetry is inherntly related to certain special differential operators ($\Delta^{(8)}$ for $AdS_5\times S^5$ and $\Delta^{(4)}$ for $AdS_5\times S^3$). We have seen that the use of these differential operators drastically simplifies the leading log singularities. Similar simplifications occur in the full reduced correlator as well. In the case of super graviton scattering in $AdS_5\times S^5$, the one-loop leading log structure
\begin{equation}\label{eq:gravitonleadinglog}
    \mathcal{H}_{AdS_5\times S^5}^{(2)}\big\vert_{\log^2u}=\Delta^{(8)}\mathcal{F}_{AdS_5\times S^5}^{(2)}
\end{equation}
(where the expression for $\mathcal{F}_{AdS_5\times S^5}^{(2)}$ can be found in \cite{Caron-Huot:2018kta}) was observed in \cite{Aprile:2019rep} to promote to the reduced correlator with a slight modification
\begin{equation}\label{eq:hs10d1l}
 \mathcal{H}_{AdS_5\times S^5}^{(2)} = \Delta^{(8)} \mathcal{L}_{AdS_5\times S^5}^{(2)} +\frac{1}{4} \mathcal{H}_{AdS_5\times S^5}^{(1)}.
\end{equation}
Here $\mathcal{L}_{AdS_5\times S^5}^{(2)}$ is a simpler object known as the {\it pre-correlator}, and $\mathcal{H}_{AdS_5\times S^5}^{(1)}$ is exactly the tree-level reduced correlator. 

Given the analogy between the leading log structure in the two theories \eqref{eq:leadinglog} and \eqref{eq:gravitonleadinglog}, it is very tempting to assume the following decomposition for the super gluon correlator $\mathcal{H}^{(2)}$
\begin{equation}\label{eq:oneloopansatz}
\mathcal{H}^{(2)}(z,\bar{z})=\Delta^{(4)} \mathcal{L}^{(2)}(z,\bar{z})+\bar{\mathcal{H}}^{(1)}(z,\bar{z}).
\end{equation}
However, in contrast to \eqref{eq:hs10d1l} we cannot simply take the modification term $\bar{\mathcal{H}}^{(1)}$ to be proportional to the tree-level correlator $\mathcal{H}^{(1)}$. This is because $\mathcal{H}^{(2)}$ is expected to depend on one-loop color factors $\{\mathtt{d}_{st},\mathtt{d}_{su},\mathtt{d}_{tu}\}$ as discussed previously, and it is impossible to relate the tree-level color factors $\{\mathtt{c}_s,\mathtt{c}_t,\mathtt{c}_u\}$ in $\mathcal{H}^{(1)}$ to $\{\mathtt{d}_{st},\mathtt{d}_{su},\mathtt{d}_{tu}\}$ in a $G_F$-independent way. Nevertheless, we can still assume $\bar{\mathcal{H}}^{(1)}$ to share some common features with $\mathcal{H}^{(1)}$. In particular, we require that as a combination of SVMPLs it has a maximal weight $2$ as the tree-level correlator. This equivalently means that the higher-weight parts can entirely be written in terms of the action of $\Delta^{(4)}$. This is a reasonable assumption since the higher-weight parts are expected to be closely related to the leading log singularities.

As a differential operator acting in the s-channel, $\Delta^{(4)}$ only preserves the Bose symmetry of exchanging operators $1\leftrightarrow 2$
\begin{equation}\label{eq:delta4ex12}
    \Delta^{(4)} = \left.\Delta^{(4)} \right\vert_{z\rightarrow\frac{z}{z-1},\zb\rightarrow\frac{\zb}{\zb-1}}.
\end{equation}
Thus we expect that $\mathcal{L}^{(2)}$ is invariant under $1\leftrightarrow 2$, but not under other Bose symmetries.
\end{itemize}
As a consequence of the above discussions, we construct an ansatz for the one-loop super gluon reduced correlator $\mathcal{H}^{(2)}$ in the form of \eqref{eq:oneloopansatz}, and assume that both $\mathcal{L}^{(2)}$ and $\bar{\mathcal{H}}^{(1)}$ receive color decomposition with respect to the color factors $\{\mathtt{d}_{st},\mathtt{d}_{su},\mathtt{d}_{tu}\}$ and further linear decomposition onto the SVMPL basis described above. More precisely we write them as
\begin{align} \label{eq:ansatz1loop}
    \mathcal{L}^{(2)} &= \sum_{w=0}^4\sum_i \frac{\hsv_{w,i}(z,\zb)}{(z-\zb)^5}\left( p^{st}_{w,i}(z,\zb)\,\mathtt{d}_{st}+p^{su}_{w,i}(z,\zb)\,\mathtt{d}_{su}+p^{tu}_{w,i}(z,\zb)\,\mathtt{d}_{tu} \right),\\
    \label{eq:barH1}\bar{\mathcal{H}}^{(1)} &= \frac{u^2}{v} \sum_{w=0}^2\sum_i \frac{\hsv_{w,i}(z,\zb)}{(z-\zb)^5}\left( q^{st}_{w,i}(z,\zb)\,\mathtt{d}_{st}+q^{su}_{w,i}(z,\zb)\,\mathtt{d}_{su}+q^{tu}_{w,i}(z,\zb)\,\mathtt{d}_{tu} \right).
\end{align}
Here $p^A_{w,i}(z,\zb)$, $q^A_{w,i}(z,\zb)$ are polynomials of $z$ and $\zb$, with degrees in each variable no higher than $5$
\begin{equation}
    p^A_{w,i}(z,\zb) = \sum_{j,k=0}^5 (c^A_{w,i})_{j,k} z^j \zb^k,\quad q^A_{w,i}(z,\zb) = \sum_{j,k=0}^5 (d^A_{w,i})_{j,k} z^j \zb^k, \quad A = st,su,tu,
\end{equation}
where all coefficients $(c^A_{w,i})_{j,k}$ and $(d^A_{w,i})_{j,k}$ are unknown rational numbers (the rationality originates from the fact that we absorb all transcendentality into the SVMPL basis, including various $\zeta$ values). Motivations for the maximal weights of $\mathcal{L}^{(2)}$ and $\bar{\mathcal{H}}^{(2)}$ was already discussed before. Some comments need to be drawn regarding the form of the function coefficients in front of the SVMPLs. The formal appearance of poles at $z-\bar{z}$ naturally descends from the structure in the leading log singularities, and its power is set in correspondence to the counting in $\mathcal{D}_{(2)}h^{(2)}$. The degrees of the numerator polynomials are then determined by transformations under crossing. In $\bar{\mathcal{H}}^{(1)}$ the appearance of an extra pole $\frac{1}{v}$ is a necessary assumption, whose role will be clarified in the discussion of pole cancellation constraints in the next subsection. The factor $u^2$ is set merely for the simplification of later computations. In principle one can of course start with a more general ansatz for these function coefficients, but in the end they all reduce to the form presented above after solving the constraints to be described in the next step.

\subsection{Constraints}\label{sec:oneloop:ansatz}

The ansatz constructed above already takes into consideration a few structural properties of the one-loop reduced correlator $\mathcal{H}^{(2)}$. In addition to these there are several generic CFT properties and theory-specific data that further constrain the ansatz. These are listed as follows.
\begin{itemize}
\item {\bf Leading logarithmic singularity.} The leading logarithmic singularity of the ansatz must agree with the prediction from Section  \ref{Hidden conformal symmetry}. In terms of the pre-correlator, it amounts to the condition  
\begin{subequations}
    \begin{align}
         &\left. \mathcal{L}^{(2)}(z,\zb)\right\vert_{\log^{n\geq 3} u} = 0,\\
         &\left. \mathcal{L}^{(2)}(z,\zb)\right\vert_{\log^2 u} \quad  = \mathcal{D}_2 h^{(2)}(z).
    \end{align}
\end{subequations}
\item {\bf Bose symmetry.} Since the external operators are bosons, the correlator should be invariant under permutations. To discuss the constraints on $\mathcal{L}^{(2)}$ and $\bar{\mathcal{H}}^{(1)}$, let us distinguish two cases.
\begin{enumerate}
    \item Exchanging 1 and 2. The operator $\Delta^{(4)}$ is symmetric under $1\leftrightarrow 2$. Therefore, we can directly impose the symmetry condition on $\mathcal{L}^{(2)}(z,\bar{z})$
\begin{subequations}\label{eq:oneloop12}
        \begin{align}
            \mathcal{L}^{(2)}(z,\zb)&=\left.\mathcal{L}^{(2)}\left(\frac{z}{z-1},\frac{\zb}{\zb-1}\right)\right\vert_{\mathtt{d}_{st} \leftrightarrow \, \mathtt{d}_{su}},\\
            \bar{\mathcal{H}}^{(1)}(z,\zb)&=\left.\bar{\mathcal{H}}^{(1)}\left(\frac{z}{z-1},\frac{\zb}{\zb-1}\right)\right\vert_{\mathtt{d}_{st} \leftrightarrow \, \mathtt{d}_{su}}.
        \end{align}
        \end{subequations}
    
    \item Exchanging 1 and 3. By contrast, $\Delta^{(4)}$ is not invariant under $1\leftrightarrow 3$. Therefore, we can only impose symmetry after acting with $\Delta^{(4)}$
    \begin{subequations}
        \begin{align}
            \left[\Delta^{(4)} \mathcal{L}^{(2)}\right](z,\zb)&=\frac{u^3}{v^3}\left.\left[\Delta^{(4)} \mathcal{L}^{(2)}\right](1-z,1-\zb)\right\vert_{\mathtt{d}_{su} \leftrightarrow \, \mathtt{d}_{tu}},\\
            \bar{\mathcal{H}}^{(1)}(z,\zb)&=\frac{u^3}{v^3}\left.\bar{\mathcal{H}}^{(1)}(1-z,1-\zb)\right\vert_{\mathtt{d}_{su} \leftrightarrow \, \mathtt{d}_{tu}}.
        \end{align}
        \end{subequations}        
\end{enumerate}
        \item {\bf Symmetry under $z\leftrightarrow\zb$.} The freedom of $z\leftrightarrow\zb$ in the change of variables from $u$, $v$ to $z$, $\zb$ is an artifact of the parameterization. Therefore,  the correlator should be invariant under its action. This leads us to impose  
        \begin{equation}
            \mathcal{L}^{(2)}(z,\zb)=\mathcal{L}^{(2)}(\zb,z), \quad \bar{\mathcal{H}}^{(1)}(z,\zb)=\bar{\mathcal{H}}^{(1)}(\zb,z).
        \end{equation}
\item {\bf Finiteness at $z=\zb$.} The condition $z=\bar{z}$ corresponds to the configuration where all the four operators are inserted on a line. This is not a singular configuration. Therefore, $\mathcal{L}^{(2)}$ and $\bar{\mathcal{H}}^{(1)}$ should remain finite at $z=\zb$ in the Euclidean region. This means that the Taylor expansion of the numerators in $\mathcal{L}^{(2)}$ and $\bar{\mathcal{H}}^{(1)}$ at $z=\zb$ should start with $(z-\zb)^5$ to cancel the $z-\zb$ poles in the ansatz.
\item {\bf Cancellation of unphysical poles.}  Our ansatz includes  $v$ poles appearing separately in $\Delta^{(4)} \mathcal{L}^{(2)}$ and $\bar{\mathcal{H}}^{(1)}$. However, their contributions should cancel in the whole reduce correlator $\mathcal{H}^{(2)}$. This is because $v$ poles correspond to operator exchanges in the t-channel with twist $\tau<4$. However, beyond tree level, no such operators are supposed to appear in the reduced correlator of super gluons.  
\end{itemize}

\subsection{Results at one loop}\label{sec:oneloop:results}

The constraints listed above determine most of the unknown variables in the ansatz. At this stage we can inspect the nature of the remaining degrees of freedom by checking the expressions that they multiply. They fall into three different types.
\begin{itemize}
    \item The first type of dof can be identified as a unique contact diagram in AdS that contributes to $\mathcal{H}^{(2)}$, and in terms of the $\bar{D}$ functions it is simply
    \begin{equation}\label{eq:H2c}
        \mathcal{H}^{(2)}_{\rm c}= \left(\mathtt{d}_{st} + \mathtt{d}_{su} + \mathtt{d}_{tu}\right)\,u^3 \bar{D}_{3333}\equiv\frac{1}{6}\left(\mathtt{d}_{st} + \mathtt{d}_{su} + \mathtt{d}_{tu}\right)\Delta^{(4)}\left[u\bar{D}_{1133}\right].
    \end{equation}
    This serves as the counterterm to the one-loop scattering amplitude of four super gluons, which is expected since the latter necessarily has UV divergence.
    \item The second type of dof are only present inside $\mathcal{L}^{(2)}$, and their contributions can be explicitly organized as
    \begin{equation}\label{eq:L2kernel}
        \mathcal{L}^{(2)}\supset(\mathtt{d}_{st}+\mathtt{d}_{su})(a_1 + a_2 u \bar{D}_{1111}) + \mathtt{d}_{tu}(a_3 + a_4 u \bar{D}_{1111} ) + a_5 (\mathtt{d}_{st}-\mathtt{d}_{su})\log v,
    \end{equation}
    where $a_i$'s are free parameters \footnote{Because our ansatz sets a maximal weight $4$ for $\mathcal{H}^{(2)}$, in the result directly obtained from our bootstrap computation these coefficients $a_i$ actually come in the form of linear combinations of zeta values with maximal weight $2$, i.e.~$a_i\equiv a_{i,1}+a_{i,2}\pi^2$, with $a_{i,1},a_{i,2}\in\mathbb{Q}$.}. These contributions turn out to be annihilated by the action of $\Delta^{(4)}$, and so they completely live inside the kernel of this differential operator and have zero physical effects on $\mathcal{H}^{(2)}$.
    \item The last type of dof can be identified as a single free parameter, showing up in both $\mathcal{L}^{(2)}$ and $\bar{\mathcal{H}}^{(1)}$. In $\bar{\mathcal{H}}^{(1)}$ it multiplies the function $u^2 v^{-1}\left(\mathtt{d}_{st} + \mathtt{d}_{su} + \mathtt{d}_{tu}\right)$, but this turns out to be exactly canceled by its corresponding contribution to $\Delta^{(4)}\mathcal{L}^{(2)}$. Hence this dof has no influence on the reduced correlator $\mathcal{H}^{(2)}$ either.
\end{itemize}
In summary, we observe that the latter two types of dof are only artificial redundancy due to the particular form of our ansatz and are totally unphysical, while the only true remaining dof (the first type) exactly relates to counterterms. In this sense our ansatz is completely solved by the constraints in the previous subsection.

\begin{figure}[h]
    \centering
        \begin{tikzpicture}
        \draw [line width=1.5pt] (0.5,0.5) -- (0.5,-0.5);
        \draw [line width=1.5pt](0.5,-0.5) -- (-0.5,-0.5);
        \draw [line width=1.5pt](-0.5,-0.5) -- (-0.5,0.5);
        \draw [line width=1.5pt](-0.5,0.5) -- (0.5,0.5);
        \draw [line width=1.5pt] (0.5,0.5) -- (1,-1.0);
        \draw [draw=white,line width=2pt] (0.63,0.11) -- (0.72,-0.16);
        
        \draw [line width=1.5pt] (0.5,-0.5) -- (1,1.0);
        \draw [line width=1.5pt] (-0.5,-0.5) -- (-1.0,-1.0);
        \draw [line width=1.5pt] (-0.5,0.5) -- (-1.0,1.0);
        \node [anchor=east] at (-1.0,1.0) {$I_1$};
        \node [anchor=west] at (1.0,1.0) {$I_4$};
        \node [anchor=west] at (1.0,-1.0) {$I_3$};
        \node [anchor=east] at (-1.0,-1.0) {$I_2$};
    \end{tikzpicture}
    \begin{tikzpicture}
        \draw [line width=1.5pt] (0.5,0.5) -- (0.5,-0.5);
        \draw [line width=1.5pt](0.5,-0.5) -- (-0.5,-0.5);
        \draw [line width=1.5pt](-0.5,-0.5) -- (-0.5,0.5);
        \draw [line width=1.5pt](-0.5,0.5) -- (0.5,0.5);
        \draw [line width=1.5pt](0.5,0.5) -- (1.0,1.0);
        \draw [line width=1.5pt](0.5,-0.5) -- (1.0,-1.0);
        \draw [line width=1.5pt] (-0.5,-0.5) -- (-1.0,-1.0);
        \draw [line width=1.5pt] (-0.5,0.5) -- (-1.0,1.0);
        \draw [line width=0.5pt] (-2.0,0) -- (-1.75,0);
        \node [anchor=east] at (-1.0,1.0) {$I_1$};
        \node [anchor=west] at (1.0,1.0) {$I_4$};
        \node [anchor=west] at (1.0,-1.0) {$I_3$};
        \node [anchor=east] at (-1.0,-1.0) {$I_2$};
    \end{tikzpicture}
    \begin{tikzpicture}
        \draw [line width=1.5pt] (-0.8,0.3) -- (-1.2,0.6);
        \draw [line width=1.5pt](-0.8,-0.3) -- (-1.2,-0.6);
        \draw [line width=1.5pt](-0.8,0.3) -- (-0.3,0);
        \draw [line width=1.5pt](-0.8,-0.3) -- (-0.3,0);
        \draw [line width=1.5pt](-0.8,0.3) -- (-0.8,-0.3);
        \draw [line width=1.5pt](-0.3,0) -- (0.6,0);
        \draw [line width=1.5pt](0.6,0) -- (1.2,-0.6);
        \draw [line width=1.5pt](0.6,0) -- (1.2,0.6);

        \node [anchor=east] at (-1.0,1.0) {$I_1$};
        \node [anchor=west] at (1.0,1.0) {$I_4$};
        \node [anchor=west] at (1.0,-1.0) {$I_3$};
        \node [anchor=east] at (-1.0,-1.0) {$I_2$};
        \node at (-2,0) {$=-$};
    \end{tikzpicture}
    \begin{tikzpicture}
        \draw [line width=1.5pt] (-0.6,0) -- (-1.2,0.6);
        \draw [line width=1.5pt](-0.6,0) -- (-1.2,-0.6);
        \draw [line width=1.5pt](-0.6,0) -- (0.6,0);
        \draw [line width=1.5pt](0.6,0) -- (1.2,-0.6);
        \draw [line width=1.5pt](0.6,0) -- (1.2,0.6);

        \node [anchor=east] at (-1.0,1.0) {$I_1$};
        \node [anchor=west] at (1.0,1.0) {$I_4$};
        \node [anchor=west] at (1.0,-1.0) {$I_3$};
        \node [anchor=east] at (-1.0,-1.0) {$I_2$};
        \node at (-2,0) {$\propto$};
    \end{tikzpicture}
    \caption{Simplifying $\bar{\mathtt{c}}_s$ to an exchange diagram with vertex correction using Jacobi identity.}
    \label{fig:cbar}
\end{figure}
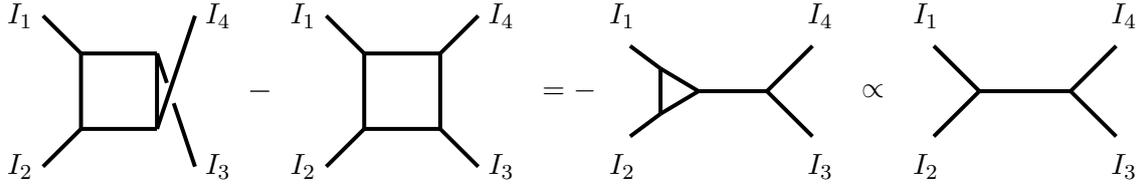

Being physically irrelevant, the latter two types of dof do have the virtue of allowing us the freedom in writing $\mathcal{H}^{(2)}$ into convenient forms while preserving the structure of \eqref{eq:oneloopansatz}. In particular, by tuning the last type of dof, i.e.~the parameter accompanying $u^2 v^{-1}\left(\mathtt{d}_{st} + \mathtt{d}_{su} + \mathtt{d}_{tu}\right)$ in $\bar{\mathcal{H}}^{(1)}$, we can make $\bar{\mathcal{H}}^{(1)}$  \textit{exactly} the same as the tree-level correlator $\mathcal{H}^{(1)}$, upon the replacement of the color structures $\mathtt{c}_{s,t,u}\mapsto \bar{\mathtt{c}}_{s,t,u}$
\begin{equation}
    \bar{\mathtt{c}}_s = \frac{1}{6}\left(\mathtt{d}_{su}-\mathtt{d}_{st}\right)\;,\quad 
    \bar{\mathtt{c}}_t = \frac{1}{6}\left(\mathtt{d}_{st}-\mathtt{d}_{tu}\right)\;,\quad
    \bar{\mathtt{c}}_u = \frac{1}{6}\left(\mathtt{d}_{tu}-\mathtt{d}_{su}\right)\;.
\end{equation}
Clearly, the new color structures $\bar{\mathtt{c}}_{s,t,u}$ has the same crossing properties as the unbarred ones ({\it e.g.}, under $1\leftrightarrow 3$,  $\bar{\mathtt{c}}_u\to -\bar{\mathtt{c}}_u$  and  $\bar{\mathtt{c}}_s\to -\bar{\mathtt{c}}_t$), and also satisfy the Jacobi identity $\bar{\mathtt{c}}_s + \bar{\mathtt{c}}_t + \bar{\mathtt{c}}_u = 0$.  In fact, $\bar{\mathtt{c}}_{s,t,u}$ are the same as $\mathtt{c}_{s,t,u}$ up to a $G_F$-dependent factor. This can be proven by using the Jacobi identity as in Figure \ref{fig:cbar}. In the first step, the Jacobi identity turns the difference of the two color box diagrams into an exchange color diagram with a vertex correction. Since the structure constant $f^{abc}$ is the only invariant tensor with three indices, the correction to the vertex is only a multiplicative factor and the color structure is the same as at the tree level. We see that even though the ansatz starts off with a modification term $\bar{\mathcal{H}}^{(1)}$ that has very minimal resemblance to the tree-level correlator $\mathcal{H}^{(1)}$, the constraints shape it into the latter.

With the above convention our result for $\mathcal{H}^{(2)}$ is thus put into the form
\begin{equation}
    \mathcal{H}^{(2)}=\Delta^{(4)}\mathcal{L}^{(2)}+\underbrace{\bar{\mathtt{c}}_s\mathcal{H}^{(1)}_{s}+\bar{\mathtt{c}}_t\mathcal{H}^{(1)}_{t}+\bar{\mathtt{c}}_u\mathcal{H}^{(1)}_{u}}_{\propto\mathcal{H}^{(1)}}+a_0\mathcal{H}^{(2)}_{\rm c},
\end{equation}
with some free parameter $a_0$. The pre-correlator receives the color decomposition
\begin{equation}
    \mathcal{L}^{(2)}(z,\zb) = \dst\mathcal{L}^{(2)}_{st}(z,\zb) + \dsu\mathcal{L}^{(2)}_{su}(z,\zb) + \dtu\mathcal{L}^{(2)}_{tu}(z,\zb),
\end{equation}
where the component $\mathcal{L}^{(2)}_{st}$ is related to $\mathcal{L}^{(2)}_{su}$ by crossing symmetry
\begin{equation}
        \mathcal{L}^{(2)}_{st}(z,\zb) = \mathcal{L}^{(2)}_{su}\left( \frac{z}{z-1},\frac{\zb}{\zb-1} \right).
\end{equation}
Before writing down the explicit expressions let us introduce several functions derived from (linear combinations of) the SVMPL basis
\begin{subequations}\label{eq:H2WQs}
\begin{align}
    W_{2}(z,\zb) =& \text{Li}_2(z)-\text{Li}_2(\zb)+\frac{1}{2} \log u\, \log \frac{1-z}{1-\zb}\;,\\
    W_{3}(z,\zb) =& \text{Li}_3(z)+\text{Li}_3(\zb)-\frac{1}{2} \log u\, \left(\text{Li}_2(z)+\text{Li}_2(\zb)\right)-\frac{1}{12} \log ^2 u\, \log v\;,\\
    \label{eq:Q3}Q_3(z,\zb) =& \text{Li}_3(z)-\text{Li}_3(\zb)-\text{Li}_3\left(\frac{z}{z-1}\right)+\text{Li}_3\left(\frac{\zb}{\zb-1}\right) \nonumber \\ &- \text{Li}_2(z) \log (1-\zb)+\text{Li}_2(\zb) \log (1-z) \nonumber \\ & + \frac{1}{4} \log \frac{1-z}{1-\zb}\, \log u\, \log \frac{u}{v}+\frac{1}{6} \log ^3(1-z)-\frac{1}{6} \log ^3(1-\zb)  \nonumber\\ &+ \frac{1}{2} \log u\, \left(G_{\frac{1}{z},\frac{1}{\zb}}(1)-G_{\frac{1}{\zb},\frac{1}{z}}(1)\right)-G_{0,\frac{1}{z},\frac{1}{\zb}}(1)+G_{0,\frac{1}{\zb},\frac{1}{z}}(1)\;, \\
    W_{4}(z,\zb) =& \text{Li}_4(z)-\text{Li}_4(\zb)-\frac{1}{2} \log u\, \left(\text{Li}_3(z)-\text{Li}_3(\zb)\right)+\frac{1}{12} \log ^2 u\, \left(\text{Li}_2(z)-\text{Li}_2(\zb)\right)\;. 
\end{align}
\end{subequations}
With these the two independent $\mathcal{L}^{(2)}$ components are
\begin{equation}
    \begin{split}
        \mathcal{L}^{(2)}_{su}(z,\zb) =&- \frac{6u^3(1+u-v)}{(z-\zb)^5} W_4(z,\zb) + \frac{u}{3(z-\zb)}W_4(1-z,1-\zb)  \\
        &+ \left[ \frac{u^2}{2(z-\zb)^2} + \frac{6u^3}{(z-\zb)^4} \right] W_{3}(z,\zb) + \left[ \frac{2u^3}{3(z-\zb)^3} + \frac{4u^3v}{(z-\zb)^5} \right] Q_{3}(z,\zb) \\
        &+ \left[ \frac{u^2(1-7u-v)}{12(z-\zb)^3} + \frac{u^3v(1-u-3v)}{(z-\zb)^5} \right]  W_2(z,\zb)\log u\\
        &+ \left[ \frac{u^3}{3(z-\zb)^3} + \frac{2u^3v}{(z-\zb)^5} \right]  W_2(z,\zb)\log v - \left[ \frac{u^2(3+3u-v)}{3(z-\zb)^3} + \frac{8u^3v}{3(z-\zb)^5} \right] W_2(z,\zb)\\
        &+ \frac{u^3(1+u-v)}{2(z-\zb)^4} \log^2 u - \left[ \frac{u(1+u-v)}{12(z-\zb)^2} - \frac{u^2(1-u)(1+u-v)}{2(z-\zb)^4} \right] \log u\,\log v \\
        & - \left[ \frac{u(1-v)}{3(z-\zb)^2} - \frac{2u^2(1-u+v)(1-u-v)}{3(z-\zb)^4} \right]\log v \\
        & - \left[ \frac{u^2}{6(z-\zb)^2} - \frac{4u^3(1-u+v)}{3(z-\zb)^4} \right] \log u + \frac{2u^2}{3(z-\zb)^2} \\
        & + a_1 + a_2 u \bar{D}_{1111} - a_5 \log v\;,
    \end{split}
\end{equation}
and
\begin{equation}
    \begin{split}
        \mathcal{L}^{(2)}_{tu}(z,\zb) =& -\left[ \frac{2u}{3(z-\zb)} + \frac{6uv}{(z-\zb)^3} - \frac{6u^2v(1-u+v)}{(z-\zb)^5} \right]W_4(1-z,1-\zb)\\
        &- \left[ \frac{u(3-4u+3v)}{2(z-\zb)^2} - \frac{6u^2v}{(z-\zb)^4} \right] W_{3}(1-z,1-\zb) \\
        &+ \left[ \frac{2u^3}{3(z-\zb)^3} + \frac{4u^3v}{(z-\zb)^5} \right] Q_{3}(z,\zb) - \left[ \frac{u^3}{3(z-\zb)^3} + \frac{2u^3v}{(z-\zb)^5} \right]  W_2(z,\zb)\log u\\
        &- \left[ \frac{u}{8(z-\zb)} + \frac{u(3-7u+9v)(1+u-v)}{24(z-\zb)^3} - \frac{u^2v(1+u-v)}{(z-\zb)^5}\right]  W_2(z,\zb)\log v \\
        &- \left[ \frac{2u(1-v)^2}{3(z-\zb)^3} + \frac{8u^3v}{3(z-\zb)^5} \right] W_2(z,\zb) - \left[ \frac{u(1-u-5v)}{12(z-\zb)^2} - \frac{u^2v(1-u+v)}{2(z-\zb)^4} \right] \log^2 v \\
        & - \left[ \frac{u(1-v)}{3(z-\zb)^2} - \frac{u^2(1-v)(1-u+v)}{2(z-\zb)^4} \right] \log u\,\log v - \left[ \frac{2u^2}{3(z-\zb)^2} - \frac{4u^3(1-u+v)}{3(z-\zb)^4} \right]\log v \\
        & - \left[ \frac{u(3-2u-3v)}{6(z-\zb)^2} - \frac{2u^2(1-u+v)(1-u-v)}{3(z-\zb)^4} \right] \log u + \frac{2u^2}{3(z-\zb)^2}\\
        &+ a_3 + a_4 u \bar{D}_{1111}\;.
    \end{split}
\end{equation}
These $a_i$s are possible ambiguities described in \eqref{eq:L2kernel}, and $\Delta^{(4)}$ maps them to 0. After applying the $\Delta^{(4)}$ operator the full $\mathcal{H}^{(2)}$ can be analytically expressed in term of the same set of functions \eqref{eq:H2WQs}, and the explicit results are presented in Appendix \ref{sec:H2analytic}.

Let us conclude our one-loop bootstrap computation by making a comment  regarding a simplification. In \cite{Aprile:2019rep} the authors observed that for the lowest KK modes there is also a ``without really trying'' way to bootstrap $AdS_5\times S^5$ one-loop correlators. Based on a similar ansatz in terms of a pre-correlator, they noticed that the exactly same result can be reproduced when replacing the constraints from the complete and theory-specific leading log data by a minimal requirement that the leading log singularities have analytic support on spin. This means that given the other conditions the leading log data turns out to be largely redundant for the determination of the one-loop super graviton correlator in $AdS_5\times S^5$. A very similar phenomenon occurs in the super gluon case in $AdS_5\times S^3$ as well. If one temporarily ignores the leading log condition during the computation in Section \ref{sec:oneloop:ansatz}, the result will be almost the same as described above, and the difference resides in only one extra remaining dof whose contribution to the leading log singularities is proportional to the color factor $\mathtt{d}_{tu}$. This contribution must vanish as it contradicts  the possible color structures in the leading logarithmic singularity following our previous analysis. This allows us to recover our result for $\mathcal{H}^{(2)}$ without inputting the precise details  of the leading logarithmic singularity.

\subsection{Comparison with the Mellin space result}\label{Subsec:compareMellin}

To further confirm the validity of the result from our position space computation, let us now compare it with the Mellin space result of \cite{Alday:2021ajh}. The latter is expressed in terms of the Mellin amplitude, which can be written in a $G_F$-independent form as
\begin{equation}\label{eq:mellinresult}
    \widetilde{\mathcal{M}}_{2222}^{\scalebox{0.65}{$\text{AdS}_5\! \times\! S^3$}}=9\big(\mathtt{d}_{st}\mathcal{B}_{st}^{\rm  8d}+\mathtt{d}_{su}\mathcal{B}_{su}^{\rm  8d}+\mathtt{d}_{tu}\mathcal{B}_{tu}^{\rm  8d}\big)\;,
\end{equation}
where
\begin{eqnarray}\label{boxAdS5S3}
\nonumber \mathcal{B}^{\rm 8d}_{st}=\sum_{m,n=2}^\infty\frac{c_{mn}^{\rm 8d}}{(s-2m)(t-2n)}\;,\\
\mathcal{B}^{\rm 8d}_{su}=\sum_{m,n=2}^\infty\frac{c_{mn}^{\rm 8d}}{(s-2m)(\tilde{u}-2n)}\;,\\
\nonumber \mathcal{B}^{\rm 8d}_{tu}=\sum_{m,n=2}^\infty\frac{c_{mn}^{\rm 8d}}{(t-2m)(\tilde{u}-2n)}\;,
\end{eqnarray}
with the coefficients given by 
\begin{equation}
c_{mn}^{\rm 8d}=\frac{4 \left(3 m^2 n-4 m^2+3 m n^2-16 m n+15 m-4 n^2+15 n-12\right)}{27 (m+n-4) (m+n-3) (m+n-2)}\;.
\end{equation}
  The $s,t,\tilde{u}$ here are Mellin variables, satisfying $s+t+\tilde{u}=6$, and $\mathtt{d}_{st}, \mathtt{d}_{su}, \mathtt{d}_{tu}$ are color structures of three types of box diagrams. However, these expressions as infinite double sums are a bit formal as the sums need regularizations. The regularized and resummed amplitude was obtained in \cite{Alday:2021ajh} and reads
\begin{align}
\mathcal{B}_{s t}^{8 \mathrm{d}}= & R_0(s, t)\left(\psi^{(1)}\left(2-\frac{s}{2}\right)+\psi^{(1)}\left(2-\frac{t}{2}\right)-\left(\psi^{(0)}\left(2-\frac{s}{2}\right)-\psi^{(0)}\left(2-\frac{t}{2}\right)\right)^2\right)\nonumber \\ 
+&R_1(s, t) \psi^{(0)}\left(2-\frac{s}{2}\right)+R_1(t, s) \psi^{(0)}\left(2-\frac{t}{2}\right)+\pi^2 R_2(s, t)-\frac{32}{27(s+t-8)}+b\;,
\end{align}
where $\psi ^{(n)}(x)$ is the ploygamma function defined by $\psi ^{(n)}(x) = \left(\log\Gamma(x)\right)^{(n)}$ and 
\begin{align}
& R_0(s, t)=-\frac{4\left(3 s^2 t-8 s^2+3 s t^2-32 s t+60 s-8 t^2+60 t-96\right)}{27(s+t-8)(s+t-6)(s+t-4)}\;, \\
& R_1(s, t)=\frac{8\left(3 s^2+3 s t-26 s-10 t+48\right)}{27(s+t-8)(s+t-4)}\;, \\
& R_2(s, t)=\frac{4\left(3 s^2 t-8 s^2+3 s t^2-32 s t+60 s-8 t^2+60 t-96\right)}{27(s+t-8)(s+t-6)(s+t-4)}\;.
\end{align}
The parameter $b$ is a regularization constant corresponding to a contact counterterm. With this expression, one could in principle just perform the the following inverse Mellin transformation
\begin{equation}\label{eq:mellintoH}
\mathcal{H}^{(2)}=\int_{-i \infty}^{i \infty} \frac{\dd s \dd t}{(4 \pi i)^2} u^{\frac{s}{2}} v^{\frac{t-4}{2}} \widetilde{\mathcal{M}}_{2222}^{\scalebox{0.65}{$\text{AdS}_5\! \times\! S^3$}} \Gamma^2\left[\frac{4-s}{2}\right]\Gamma^2\left[\frac{4-t}{2}\right]\Gamma^2\left[\frac{4-\tilde{u}}{2}\right]
\end{equation}
to translate the Mellin amplitude into a position space expression. However, technically this is difficult and we do not have a general understanding of how to convert the integral into our basis functions. Instead, we find it easier to seek for series expansions on both sides and compare term by term in the series. On the one hand, for $\mathcal{H}^{(2)}$ we can expand our position space result around $z=0$ and $\bar{z}=1$, and rewrite the resulting series in terms of small $u$ and $v$. On  the other hand, in the inverse Mellin transformation \eqref{eq:mellintoH} we can close the contours for both $s$ and $t$ to the right so that the integrals effectively transform into a sum over residues at $s=2m$ and $t=2n$ for $m,n\in\mathbb{Z}_{\geq 2}$. This also gives rise to an expansion in small $u$ and $v$. We find perfect agreement between the two series upon identifying the counterterm parameters as $b=4(-2a_0+6 \gamma-25)/27$ ($\gamma$ being the Euler--Mascheroni constant).

\section{Two-loop correlator}\label{sec:twoloops}

In the previous section we saw how the hidden symmetry structure \eqref{eq:oneloopansatz} helped us to bootstrap the one-loop correlator. In particular, we found that the tree-like piece $\bar{\mathcal{H}}^{(1)}$ is dynamically the same as the tree-level correlator $\mathcal{H}^{(1)}$, up to a simple modification in the color structures. we now turn to the two-loop level and further make use of this encouraging fact. Comparing with the super graviton correlator at two loops (see \eqref{eq:hs10d:2}), we will need a tree-like function $\widetilde{\mathcal{H}}^{(1)}$ and a one-loop-like function $\widetilde{\mathcal{H}}^{(2)}$ in the ansatz for the super gluon two-loop correlator $\mathcal{H}^{(3)}$. These extra pieces will be related to the corresponding correlators in the same way as at one loop. 

\subsection{Color structures at two loops}\label{sec:color2loop}

Parallel to the one-loop computation we begin by analyzing the color structures at two loops. Again we use the s-channel projectors for the $E_8$ group as an efficient tool to find out the linear relations among color factors of two-loop diagrams as well as their transformations under crossing. It is worth noting that despite of this $G_F$-specific technique, these resulting relations are in fact independent of the choice of the gauge group $G_F$.

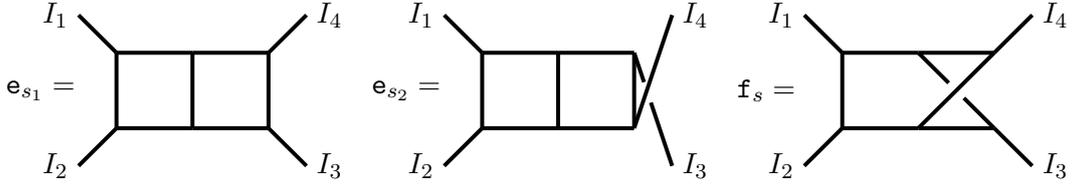
\begin{figure}[h]
    \centering
   \begin{tikzpicture}
        \draw [line width=1.5pt] (0,0.5) -- (0,-0.5);
        \draw [line width=1.5pt](0,0.5) -- (1,0.5);
        \draw [line width=1.5pt](1,0.5) -- (1,-0.5);
        \draw [line width=1.5pt](0,-0.5) -- (1,-0.5);
        \draw [line width=1.5pt](0,-0.5) -- (-1.0,-0.5);
        \draw [line width=1.5pt](-1,-0.5) -- (-1,0.5);
        \draw [line width=1.5pt] (-1,0.5) -- (0,0.5);
        \draw [line width=1.5pt] (1,0.5) -- (1.5,1.0);
        \draw [line width=1.5pt] (-1,0.5) -- (-1.5,1.0);
        \draw [line width=1.5pt] (1,-0.5) -- (1.5,-1.0);
        \draw [line width=1.5pt] (-1,-0.5) -- (-1.5,-1.0);
        \node at (-2,0) {$\mathtt{e}_{s_1} = $};
        \node [anchor=east] at (-1.5,1) {$I_1$};
        \node [anchor=west] at (1.5,1.0) {$I_4$};
        \node [anchor=west] at (1.5,-1.0) {$I_3$};
        \node [anchor=east] at (-1.5,-1.0) {$I_2$};
    \end{tikzpicture}
    \begin{tikzpicture}
        \draw [line width=1.5pt] (0,0.5) -- (0,-0.5);
        \draw [line width=1.5pt](0,0.5) -- (1,0.5);
        \draw [line width=1.5pt](1,0.5) -- (1,-0.5);
        \draw [line width=1.5pt](0,-0.5) -- (1,-0.5);
        \draw [line width=1.5pt](0,-0.5) -- (-1.0,-0.5);
        \draw [line width=1.5pt](-1,-0.5) -- (-1,0.5);
        \draw [line width=1.5pt] (-1,0.5) -- (0,0.5);
        \draw [line width=1.5pt] (1,0.5) -- (1.5,-1.0);
        \draw [draw=white,line width=2pt] (1.13,0.11) -- (1.22,-0.16);
        \draw [line width=1.5pt] (1,-0.5) -- (1.5,1.0);
        \draw [line width=1.5pt] (-1,0.5) -- (-1.5,1.0);
        \draw [line width=1.5pt] (-1,-0.5) -- (-1.5,-1.0);
        \node at (-2,0) {$\mathtt{e}_{s_2} = $};
        \node [anchor=east] at (-1.5,1) {$I_1$};
        \node [anchor=west] at (1.5,1.0) {$I_4$};
        \node [anchor=west] at (1.5,-1.0) {$I_3$};
        \node [anchor=east] at (-1.5,-1.0) {$I_2$};
    \end{tikzpicture}
    \begin{tikzpicture}
        \draw [line width=1.5pt](0,0.5) -- (1,0.5);
        \draw [line width=1.5pt](0,-0.5) -- (1,-0.5);
        \draw [line width=1.5pt](0,-0.5) -- (-1.0,-0.5);
        \draw [line width=1.5pt](-1,-0.5) -- (-1,0.5);
        \draw [line width=1.5pt] (-1,0.5) -- (0,0.5);
        \draw [line width=1.5pt] (1,0.5) -- (1.5,1.0);
        \draw [line width=1.5pt] (-1,0.5) -- (-1.5,1.0);
        \draw [line width=1.5pt] (1,-0.5) -- (1.5,-1.0);
        \draw [line width=1.5pt] (-1,-0.5) -- (-1.5,-1.0);
        \draw [line width=1.5pt] (0,0.5) -- (1,-0.5);
        \draw [draw=white,line width=2pt] (0.4,0.1) -- (0.6,-0.1);
        \draw [line width=1.5pt] (0,-0.5) -- (1,0.5);
        \node at (-2,0) {$\mathtt{f}_s = $};
        \node [anchor=east] at (-1.5,1) {$I_1$};
        \node [anchor=west] at (1.5,1.0) {$I_4$};
        \node [anchor=west] at (1.5,-1.0) {$I_3$};
        \node [anchor=east] at (-1.5,-1.0) {$I_2$};
    \end{tikzpicture}
    \caption{2-loop color structures $\mathtt{e}_{s_1}$, $\mathtt{e}_{s_2}$ and $\mathtt{f}_s$. }
    \label{fig:fourpt2loop}
\end{figure}

At two-loop level we encounter planar and non-planar double-box diagrams. In Figure \ref{fig:fourpt2loop} we show these diagrams in the s-channel, and they can be obtained by gluing lower-loop diagrams as
\begin{align}
    \mathtt{e}_{s_1} &= (-\mathtt{c}_t)^3 = \left( \psi^2 h^\vee \right)^3 \left(1,\frac{1}{125},-\frac{1}{27000},\frac{1}{8},0 \right),\\
    \mathtt{e}_{s_2} &= (-\mathtt{c}_t)^2 \mathtt{c}_u = \left( \psi^2 h^\vee \right)^3 \left(1,\frac{1}{125},-\frac{1}{27000},-\frac{1}{8},0 \right),\\
    \mathtt{f}_s &= -\mathtt{c}_t \mathtt{d}_{tu} = \left( \psi^2 h^\vee \right)^3 \left(\frac{1}{2},-\frac{3}{250},\frac{2}{3375},0,0 \right).
\end{align}
The t- and u-channel diagrams can be obtained by applying the color crossing matrices defined in \eqref{eq:E8FtFu}
\begin{align}
    \left( \mathtt{e}_{t_1}, \mathtt{e}_{t_2}, \mathtt{f}_t \right) &= \left( {\rm F}_t\, \mathtt{e}_{s_1}, {\rm F}_t\, \mathtt{e}_{s_2}, {\rm F}_t\, \mathtt{f}_s \right),\\
    \left( \mathtt{e}_{u_1}, \mathtt{e}_{u_2}, \mathtt{f}_u \right) &= \left( {\rm F}_u\, \mathtt{e}_{s_1}, {\rm F}_u\, \mathtt{e}_{s_2}, {\rm F}_u\, \mathtt{f}_s \right).
\end{align}
The two-loop color structures described above are not linearly independent. For concreteness in the ansatz construction, we choose $\mathtt{e}_{s_1}$, $\mathtt{e}_{s_2}$, $\mathtt{f}_s$, $\mathtt{e}_{t_1}$ and $\mathtt{e}_{u_1}$ to be our basis. The other two-loop color structures can be decomposed onto this basis as
\begin{subequations}\label{eq:colorbasis2loop}
    \begin{align}
        \mathtt{e}_{t_2} &= -2 \mathtt{e}_{s_1} + 3 \mathtt{f}_s + 2 \mathtt{e}_{t_1} + \mathtt{e}_{u_1}\;,\\
        \mathtt{f}_t &= - \mathtt{e}_{s_1} + \mathtt{f}_s + \mathtt{e}_{t_1}\;,\\
        \mathtt{e}_{u_2} &= \mathtt{e}_{s_1} + \mathtt{e}_{s_2} - 3 \mathtt{f}_s - \mathtt{e}_{t_1}\;,\\
        \mathtt{f}_u &= \mathtt{e}_{s_1} - 2 \mathtt{f}_s - \mathtt{e}_{t_1}\;.
    \end{align}
\end{subequations}
Note these relations \eqref{eq:colorbasis2loop} can also be proved by using Jacobi identity, and are therefore  $G_F$-independent. Together with the associated color diagrams, these relations also completely determine the behavior of each color factor under exchanging operators.

The presence of the planar double box diagrams is already suggested by the leading log singularities following the same logic as that at one loop. The need of the non-planar double box diagrams can be expected from the flat space limit, where $\mathcal{H}^{(3)}$ should reduce to the two-loop scattering amplitude in 8d maximal super Yang--Mills, whose result decomposes into both planar and non-planar diagrams as shown in \eqref{eq:tildeA2loop}. One may also wonder if we need to introduce more color structures at two loops to write down the reduced correlator $\mathcal{H}^{(3)}$. The answer is no. 
It is a simple exercise to check that any color factors in correspondence to two-loop four-point diagrams can always be linearly decomposed onto our choice of basis $\{\mathtt{e}_{s_1},\mathtt{e}_{s_2},\mathtt{f}_s,\mathtt{e}_{t_1},\mathtt{e}_{u_1}\}$. Therefore this basis is already sufficient for uniquely setting up the ansatz.

\subsection{Ansatz and constraints}\label{sec:H3constraints}

For IIB supergravity on $AdS_5\times S^5$, the hidden symmetry structure \eqref{eq:hs10d1l} extends to two loops in the form \cite{Huang:2021xws}
\begin{equation}\label{eq:hs10d:2}
    \mathcal{H}_{AdS_5\times S^5}^{(3)} = \left[\Delta^{(8)}\right]^2 \mathcal{L}_{AdS_5\times S^5}^{(3)} +\frac{5}{4} \mathcal{H}_{AdS_5\times S^5}^{(2)} - \frac{1}{16} \mathcal{H}_{AdS_5\times S^5}^{(1)},
\end{equation}
where $\mathcal{H}_{AdS_5\times S^5}^{(1)}$ and $\mathcal{H}_{AdS_5\times S^5}^{(2)}$ are precisely the tree and one-loop reduced correlator. Given the analogy at one loop, this naturally suggests a similar structure for super gluons in ${AdS}_5 \times S^3$ at two loops. In other words, the correlator $\mathcal{H}^{(3)}$ is constructed from a two-loop pre-correlator $\mathcal{L}^{(3)}$, and is completed by a one-loop-like function $\widetilde{\mathcal{H}}^{(2)}$ and a tree-like function $\widetilde{\mathcal{H}}^{(1)}$
\begin{equation}\label{eq:H3L3}
\mathcal{H}^{(3)}=\left[\Delta^{(4)}\right]^2 \mathcal{L}^{(3)}+\widetilde{\mathcal{H}}^{(2)}+\widetilde{\mathcal{H}}^{(1)}\;.
\end{equation}
As we have mentioned in \eqref{eq:delta4ex12}, the operator $\Delta^{(4)}$ only preserves the Bose symmetry of $1\leftrightarrow 2$. However, for $\left[\Delta^{(4)}\right]^2$ there is an enhanced Bose symmetry\footnote{This enhanced symmetry was first observed on $[\Delta^{(8)}]^2$ in \cite{Drummond:2022dxw}.}. Under $1\leftrightarrow 3$, the operator $\left[\Delta^{(4)}\right]^2$ transforms as
\begin{equation}\label{eq:delta4ex13}
    \left(\frac{u}{v}\right)^{-3}\left[\Delta^{(4)}\right]^2 \frac{u}{v} = \left.\left[\Delta^{(4)}\right]^2 \right\vert_{z\rightarrow 1-z,\zb\rightarrow 1-\zb}.
\end{equation}
This allows us to directly impose the crossing property under $1\leftrightarrow 3$ on the pre-correlator $\mathcal{L}^{(3)}$. Together with the $1\leftrightarrow 2$ crossing, we conclude that $\mathcal{L}^{(3)}$ is fully crossing symmetric. We have the following ansatz for the pre-correlator
\begin{equation}
\mathcal{L}^{(3)} = \sum_CC\sum_{w=0}^6\sum_i \frac{\hsv_{w,i}(z,\zb)}{(z-\zb)^5} r^{C}_{w,i}(z,\zb)\,,
\end{equation}
where $C$ is the two-loop color factor taking values in the basis $\{\mathtt{e}_{s_1},\mathtt{e}_{s_2},\mathtt{f}_s,\mathtt{e}_{t_1},\mathtt{e}_{u_1}\}$. $r^{C}_{w,i}(z,\zb)$ are polynomials of $z$ and $\zb$ with the power in each variable no higher than $5$
\begin{equation}
     r^{C}_{w,i}(z,\zb) = \sum_{j,k=0}^5 (\rho^{C}_{w,i})_{j,k} z^j \zb^k,
\end{equation}
where all the coefficients $(\rho^{C}_{w,i})_{j,k}$ are unknown rational numbers. The pattern of this ansatz follows exactly the same considerations as that discussed at one loop.

At one loop we observed that the modification term $\bar{\mathcal{H}}^{(1)}$ there can be tuned such that it descends from the tree-level correlator $\mathcal{H}^{(1)}$ by keeping its dynamical part while replacing the original color factors $\mathtt{c}_{s,t,u}$ with another set of factors $\bar{\mathtt{c}}_{s,t,u}$. The $\bar{\mathtt{c}}$ factors are linear combinations of the one-loop color factors $\mathtt{d}$, but obey the same algebras as the tree-level $\mathtt{c}$. Inspired by this result, at two loops we directly make the assumption that the modification terms $\widetilde{\mathcal{H}}^{(2)}$ and $\widetilde{\mathcal{H}}^{(1)}$ descend from the one-loop $\mathcal{H}^{(2)}$ and tree-level $\mathcal{H}^{(1)}$ respectively, by merely replacing the original color factors into new ones formed by the basis $\{\mathtt{e}_{s_1},\mathtt{e}_{s_2},\mathtt{f}_s,\mathtt{e}_{t_1},\mathtt{e}_{u_1}\}$ while preserving the algebra of the color factors. Specifically, for the tree-like piece $\widetilde{\mathcal{H}}^{(1)}$ we perform the replacement $\mathtt{c}_{s,t,c}\mapsto\tilde{\mathtt{c}}_{s,t,u}$, requiring that the $\tilde{\mathtt{c}}$'s are related by crossing and sum up to zero $\tilde{\mathtt{c}}_s + \tilde{\mathtt{c}}_t + \tilde{\mathtt{c}}_u = 0$. The most general combinations under these conditions read
\begin{subequations}\label{eq:treeto2loop}
    \begin{align}
        \tilde{\mathtt{c}}_s &= a_1\left( 2\mathtt{e}_{s_1}\!-3\mathtt{f}_s\!-2 \mathtt{e}_{t_1}\right) + a_2\left( 2\mathtt{e}_{s_2}\!- 3\mathtt{f}_s\!- 2\mathtt{e}_{t_1}\right),\\
        \tilde{\mathtt{c}}_t &= a_1\left( -\mathtt{e}_{s_1}\! + 3\mathtt{f}_s\!+ \mathtt{e}_{t_1}\right) + a_2\left( 3\mathtt{e}_{s_1}\!- 3\mathtt{f}_s\!- \mathtt{e}_{t_1}\! - 2\mathtt{e}_{u_1}\right),\\
        \tilde{\mathtt{c}}_u &= a_1\left( -\mathtt{e}_{s_1}+ \mathtt{e}_{t_1}\right) + a_2\left( -3\mathtt{e}_{s_1}-2\mathtt{e}_{s_2}+6\mathtt{f}_s + 3\mathtt{e}_{t_1} + 2\mathtt{e}_{u_1}\right),
    \end{align}
\end{subequations}
which contain two undetermined parameters $a_1$ and $a_2$, and these are the only dof in $\widetilde{\mathcal{H}}^{(1)}$.  For the one-loop-like piece $\widetilde{\mathcal{H}}^{(2)}$ we replace $\mathtt{d}_{st,su,tu}\mapsto\tilde{\mathtt{d}}_{st,su,tu}$, but this time the new factors are only required to obey the crossing constraints. Correspondingly the most general solution reads
\begin{subequations}\label{eq:1loopto2loop}
    \begin{align}
        \tilde{\mathtt{d}}_{st} &= b_1\left( \mathtt{e}_{s_1}+ \mathtt{e}_{t_1}\right) + b_2\left( \mathtt{e}_{s_2}+\mathtt{e}_{t_1}+\mathtt{e}_{u_1}\right) + b_3\left( \mathtt{e}_{t_1} + \mathtt{f}_s\right),\\
        \tilde{\mathtt{d}}_{su} &= b_1\left( \mathtt{e}_{s_1}\!+ 2\mathtt{e}_{s_2}\!  -\mathtt{e}_{t_1} - 3\mathtt{f}_s\right) + b_2\left( \mathtt{e}_{s_2}\!+\mathtt{e}_{t_1}\!+\mathtt{e}_{u_1}\right)  + b_3\left(\mathtt{e}_{s_1}+\mathtt{e}_{s_2}-\mathtt{e}_{t_1}-2\mathtt{f}_s\right),\\
        \tilde{\mathtt{d}}_{tu} &= b_1\left(\! -2\mathtt{e}_{s_1}\!  +\! 2\mathtt{e}_{t_1}\! +\! 2\mathtt{e}_{u_1} \!+\! 3\mathtt{f}_s\right)\! +\! b_2\left( \mathtt{e}_{s_2}\!+\mathtt{e}_{t_1}\!+\mathtt{e}_{u_1}\right)\!+\! b_3\left(\!-\mathtt{e}_{s_1}\!+\mathtt{e}_{t_1}\!+\mathtt{e}_{u_1}\!+\mathtt{f}_s\right),
    \end{align}
\end{subequations}
and so the only dof in $\widetilde{\mathcal{H}}^{(2)}$ are $b_1$, $b_2$ and $b_3$. 

With the above ansatz constructed, let us list all the constraints which should be imposed on the reduced correlator. Some of these constraints have already appeared in the one-loop case and we will simply enumerate them without additional comments. These constraints include
\begin{itemize}
    \item {\bf Leading logarithmic singularity.} The leading logarithmic singularity of $\mathcal{L}^{(3)}$ should match $\mathcal{D}_2 h^{(3)}(z)$ given in (\ref{eq:h:3})
        \begin{align}
            &\left. \mathcal{L}^{(3)}(z,\zb)\right\vert_{\log^{n\geq 4} u} = 0,\\
            \label{eq:leadinglogcompareH3}&\left. \mathcal{L}^{(3)}(z,\zb)\right\vert_{\log^3 u} \quad  = \mathcal{D}_2 h^{(3)}(z).
        \end{align}
    \item {\bf Bose symmetry.}
    \begin{itemize}
        \item Exchanging 1 and 2. Invariance under $1\leftrightarrow2$ leads to the condition 
    \begin{equation}
     \mathcal{L}^{(3)}(z,\zb)= \left.\mathcal{L}^{(3)}\left(\frac{z}{z-1},\frac{\zb}{\zb-1}\right)\right\vert_{\mathtt{e}_{s_1} \leftrightarrow \, \mathtt{e}_{s_2},\ \mathtt{e}_{t_1} \rightarrow \, \mathtt{e}_{u_2},\ \mathtt{e}_{u_1} \rightarrow \, \mathtt{e}_{t_2}}.
        \end{equation}
        \item Exchanging 1 and 3. As we pointed out in (\ref{eq:delta4ex13}), although $\Delta^{(4)}$ is not invariant under $1\leftrightarrow 3$, the operator $(\Delta^{(4)})^2$ is. This leads to the following condition on $\mathcal{L}^{(3)}$
\begin{equation}
    \mathcal{L}^{(3)}(z,\zb)=\left.\frac{u}{v}\, \mathcal{L}^{(3)}(1-z,1-\zb)\right\vert_{\mathtt{e}_{s_1} \leftrightarrow \, \mathtt{e}_{t_1},\ \mathtt{e}_{s_2} \rightarrow \, \mathtt{e}_{t_2},\ \mathtt{e}_{u_1} \rightarrow \, \mathtt{e}_{u_2},\ \mathtt{f}_s \rightarrow \, \mathtt{f}_t}\;.
        \end{equation}
    \end{itemize}
    \item {\bf Symmetry under $z\leftrightarrow\zb$.} $\mathcal{L}^{(3)}$ should be invariant under $z\leftrightarrow\zb$
    \begin{equation}
  \mathcal{L}^{(3)}(z,\zb)=\mathcal{L}^{(3)}(\zb,z)\;.
        \end{equation}
    \item {\bf Finiteness at $z=\zb$.} $\mathcal{L}^{(3)}$ should remain finite at $z=\zb$ in Euclidean region. This means that the Taylor expansion of the numerators in $\mathcal{L}^{(3)}$ at $z=\zb$ should start with $(z-\zb)^5$ to cancel the $z-\zb$ poles in the ansatz.
    \item {\bf Cancellation of unphysical poles.} The $v$ pole appearing separately in $\left[\Delta^{(4)}\right]^2 \mathcal{L}^{(3)}$ and $\widetilde{\mathcal{H}}^{(1)}$ should cancel in the full reduced correlator $\mathcal{H}^{(3)}$, for the same reason that operators with twist $\tau<4$ are not supposed to appear in $\mathcal{H}$ beyond tree level.
\end{itemize}
However, it turns out that these constraints are not yet sufficient to completely fix the two-loop correlator. One can understand this from the fact that, by the dispersion relations the data necessary to determine the full two-loop correlator are encoded not only in the leading log singularities, but also in the subleading log singularities at $\log^2u$. Unfortunately this part of the correlator is in general expected to receive contributions from triple-trace operators, whose data are for the time being beyond our reach. Therefore we have to seek for other available physical constraints to help complete the bootstrap computation.  These extra constraints are
\begin{itemize}
    \item {\bf Bulk-point limit.} Under proper conditions scattering in AdS can reduce to the corresponding process in flat space. One such prescription that is convenient to carry out directly at the level of position-space correlators is the so-called bulk-point limit \cite{Maldacena:2015iua,Alday:2017vkk}. By approaching the limit $z=\bar{z}$ in the Lorentzian region, it forces the dominant contribution to be concentrated at the center of AdS which is locally flat. Therefore the leading divergence (i.e.~terms with the highest-order $z-\bar{z}$ pole) of the holographic correlator should match the flat-space scattering amplitude. In the case of $\mathcal{H}^{(3)}$ the flat-space counterpart is the two-loop four-gluon amplitude $\mathcal{A}^{\text{2-loop}}$ in 8 dimensions. 
    
    Two simplifications occur when analyzing $\mathcal{H}^{(3)}$ in the form of decomposition \eqref{eq:H3L3}. On the one hand, $[\Delta^{(4)}]^2\mathcal{L}^{(3)}$, $\widetilde{\mathcal{H}}^{(2)}$ and $\widetilde{\mathcal{H}}^{(1)}$ each scale as $(z-\bar{z})^{-13}$, $(z-\bar{z})^{-9}$ and $(z-\bar{z})^{-5}$ respectively, and so $[\Delta^{(4)}]^2\mathcal{L}^{(3)}$ dominates over the two modification terms in this limit. On the other hand, for any function $f(z,\bar{z})$
    \begin{equation}
        \Delta^{(4)}\frac{f(z,\zb)}{(z-\zb)^n} = \frac{\Gamma(n+4)}{\Gamma(n)} \frac{f(z,\zb)}{(z-\zb)^{n+4}} + \order{\frac{1}{(z-\zb)^{n+3}}},
    \end{equation}
    where the leading divergent term is obtained by applying all derivatives in $\Delta^{(4)}$ on $(z-\zb)^{-n}$. This means $\Delta^{(4)}$ acts on the dominant part in the bulk-point limit merely as a multiplication factor. Therefore it suffices to directly consider the connection between the pre-correlator $\mathcal{L}^{(3)}$ and the flat-space amplitude $\mathcal{A}^{\rm 2-loop}$.
    
    For the computation we follow the prescription described in \cite{Drummond:2022dxw} where we first analytically continue $z$ around $0$ and $\zb$ around $1$ counter-clock-wisely, and then set $z=\zb+2 \omega \zb \sqrt{1-\zb}$, with $\omega\to0$. Then the precise relation reads
        \begin{equation}\label{eq:flatspacelimit}
            \lim_{\omega\to0}\left.(z-\zb)^5 \mathcal{L}^{(3)}\left(z^{\circlearrowleft_0}, \zb^{\circlearrowleft_1} \right)\right|_{z=\zb+2 \omega \zb \sqrt{1-\zb}} 
            = \left.  \frac{48\pi^2 (1-\zb)^2 \zb^6}{s^{2}}  \widetilde{\mathcal{A}}^{\,\text{2-loop}}(x)\right|_{x=1 / \zb},
        \end{equation}
        where $\widetilde{\mathcal{A}}^{\,\text{2-loop}}$ is a quantity closely related to the amplitude $\mathcal{A}^{\text{2-loop}}$. Details of this relation and the analytic expression for $\widetilde{\mathcal{A}}^{\,\text{2-loop}}$ are discussed in Appendix \ref{sec:flatspacelimit}. In taking this limit, we will encounter singularities like $\log(z-\zb)$ due to the existence of the symbol letter $z-\zb$ in the basis. These $\log(z-\zb)$ singularities can be matched with the regulator $\epsilon^{-1}$ in  dimensional regularization by taking\footnote{We managed to match with the $\epsilon^0$ and $\epsilon^{-1}$ terms in $\mathcal{A}^{(2)}$ in the flat-space limit \eqref{eq:flatspacelimit}, but not the leading divergent part $\epsilon^{-2}$. However, this will not cause any problems since the leading divergent part in $\mathcal{A}^{(2)}$ is related to the UV divergence at the one-loop level, and can be absorbed by a suitable subtraction at one loop.}
        \begin{equation}\label{eq:bplzzb}
            \log(z-\zb) \to \log\left( \zb\sqrt{1-\zb} \right) + \log(s) + \frac{1}{4\epsilon}.
        \end{equation}
        Reversely, one may use \eqref{eq:bplzzb} to predict whether functions with symbol $z-\zb$ will show up in the correlator by the corresponding flat-space amplitude. If the flat-space amplitude contains UV divergence $\epsilon^{-n}$, then there must be functions with symbol $z-\zb$ at weight $n+2$ in the original correlator, in order to recover the $\epsilon^{-n}$ divergence in the bulk-point limit.
\item {\bf Data of twist-4 operators.} At twist $4$ it is known that the only long operators are double-trace operators and they are free of degeneracy at the classical level. Hence in this specific case we can safely use the data from the lower-order correlators to recursively determine their contributions to the subleading log terms at two loops, i.e.~coefficients of $\log^2u$ at small $z$ and $\bar{z}$
\begin{equation}\label{eq:H3logu2twist4}
    \begin{split}
        \mathcal{H}^{(3)}\Big\vert_{\substack{\log^2u\\\text{twist }4}}=&\sum_{\ell=0}^\infty\frac{a_{\mathbf{a}}^{(0)}(\gamma_{\mathbf{a}}^{(1)})^3}{8}\left(\partial_\tau g_{\tau+2,\ell(z,\bar{z})}\Big\vert_{\substack{\log{u}\to0\\\tau\to4}}\right)\\
        &+\sum_{\ell=0}^\infty\frac{1}{8}\left(a_{\mathbf{a}}^{(1)}(\gamma_{\mathbf{a}}^{(1)})^2+2 a_{\mathbf{a}}^{(0)} \gamma_{\mathbf{a}}^{(1)} \gamma_{\mathbf{a}}^{(2)}\right)\,g_{6,\ell}(z,\bar{z}),
    \end{split}
\end{equation}
where $a_{\mathbf{a}}^{(0)}$, $\gamma_{\mathbf{a}}^{(1)}$, $\gamma_{\mathbf{a}}^{(2)}$ arise in the disconnected, tree-level and one-loop correlators respectively. The coefficients in the first line can be determined following the discussion in Section \ref{sec:recursionunitarity}. They already make an appearance in the leading logarithmic singularities, and so effectively we have used them. The data that actually generate new constraints are the coefficients in the second line \footnote{The validity of \eqref{eq:twist4data} is for $\ell \geq1$ as the one-loop counterterm spoils the analyticity to $\ell=0$. }
    \begin{equation}\label{eq:twist4data}
     \begin{split}
              & a_{\mathbf{a}}^{(1)}(\gamma_{\mathbf{a}}^{(1)})^2+2 a_{\mathbf{a}}^{(0)} \gamma_{\mathbf{a}}^{(1)} \gamma_{\mathbf{a}}^{(2)}\\
             =&- (1+(-1)^\ell) (\mathtt{f}_s)_{\mathbf{a}}\ \frac{\Gamma (\ell+3)^2}{\Gamma (2 \ell+5)}\frac{16 \left(5 \ell^2+25 \ell+24\right) }{3 \ell (\ell+1)^2 (\ell+4)^2 (\ell+5)}\\
             &+(\mathtt{e}_{s_1}+(-1)^\ell \mathtt{e}_{s_2})_{\mathbf{a}}\ \frac{\Gamma(\ell+3)^2}{\Gamma(2\ell+5)}\left[-\frac{32 \left(2\ell^3+23\ell^2+65\ell+32\right)}{\ell (\ell+1)^3 (\ell+4)^3 (\ell+5)}\right.\\
             &\qquad + \left.  \frac{8 \left(12\psi ^{(0)}(\ell+3)-12\psi ^{(0)}(2 \ell+5)+18-(-1)^\ell \right)}{3(\ell+1)^2 (\ell+4)^2} \right]\;.
        \end{split}
        \end{equation}
The detailed recursive calculation of \eqref{eq:twist4data} is presented in Appendix \ref{sec:recursion}. The same coefficients can alternatively be obtained from the ansatz with the help of the Lorentzian inversion formula \cite{Caron-Huot:2017vep,Alday:2017vkk}, and we require that the resulting expression should match \eqref{eq:twist4data}.
\end{itemize}

\subsection{Results at two loops}
Imposing all the constraints described above fixes the ansatz down to a bunch of free parameters. Like the situation at one loop, some of them have no effects on the reduced correlator while the others can be identified as ambiguities in correspondence to the UV divergence at two loops. In this sense the correlator $\mathcal{H}^{(3)}$ is again completely solved. The full expressions of both $\mathcal{L}^{(3)}$ and $\mathcal{H}^{(3)}$ are too long to fit into the paper, so instead we record them in an ancillary file included in the arXiv submission of this paper. For the readers to have a glimpse of the structures of these quantities, here we just present the terms in $\mathcal{L}^{(3)}$ with the highest transcendental weight, which are simple and intuitive
\begin{equation}
    \begin{split}
        \mathcal{L}^{(3)}(z,\zb) = \mathtt{e}_{u_1}&\left[ \left(\frac{u}{6 (z-\zb)^3}+\frac{u^2 (7+u-v)}{6 (z-\zb)^5}\right)W_{6,1}(z,\zb) \right.\\
        &\left. + \left( \frac{u}{216 (z-\zb)}-\frac{u (1-u+v) (7-5 u-7 v)}{540 (z-\zb)^3} \right)W_{6,2}(z,\zb) \right] \\
        +&   \mathtt{e}_{s_1}\left[\left(z\to \frac{1}{z},\zb\to \frac{1}{\zb}\right)\right]
        +  \mathtt{e}_{s_2}\left[\left(z\to 1-\frac{1}{z},\zb\to 1-\frac{1}{\zb}\right)\right]\\
        +& \mathtt{e}_{t_1}\left[\left(z\to \frac{1}{1-z},\zb\to \frac{1}{1-\zb}\right)\right]
        +  \mathtt{e}_{t_2}\left[\left(z\to \frac{z}{z-1},\zb\to \frac{\zb}{\zb-1}\right)\right]\\
        +& \mathtt{e}_{u_2}\left[\left(z\to 1-z,\zb\to 1-\zb\right)\right] +(\text{lower-weight parts}).
    \end{split}
\end{equation}
Using the abbreviations $G_{\vec{a}} \equiv G_{\vec{a}}(z)$ and $\bar{G}_{\vec{a}} \equiv G_{\vec{a}}(\zb)$, the two $W$ functions appearing above are defined as
\begin{equation}
    \begin{split}
        W_{6,1}(z,\zb) = &\quad G_{0,1,0,1,1,0}+ G_{0,1,0,1,1}\bar{G}_0+G_{0,1,0,1} \bar{G}_{0,1}+G_{0,1,0} \bar{G}_{0,1,1} +G_{0,1} \bar{G}_{0,1,1,0}\\&+G_0\bar{G}_{0,1,1,0,1}+\bar{G}_{0,1,1,0,1,0} +2 \zeta_3 \left(2 G_{0,1,1} + 3 G_0 \bar{G}_{0,1}+3 \bar{G}_{0,1,0} \right) \\
        &- (z\leftrightarrow \zb),
    \end{split}
\end{equation}
and
\begin{equation}
    \begin{split}
        W_{6,2}(z,\zb) =& \quad G_{1,0,1} \bar{G}_{0,1,0}+G_{0,1,0} \bar{G}_{0,1,1}+G_{0,1,1} \bar{G}_{1,0,0}+G_{1,0,0} \bar{G}_{1,0,1} +G_{0,1,0,1}\bar{G}_{0,1}\\&+G_{0,1,1,0} \bar{G}_{1,0}+G_{1,0,0,1} \bar{G}_{1,0}+G_{1,0,1,0} \bar{G}_{0,1}+G_{1,0} \bar{G}_{0,1,0,1}+G_{0,1}\bar{G}_{0,1,1,0}\\&+G_{0,1} \bar{G}_{1,0,0,1}+G_{1,0} \bar{G}_{1,0,1,0}+\bar{G}_0 G_{0,1,0,1,1}+\bar{G}_1 G_{0,1,1,0,0}+\bar{G}_1 G_{1,0,0,1,0}\\&+\bar{G}_0 G_{1,0,1,0,1}+G_1 \bar{G}_{0,1,0,1,0}+G_0 \bar{G}_{0,1,1,0,1}+G_0 \bar{G}_{1,0,0,1,1}+G_1 \bar{G}_{1,0,1,0,0}\\&+\bar{G}_{0,1,0,1,0,1}+\bar{G}_{0,1,1,0,1,0}+\bar{G}_{1,0,0,1,1,0}+\bar{G}_{1,0,1,0,0,1}+G_{0,1,0,1,1,0}\\&+G_{0,1,1,0,0,1}+G_{1,0,0,1,0,1}+G_{1,0,1,0,1,0} + 6\zeta_3 \left(G_{0,1,1}+G_{1,0,1}+2 \bar{G}_{0,0,1} \right. \\
        &+\left. 2 \bar{G}_1 G_{0,1} + (G_0 + \bar{G}_0) G_{\frac{1}{z},\frac{1}{\zb}}(1)+2G_{0,\frac{1}{\zb},\frac{1}{z}}(1)\right) + 15\zeta_5 G_{1}\\
        &- (z\leftrightarrow \zb).
    \end{split}
\end{equation}

Apart from the determined part of $\mathcal{L}^{(3)}$, there are still 12 unconstrained parameters left in $\mathcal{L}^{(3)}$, which fall into three types. 
\begin{itemize}
    \item The first and the simplest type resides in the kernel of $\left[\Delta^{(4)}\right]^2$, with 3 free parameters $c_1$, $c_2$ and $c_3$
    \begin{equation}
        \begin{split}
           \mathcal{L}^{(3)}\supset &\  c_1  (\mathtt{e}_{s_2}+\mathtt{e}_{t_1}+\mathtt{e}_{u_1}) u \bar{D}_{1111}+c_2 (\mathtt{f}_{s}\log u + \mathtt{f}_{t}\log v) u \bar{D}_{1111} \nonumber \\
            &+ c_3 \left[(3\mathtt{e}_{t_1}\!-2\mathtt{e}_{t_2}\!-2\mathtt{e}_{s_1}\!+\mathtt{e}_{u_2})\log u+(3\mathtt{e}_{s_1}\!-2\mathtt{e}_{s_2}\!-2\mathtt{e}_{t_1}\!+\mathtt{e}_{u_1})\log v \right] u \bar{D}_{1111}.
        \end{split}
    \end{equation}
    They do not affect the final result of $\mathcal{H}^{(3)}$, since they are mapped to $0$ under the action of $\left[\Delta^{(4)}\right]^2$.
\end{itemize}
 The other two types of free parameters are related to the counterterms for the UV divergence two-loop scattering in AdS. At two-loop level, there are two types of diagrams containing counterterm vertices, contact diagrams and one-loop diagrams, each corresponding to one type of free parameters. We call them tree-like ambiguities and one-loop-like ambiguities.
 \begin{itemize}
     \item There are 3 free parameters for tree-like ambiguities. In $\mathcal{L}^{(3)}$ they are
    \begin{equation}\label{eq:contact2loop}
       \begin{split}
          \mathcal{L}^{(3)}\supset &\ c_4\left[\mathtt{f}_{t}u\bar{D}_{1111}+(\mathtt{e}_{s_1}-\mathtt{e}_{t_1})u\bar{D}_{1122}+(\mathtt{e}_{u_1}-\mathtt{e}_{t_2})u\bar{D}_{1212}\right] \\
           &+ c_5  \left[(\mathtt{e}_{t_1}\!+\mathtt{e}_{u_1}\!-\mathtt{f}_{u})u\bar{D}_{1111}+(\mathtt{e}_{s_2}\!-\mathtt{e}_{t_2})u\bar{D}_{1122}+(\mathtt{e}_{u_2}\!-\mathtt{e}_{t_1})u\bar{D}_{1212} \right]\\
          &+ c_6 (\mathtt{e}_{s_2}+\mathtt{e}_{t_1}+\mathtt{e}_{u_1}) \frac{u}{z-\zb}\left( Q_3(z,\zb) - \frac{1}{3}\log \frac{u^2}{v}\, W_2(z,\zb) \right),
        \end{split}
    \end{equation}
    and the action of  $\left[\Delta^{(4)}\right]^2$ maps them to 
    \begin{equation}
        \begin{split}
            \mathcal{H}^{(3)}\supset &\ 24\, c_4 \left[3 \mathtt{f}_{u} u^3\bar{D}_{3333}+(\mathtt{f}_{s}-\mathtt{f}_{u})u^3\bar{D}_{3344}+(\mathtt{f}_{t}-\mathtt{f}_{u})u^3\bar{D}_{4334} \right] \\
            &+24\, c_5  \left[(\mathtt{e}_{u_1}+\mathtt{e}_{u_2}-\mathtt{e}_{s_1}-\mathtt{e}_{t_1})u^3\bar{D}_{3333}+(\mathtt{e}_{s_1}-\mathtt{e}_{u_1})u^3\bar{D}_{3344}\right.\\
            &\left. +(\mathtt{e}_{t_1}-\mathtt{e}_{u_2})u^3\bar{D}_{4334}\right] -3\, c_6 (\mathtt{e}_{s_2}+\mathtt{e}_{t_1}+\mathtt{e}_{u_1}) u^3 \bar{D}_{3333}\;.
        \end{split}
    \end{equation}
    One can check that they are indeed contact diagrams which exchange spin $\ell=0,1$ operators only.
    \item The rest 6 free parameters are all one-loop-like ambiguities. The actual leading log singularity of these one-loop-like ambiguities are $\log^2 u$ terms, which have non-zero support only on spin $\ell=0$. This characteristic can be explained from the CFT origin of one-loop-like ambiguities. We have already seen that the $\mathcal{H}^{(2)}$ contains an unfixed contact diagram, which causes ambiguity in the $\ell=0$ data of $\log u$. This ambiguity on $\ell=0$ $\log u$ data passes on to the two-loop $\log^2 u$ data, via the unitarity recursion described in Section \ref{Sec:unitarityrecursion}. As a consequence, the one-loop-like ambiguities have $\log^2 u$ data with support on $\ell=0$. Despite of this simple characteristic on $\log^2 u$ data, the full expressions of these ambiguities are too lengthy, and we record them in the ancillary file as well.
\end{itemize}

Now we move on to discuss the color structure that was not completely fixed in the ansatz at the beginning. Imposing the cancellation of unphysical poles fixes parameters in the tree-like modification $\widetilde{\mathcal{H}}^{(1)}$ to $a_1=-\frac{1}{72}$ and $a_2=\frac{1}{72}$, and so
\begin{subequations}\label{eq:resulttreeto2loop}
    \begin{align}
        \tilde{\mathtt{c}}_s &= \frac{1}{36}\left(\mathtt{e}_{s_2}\! -\mathtt{e}_{s_1}  \right),\\
        \tilde{\mathtt{c}}_t &= \frac{1}{36}\left( 2\mathtt{e}_{s_1}\! - \mathtt{e}_{t_1}\! -\mathtt{e}_{u_1}\! - 3\mathtt{f}_s\right) \equiv \frac{1}{36}\left(\mathtt{e}_{t_1}\! -\mathtt{e}_{t_2}  \right),\\
        \tilde{\mathtt{c}}_u &= \frac{1}{36}\left( -\mathtt{e}_{s_1}\!-\mathtt{e}_{s_2}\!+ \mathtt{e}_{t_1}\!+\mathtt{e}_{u_1}\!+\mathtt{f}_s \right) \equiv \frac{1}{36}\left(\mathtt{e}_{u_1}\! -\mathtt{e}_{u_2}  \right).
    \end{align}
\end{subequations}
Note that the condition $\tilde{\mathtt{c}}_s+\tilde{\mathtt{c}}_t+\tilde{\mathtt{c}}_u=0$ is preserved. Like $\bar{\mathtt{c}}_{s,t,u}$ at one loop, these $\tilde{\mathtt{c}}_{s,t,u}$ factors are again proportional to the actual tree-level factors $\mathtt{c}_{s,t,u}$, through a similar process of transformations using the Jacobi identity and shrinking triangles as shown in Figure \ref{fig:cbar}.

Furthermore, the comparison with the data from twist-4 operators fixes parameters in the one-loop-like modification $\widetilde{\mathcal{H}}^{(2)}$ to $b_1=\frac{1}{6}$, $b_2=0$ and $b_3=-\frac{7}{6}$. Hence $\tilde{\mathtt{d}}_{st,su,tu}$ reads
\begin{subequations}\label{eq:result1loopto2loop}
    \begin{align}
        \tilde{\mathtt{d}}_{st} &= -\frac{1}{6}\left( -\mathtt{e}_{s_1}\!+ 6\mathtt{e}_{t_1}\!+7\mathtt{f}_s\! \right) \equiv \mathtt{f}_u+\frac{5}{6}\left( \mathtt{f}_s\! -\mathtt{e}_{s_1}\right) ,\\
        \tilde{\mathtt{d}}_{su} &=  -\frac{1}{6}\left(6\mathtt{e}_{s_1}\!+5 \mathtt{e}_{s_2}\!  -6\mathtt{e}_{t_1}\! -11 \mathtt{f}_s\right)\equiv \mathtt{f}_t+\frac{5}{6}\left(\mathtt{f}_s\! -\mathtt{e}_{s_2}\right),\\
        \tilde{\mathtt{d}}_{tu} &= -\frac{1}{6}\left(-5\mathtt{e}_{s_1}\!  +\! 5\mathtt{e}_{t_1}\!+\! 5\mathtt{e}_{u_1}+\! 4\mathtt{f}_s \right)\equiv \mathtt{f}_s+\frac{5}{6}\left(\mathtt{e}_{s_1}\!  -2\mathtt{f}_s-\! \mathtt{e}_{t_1}\!-\! \mathtt{e}_{u_1} \right).
    \end{align}
\end{subequations}
Unlike the tree-like modification, these new color factors turn out to be not simply proportional to the actual one-loop factors 
$\mathtt{d}_{st,su,tu}$. By applying the linear relations \eqref{eq:colorbasis2loop} the form closest to this goal that we manage to reach is shown in the last identity in each of the above equations. Again using the reasoning in Figure \ref{fig:cbar} one can explicitly check that $\{\tilde{\mathtt{d}}_{st}-\mathtt{f}_u,\tilde{\mathtt{d}}_{su}-\mathtt{f}_t,\tilde{\mathtt{d}}_{tu}-\mathtt{f}_s\}$ are proportional to $\{\mathtt{d}_{st},\mathtt{d}_{su},\mathtt{d}_{tu}\}$ with the same $G_F$-dependent proportionality factor. Hence in this form one can think about the one-loop-like modification $\widetilde{\mathcal{H}}^{(2)}$ as deviating from the one-loop correlator $\mathcal{H}^{(2)}$ by a crossing-symmetric shift in its color factors. Possible implications of such structure as well as the other modification terms at both one- and two-loops clearly call for a better understanding. We leave it for future investigations.

\section{Outlook}\label{Sec:discussions}

By introducing an ansatz inspired by hidden symmetries in the leading log singularities and utilizing the position space bootstrap method, in this paper we obtained analytic results of four-point functions of super gluons in $AdS_5\times S^3$ up to two loops. There are many interesting questions to be further investigated in relation to these results. Here we briefly comment on a few:
\begin{itemize}
\item Combined with the supergravity results, our results provide the necessary data for extending the tree-level observation of double copy structures \cite{Zhou:2021gnu} to loop levels. Using the flat-space case as an inspiration, it seems the first step to achieve this is to rewrite our result in a suitable form so that an appropriate integrand can be defined.  
    \item At the one-loop level, it is clear that the Mellin space expression has a much simpler form than the position space result. Therefore it would be important to translate the two-loop result into Mellin space and understand its structure. It should also be noted that the Mellin space approach and the position space approach, at one loop where they overlap, are not exactly equivalent to each other, even though they both use the leading logarithmic singularities as an input. It would be interesting to see one can combine the strengths of both approaches and to obtain a more powerful method. 
    \item Another important future direction is to extend our results to correlators of operators with higher KK weights at both one and two loops. In the supergravity case, the general pattern of such correlators with higher weights still remains elusive. For super gluons, however, the results are in general much simpler and the various color structures also allow us to distinguish different parts of the correlators, instead of studying them as a whole. Therefore, we might expect that we can first develop a more refined understanding of the structure of loop-level correlators in the super gluon case, in particular in relation with the full implication of the hidden conformal symmetry. 
    \item It would also be interesting to study loop-level correlators of super gluons in other theories. In \cite{Alday:2021odx}, all tree-level four-point functions have been computed for SYM on backgrounds of the form $AdS_{d+1}\times S^3$ with $d=3,4,5,6$, which provide the necessary data to initiate the bootstrap calculations at higher genus. However, for $d\neq 4$ there is not a convenient definition of reduced correlators and one has to work with the full correlator. Therefore, it will be important to see how our algorithm needs to modified in order to compute correlators in these theories. 
    \item On a different note, we point out that similar basis functions in the loop level correlator bootstrap also appear in the correlator of ``bound state'' operators \cite{Ceplak:2021wzz,Ma:2022ihn}. An interesting problem to explore if one can establish similar position space methods to bootstrap such bound state correlators. 
\end{itemize}

\acknowledgments

The authors would like to thank Lilin Yang for useful discussions and for sharing with us data of integrals that are needed in the flat-space computation. ZH, BW and EYY are supported by National Science Foundation of China under Grant No.~12175197 and Grand No.~12147103. EYY is also supported by National Science Foundation of China under Grant No.~11935013, and by the Fundamental Research Funds for the Chinese Central Universities under Grant No.~226-2022-00216. X.Z. is supported by funds from University of Chinese Academy of Sciences (UCAS), funds from the Kavli Institute for Theoretical Sciences (KITS), the Fundamental Research Funds for the Central Universities, and the NSFC Grant No.~12275273.

\appendix

\section{Single-valued multiple polylogarithms as basis functions}\label{sec:mpls}

Multiple polylogarithms (MPLs) are in some sense the simplest type of functions beyond rational functions, and can be defined by iterated integrals with rational integrands \cite{Goncharov:1998kja,Goncharov:2001iea}
\begin{align}
G(z)=&1,\nonumber\\
G_{\vec{a}}(z) \equiv& G_{a_1,a_2,...,a_n}(z)= \int_{0}^{z} \frac{\dd t}{t-a_1} G_{a_2,a_3,...,a_n}(t).\nonumber
\end{align}
Components of the vector $\vec{a}\equiv(a_1,a_2,...,a_n)$ as well as $z$ are complex variables. The length $|\vec{a}|$ of $\vec{a}$ is called the weight of $G_{\vec{a}}(z)$. These are generalizations of classical polylogarithms, as can be seen in a few simple examples used in the expressions for the leading log singularities \eqref{eq:h}
\begin{subequations}\label{eq:Gexamples}
\begin{align}
    G_{1}(z) &= \log(1-z), \\
    G_{0,1}(z) &= -\text{Li}_2(z), \\
    G_{1,1}(z)&=\frac{1}{2} \log ^2(1-z), \\
    G_{0,0,1}(z)&= -\text{Li}_3(z), \\
    G_{0,1,1}(z)&= -\text{Li}_3(1-z)+\text{Li}_2(1-z) \log (1-z)+\frac{1}{2} \log (z) \log ^2(1-z)+\zeta_3, \\
    G_{1,0,1}(z)&=2\text{Li}_3(1-z)-\left(\text{Li}_2(1-z)+\frac{1}{6} \pi ^2 \right) \log (1-z)-2 \zeta_3 .
\end{align}
\end{subequations}
For completeness, $G$ with the $n$-dimensional zero vector $\vec{0}_n=(0,0,\dots,0)$ is define to be
\begin{align}
G_{\vec{0}_n}(z)=\frac{1}{n!}\log{z}^n.
\end{align}
More generally, any products or linear combinations of $G$'s (with rational coefficients) are also treated as MPLs.

These functions widely appear in perturbative computations of scattering amplitudes and correlators in many theories. In the problem investigated in this paper, we already observe that the tree-level correlator $\mathcal{H}^{(1)}$
\eqref{eq:tree1} as well as the leading log singularities at loop levels \eqref{eq:h} are combinations of MPLs with rational coefficients. Hence it is natural to use these functions similarly to build an ansatz for $\mathcal{H}^{(2)}$ and $\mathcal{H}^{(3)}$, and test the existence of a solution by bootstrap.

The complete set of MPLs are too redundant, accompanied by numerous complicated linear and functional relations among $G$'s. In order to properly set up an ansatz we need to figure out a finite set of linearly independent MPLs to play as a basis. It is impossible to obtain such basis without imposing additional conditions to carve out a proper subspace. Fortunately these conditions come along with the bootstrap problem itself. 

Firstly, the target correlator should be single-valued on the Euclidean sheet $\bar{z}=z^*$, so we can require that each element in the basis is already a single-valued combination of $G$'s, which are called single-valued multiple polylogarithms (SVMPL).

Secondly, a correlator can in general become singular as two operators are light-like separated in the Lorentzian region. In terms of MPLs this means the basis functions may have singularities at either $z$, $\bar{z}$, $1-z$ or $1-\bar{z}$ being zero or infinite. An unambiguous way to analyze singularities of MPLs is to utilize an algebraic system called \emph{symbol}. The symbol $\mathcal{S}[G]$ of a function $G$ can be obtained by applying differentiation repeatedly. If
\begin{equation}
    \dd G = \sum_i G'_i\  \dd \log R_i,
\end{equation}
where $R_i$s are algebraic functions, then we assign a formal product $\otimes$
\begin{equation}
    \mathcal{S}[G] = \sum_i \mathcal{S}[G'_i] \otimes  R_i.
\end{equation}
So a symbol is in general a linear combination of $\otimes$ products, where the length of each $\otimes$ product is the same as the transcendental weight of its corresponding function. From the differential definition above it is natural that the $\otimes$ product satisfies algebraic relations
\begin{subequations}
\begin{align}
    A\otimes(xy)\otimes B&=A\otimes x\otimes B+A\otimes y\otimes B,\\
    A\otimes(x^p)\otimes B&=p\,\qty(A\otimes x\otimes B),\\
    A\otimes c\otimes B&=0,\quad\text{any numeric }c,
\end{align}
\end{subequations}
where $A$ and $B$ can be any $\otimes$ products. For a few examples, symbols of the functions listed in \eqref{eq:Gexamples} are
\begin{subequations}\label{eq:Sexamples}
\begin{align}
    \mathcal{S}[G_{1}(z)] &=\otimes(1-z), \\
    \mathcal{S}[G_{0,1}(z)] &= (1-z)\otimes z, \\
    \mathcal{S}[G_{1,1}(z)]&= (1-z)\otimes(1-z), \\
    \mathcal{S}[G_{0,0,1}(z)]&= (1-z)\otimes z\otimes z,\\
    \mathcal{S}[G_{0,1,1}(z)]&= (1-z)\otimes(1-z)\otimes z, \\
    \mathcal{S}[G_{1,0,1}(z)]&= (1-z)\otimes z\otimes (1-z) .
\end{align}
\end{subequations}
Expression in each entry of the $\otimes$ product is called a \emph{symbol letter}, and the collection of all letters the \emph{alphabet} of the symbol. Roughly speaking, by imposing that a symbol letter equals zero or infinity one learns the location of singularities of the original function. With \eqref{eq:Sexamples} we observe that the leading log singularities have an alphabet $\{z,1-z\}$, and hence also $\{\bar{z},1-\bar{z}\}$ if viewed in different channels. Therefore, in the minimal setup we can assume that the symbol alphabet of the entire reduced correlator at each perturbative order is just $\{z,\bar{z},1-z,1-\bar{z}\}$, which is consistent with the expectation on singularities in the Lorentzian region. 

In practice it turns out an extra letter $z-\bar{z}$ is required as well. This factor already make an appearance as poles in the coefficients in front of MPLs, as can be seen in the leading log singularities \eqref{eq:leadinglog}. Its corresponding singularity is very special. In the Euclidean region this singularity has to be absent so that the correlator remains finite, which in fact serves as one of the main constraints in our bootstrap computation. However, it is present in the Lorentzian region after analytic continuation, and is tied to the so-called bulk-point limit \cite{Maldacena:2015iua}, at which the perturbative scattering is expected to reduce to that in flat space \cite{Alday:2017vkk}. The need of letter $z-\bar{z}$ in the SVMPL basis was already observed in the supergravity computation in \cite{Drummond:2019hel,Huang:2021xws,Drummond:2022dxw}, and in the case of super gluon scattering the same phenomenon occurs. This is further discussed around \eqref{eq:bplzzb}.

By restricting the symbol alphabet to $\{z,\bar{z},1-z,1-\bar{z},z-\bar{z}\}$ and setting a maximal transcendental weight, there is a systematic procedure to work out a finite linear basis of SVMPLs. The rough idea is to first enumerate a basis for SVMPLs at weight one, which can be just $\log\,u$ and $\log\,v$ (weight zero is trivial), and then extent them to a basis of Hopf algebra coproducts at one higher weight using basis for weight-one MPLs \footnote{The space of MPLs is naturally accompanied by a Hopf algebra structure. A symbol can alternatively be viewed as the maximal iteration of Hopf coproducts acting on an MPL. See e.g., Section 6 of \cite{Duhr:2019tlz}}, and lift this to a basis of SVMPLs at weight two by imposing certain integrability conditions, and then repeat this analysis till the maximal weight. We refer interested readers to \cite{Chavez:2012kn} for details of the algorithm. For clarity of presentations here we name each element in the basis by $G^{\rm SV}_{w,i}(z,\bar{z})$, where $w$ labels its element, and an extra index $i$ distinguishes different basis elements with the same weight. The range of $i$ depends on the value of $w$, and its specific counting up to weight $6$ can be found in, e.g.~Table 1 of \cite{Huang:2021xws}.

We also further divide the basis into two disjoint sets, according to whether the letter $z-\bar{z}$ is contained in their symbols. Those without $z-\bar{z}$ are in fact known as single-valued harmonic polylogarithms (SVHPLs) \cite{Brown:2004,Dixon:2012yy}, and correspondingly we also denote them as $H^{\rm SV}_{w,i}\equiv G^{\rm SV}_{w,i}$. Up to weight $2$ these elements can be selected as
\begin{equation}\label{eq:basisexamples}
    \begin{split}
        &H_{0,1}^{\rm SV}=1,\quad
        H_{1,1}^{\rm SV}=\log{u},\quad
        H_{1,2}^{\rm SV}=\log{v},\\
        &H_{2,1}^{\rm SV}=\zeta_2,\quad
        H_{2,2}^{\rm SV}=\log^2 {u},\quad
        H_{2,3}^{\rm SV}=\log{u}\log{v},\quad
        H_{2,4}^{\rm SV}=\log^2 {v},\\
        &H_{2,5}^{\rm SV}\equiv W_2(z,\bar{z})=\mathrm{Li}_2(z)-\mathrm{Li}_2(\bar{z})+\frac{1}{2}\log{u}\log\frac{1-z}{1-\bar{z}}.
    \end{split}
\end{equation}
In the above expression we explicitly see that some of the basis elements at a given weight can be simply constructed from products of $H^{\rm SV}$'s at lower weights. On the other hand, we name elements with $z-\bar{z}$ as $Q_{w,i}$. They do not show up until weight $3$. At weight $3$ there is a unique element of this type, $Q_3\equiv Q_{3,1}$ (up to additive terms whose symbols are free of the letter $z-\bar{z}$). A choice of $Q_3$ that is convenient for the presentation of $\mathcal{H}^{(2)}$ was already listed in \eqref{eq:Q3}. At weight $4$ two of the $Q$ elements can be identified as products
\begin{equation}
    Q_{4,1}=Q_{3}H^{\rm SV}_{1,1},\quad
    Q_{4,2}=Q_{3}H^{\rm SV}_{1,2},
\end{equation}
and in addition to these there are three new elements $Q_{4,3}$, $Q_{4,4}$ and $Q_{4,5}$. $Q$ elements at higher weights are not needed except for $Q_{6,1}=\zeta_3 \mathcal{Q}_{3}$. In summary, our SVMPL basis $G^{\rm SV}$ includes 
\begin{equation}
    G^{\rm SV} \equiv H^{\rm SV} \sqcup \{ \mathcal{Q}_{3}, \mathcal{Q}_{4,1}, \mathcal{Q}_{4,2}, \mathcal{Q}_{4,3}, \mathcal{Q}_{4,4}, \mathcal{Q}_{4,5}, \mathcal{Q}_{6,1} \}.
\end{equation}
We use $G^{\rm SV}_A(z,\bar{z})$ to denote functions in $G^{\rm SV}$, and we use $\sum_{A\leq n}$ to represent sum over all possible $G^{\rm SV}_A(z,\bar{z})$ with weight $w\leq n$.

Our computation of both MPLs and their symbols utilizes the PolyLogTools Mathematica package introduced in \cite{Duhr:2019tlz}.

\section{Analytic result of the one-loop reduced correlator}\label{sec:H2analytic}

This appendix contains the analytic expression for the full one-loop reduced correlator $\mathcal{H}^{(2)}$. In terms of the color factors defined in Section \ref{sec:oneloopansatz} it is decomposed as
\begin{equation}
    \mathcal{H}^{(2)}(z,\zb) = \dst\mathcal{H}^{(2)}_{st}(z,\zb) + \dsu\mathcal{H}^{(2)}_{su}(z,\zb) + \dtu\mathcal{H}^{(2)}_{tu}(z,\zb)+a_0\mathcal{H}^{(2)}_{\rm c},
\end{equation}
where the counterterm $\mathcal{H}^{(2)}_{\rm c}$ was already recorded in \eqref{eq:H2c}. Bose symmetry implies that the three components are related by
\begin{subequations}
    \begin{align}
        \mathcal{H}^{(2)}_{st}(z,\zb)&=\mathcal{H}^{(2)}_{su}\left(\frac{z}{z-1},\frac{\zb}{\zb-1}\right),\\
        \mathcal{H}_{tu}^{(2)}(z,\zb)&=\frac{u^3}{v^3}\mathcal{H}_{su}^{(2)}(1-z,1-\zb).
    \end{align}
\end{subequations}
Therefore it suffices to explicitly give the analytic result for one of the components. In the following we provide the expression for $\mathcal{H}_{su}^{(2)}(z,\bar{z})$, organized by transcendental weights. For convenience we use the variable $y\equiv u-v$ and $\delta\equiv\bar{z}-z$. At weight $4$
\begin{equation}
    \begin{split}
        &\left.u^{-3}\mathcal{H}_{su}^{(2)}(z,\bar{z})\right\vert_{\text{weight }4}
        =\\
        &\left(-\frac{7+11y}{4\delta^3}+\frac{67+165y+261y^2+163y^3}{12\delta^5}-\frac{5(1+y)^3(27-30y+47y^2)}{12\delta^7}\right.\\
        &\left.\quad+\frac{35(1-y)^2(1+y)^5}{4\delta^9}\right)W_4(z,\bar{z}).
    \end{split}
\end{equation}
At weight $3$
\begin{equation}
    \begin{split}
        &\left.u^{-3}\mathcal{H}_{su}^{(2)}(z,\bar{z})\right\vert_{\text{weight }3}
        =\\
        &\left(-\frac{8}{9\delta^2}+\frac{129+318y+317y^2}{36\delta^4}-\frac{5(1+y)^2(5-4y+10y^2)}{3\delta^6}+\frac{35(1-y)^2(1+y)^4}{4\delta^8}\right)W_{3}(z,\bar{z})\\
        &+\left(-\frac{1}{8\delta}-\frac{1-27y^2}{18\delta^3}-\frac{1-26y^2+49y^4}{12\delta^5}+\frac{5(1-y^2)^2(1+5y^2)}{6\delta^7}-\frac{35(1-y^2)^4}{24\delta^9}\right)\\
        &\quad\times\left(Q_{3}(z,\bar{z})-W_2(z,\bar{z})\log{u}+\frac{1}{2}W_2(z,\bar{z})\log{v}\right)\\
        &+\left(-\frac{2+15y}{36\delta^3}-\frac{7+27y-63y^2-155y^3}{72\delta^5}+\frac{5(1+y)^2(6-13y+30y^2-23y^3)}{36\delta^7}\right.\\
        &\left.\qquad-\frac{35(1-y)^3(1+y)^4}{24\delta^9}\right)W_2(z,\bar{z})\log{u}.
    \end{split}
\end{equation}
At weight $2$
\begin{equation}
    \begin{split}
        &\left.u^{-3}\mathcal{H}_{su}^{(2)}(z,\bar{z})\right\vert_{\text{weight }2}
        =\\
        &\left(-\frac{19}{96\delta}+\frac{271+486y+405y^2}{216\delta^3}-\frac{407+364y+654y^2+884y^3+643y^4}{144\delta^5}\right.\\
        &\left.\quad+\frac{(1-y^2)^2(411+280y+295y^2)}{72\delta^7}-\frac{377(1-y^2)^4}{288\delta^9}\right)W_2(z,\bar{z})\\
        &+\left(-\frac{83+165y}{432\delta^2}+\frac{231+479y+893y^2+645y^3}{432\delta^4}-\frac{5(1+y)^3(33-42y+53y^2)}{144\delta^6}\right.\\
        &\left.\qquad+\frac{35(1-y)^2(1+y)^5}{48\delta^8}\right)\log{u}\left(\log{u}-\log{v}\right)\\
        &+\left(\frac{75+157y}{108\delta^4}-\frac{5(3+3y+8y^2+8y^3)}{9\delta^6}+\frac{35(1-y)^2(1+y)^3}{12\delta^8}\right)\log{u}\log{v}.
    \end{split}
\end{equation}
At weight $1$
\begin{equation}
    \begin{split}
        &\left.u^{-3}\mathcal{H}_{su}^{(2)}(z,\bar{z})\right\vert_{\text{weight }1}
        =\\
        &\left(\frac{31}{48\delta^2}+\frac{365+160y-3297y^2}{432\delta^4}-\frac{365+1858y-2223y^4}{144\delta^6}+\frac{1217(1-y^2)^3}{144\delta^8}\right)\log{u}\\
        &+\left(-\frac{163-1941y}{864\delta^2}-\frac{773-1105y-2761y^2+7533y^3}{864\delta^4}\right.\\
        &\left.\qquad+\frac{(1-y)^2(505+1503y+4079y^2+3081y^3)}{288\delta^6}-\frac{1217(1-y)^4(1+y)^3}{288\delta^8}\right)\left(\log{u}-\log{v}\right).
    \end{split}
\end{equation}
And finally at weight $0$
\begin{equation}
    \begin{split}
        &\left.u^{-3}\mathcal{H}_{su}^{(2)}(z,\bar{z})\right\vert_{\text{weight }0}
        =
        \frac{43}{24\delta^2}+\frac{56-28y-471y^2}{54\delta^4}+\frac{499(1-y^2)^2}{72\delta^6}.
    \end{split}
\end{equation}

\section{Bulk-point limit} \label{sec:flatspacelimit}

In the bulk-point limit $\mathcal{H}^{(3)}$ is expected to match the flat-space four-point scattering amplitude $\mathcal{A}^{\text{2-loop}}$ of the maximal supersymmetric Yang--Mills (SYM) in 8d \cite{Alday:2021odx}. This comparison in the physical region should provide non-trivial constraints to our bootstrap computation at two loops. For convenience of this comparison we define a reduced amplitude $\tilde{\mathcal{A}}^{\text{2-loop}}$ by
\begin{equation}
    \mathcal{A}^{\text{2-loop}}=stA_{1234}^{\rm tree}\tilde{\mathcal{A}}^{\text{2-loop}},
\end{equation}
where $A_{1234}^{\rm tree}$ is the so-called partial amplitude with cyclic ordering $(1234)$ appearing in the color trace decomposition of the tree-level SYM amplitude ($T^I$ being generators in the adjoint representation of the color group)
\begin{equation}
    \mathcal{A}^{\rm tree}=\sum_{\rho\in S_4/Z_4}\mathrm{Tr}\qty(T^{I_{\rho(1)}}T^{I_{\rho(2)}}T^{I_{\rho(3)}}T^{I_{\rho(4)}})A_{\rho}^{\rm tree}.
\end{equation}
The combination $stA_{1234}^{\rm tree}$ is invariant under $S_4$ permutation of the particles and encodes the full dependence on the polarization vectors, so that the reduced amplitude $\tilde{\mathcal{A}}^{\text{2-loop}}$ is a permutation invariant scalar quantity.

As studied in \cite{Bern:1997nh} by unitarity cuts in generic dimensions, the two-loop four-point maximal SYM amplitude receives a decomposition onto planar and non-planar double boxes. In terms of the reduced amplitude $\tilde{\mathcal{A}}^{\text{2-loop}}$ and the color structures defined in Section \ref{sec:color2loop} this decomposition reads
\begin{equation}\label{eq:tildeA2loop}
    \begin{split}
        \tilde{\mathcal{A}}^{\text{2-loop}}=&\mathtt{e}_{s_1}\,sI_{1234}^{\rm pdb}+\mathtt{e}_{s_2}\,sI_{1243}^{\rm pdb}+\mathtt{e}_{t_1}\,tI_{1432}^{\rm pdb}+\mathtt{e}_{t_2}\,tI_{1423}^{\rm pdb}+\mathtt{e}_{u_1}\,uI_{1324}^{\rm pdb}+\mathtt{e}_{u_2}\,uI_{1342}^{\rm pdb}\\
        &+2\mathtt{f}_{s}\,sI_{1234}^{\rm ndb}+2\mathtt{f}_{t}\,tI_{1423}^{\rm ndb}+2\mathtt{f}_u\,uI_{1324}^{\rm nbd}.
    \end{split}
\end{equation}
Here $I_{abcd}^{\rm pdb}$ and $I_{abcd}^{\rm ndb}$ are Feynman integrals associated to planar and non-planar double boxes, defined by
\begin{equation}
    \begin{split}
        I_{abcd}^{\rm pdb}&=\parbox{3cm}{\tikz{\begin{scope}[scale=0.6]
        \draw [line width=1.5pt] (0,0.5) -- (0,-0.5);
        \draw [line width=1.5pt](0,0.5) -- (1,0.5);
        \draw [line width=1.5pt](1,0.5) -- (1,-0.5);
        \draw [line width=1.5pt](0,-0.5) -- (1,-0.5);
        \draw [line width=1.5pt](0,-0.5) -- (-1.0,-0.5);
        \draw [line width=1.5pt](-1,-0.5) -- (-1,0.5);
        \draw [line width=1.5pt] (-1,0.5) -- (0,0.5);
        \draw [line width=1.5pt] (1,0.5) -- (1.5,1.0);
        \draw [line width=1.5pt] (-1,0.5) -- (-1.5,1.0);
        \draw [line width=1.5pt] (1,-0.5) -- (1.5,-1.0);
        \draw [line width=1.5pt] (-1,-0.5) -- (-1.5,-1.0);
        \node [anchor=east] at (-1.5,1) {$a$};
        \node [anchor=west] at (1.5,1.0) {$d$};
        \node [anchor=west] at (1.5,-1.0) {$c$};
        \node [anchor=east] at (-1.5,-1.0) {$b$};
        \end{scope}}}\\
        &\equiv\int\frac{\mathrm{d}^{8-2\epsilon}\ell_1}{(2\pi)^{8-2\epsilon}}\frac{\mathrm{d}^{8-2\epsilon}\ell_2}{(2\pi)^{8-2\epsilon}}\frac{1}{\ell_1^2(\ell_1+p_a)^2(\ell_1+p_a+p_b)^2\ell_2^2(\ell_1-\ell_2)^2(\ell_2-p_d)^2(\ell_2-p_c-p_d)^2},
    \end{split}
\end{equation}
and
\begin{equation}
    \begin{split}
        I_{abcd}^{\rm ndb}&=\parbox{3cm}{\tikz{\begin{scope}[scale=0.6]
        \draw [line width=1.5pt](0,0.5) -- (1,0.5);
        \draw [line width=1.5pt](0,-0.5) -- (1,-0.5);
        \draw [line width=1.5pt](0,-0.5) -- (-1.0,-0.5);
        \draw [line width=1.5pt](-1,-0.5) -- (-1,0.5);
        \draw [line width=1.5pt] (-1,0.5) -- (0,0.5);
        \draw [line width=1.5pt] (1,0.5) -- (1.5,1.0);
        \draw [line width=1.5pt] (-1,0.5) -- (-1.5,1.0);
        \draw [line width=1.5pt] (1,-0.5) -- (1.5,-1.0);
        \draw [line width=1.5pt] (-1,-0.5) -- (-1.5,-1.0);
        \draw [line width=1.5pt] (0,0.5) -- (1,-0.5);
        \draw [draw=white,line width=2pt] (0.4,0.1) -- (0.6,-0.1);
        \draw [line width=1.5pt] (0,-0.5) -- (1,0.5);
        \node [anchor=east] at (-1.5,1) {$a$};
        \node [anchor=west] at (1.5,1.0) {$d$};
        \node [anchor=west] at (1.5,-1.0) {$c$};
        \node [anchor=east] at (-1.5,-1.0) {$b$};
        \end{scope}}}\\
        &\equiv\int\frac{\mathrm{d}^{8-2\epsilon}\ell_1}{(2\pi)^{8-2\epsilon}}\frac{\mathrm{d}^{8-2\epsilon}\ell_2}{(2\pi)^{8-2\epsilon}}\frac{1}{\ell_1^2(\ell_1+p_a)^2(\ell_1+p_a+p_b)^2\ell_2^2(\ell_1-\ell_2)^2(\ell_2-p_d)^2(\ell_1-\ell_2-p_c)^2},
    \end{split}
\end{equation}
These integrals can be computed by 
solving similar integrals by in $4-2\epsilon$ dimensions using differential equations \cite{Kotikov:1990kg,Kotikov:1991pm,Remiddi:1997ny,Henn:2013pwa,Papadopoulos:2014lla}, and then lifting to $8-2\epsilon$ dimensions by dimensional recurrence relations \cite{Tarasov:1996br,Lee:2009dh,Lee:2010wea} \footnote{The authors are grateful to Lilin Yang for sharing data for these integrals in $4-2\epsilon$ dimensions, together with a set of master integrals needed for dimensional recursion which were selected following \cite{Chen:2020uyk,Chen:2022lzr}. The dimensional recursion is performed using the Mathematica package \texttt{LiteRed} \cite{Lee:2013mka}.}. In the region $s,t<0$ and $u>0$, the final result for the planar double box integral $I_{1234}^{\rm pdb}$ in $8-2\epsilon$ dimensions reads
\begin{equation}\label{eq:IpdbResult}
\begin{split}
    I_{1234}^{\rm pdb}=&(-s)^{1-2\epsilon}\Bigg(\frac{1}{144}\frac{1}{\epsilon^2}+\frac{79+2x}{1728}\frac{1}{\epsilon}+\frac{-48+3303x+242x^2}{20736x}\\
    &+\frac{(1+x)(2-15x+x^2)}{432x^2}\qty(G_0-G_1)+\frac{(1-6x+34x^2+67x^3)\pi^2}{2592(x-1)x^3}\\
    &-\frac{1-5x+25x^2+3x^3}{216x^3}\qty(G_{0,0}-G_{0,1}-G_{1,0}+G_{1,1}+\frac{5\pi^2}{12})\\
    &-\frac{-1+8x+17x^2}{108(x-1)x^2}\qty(G_{0,0,0}-G_{0,0,1}-G_{0,1,0}+G_{0,1,1}+\frac{\pi^2}{6}\qty(2G_0+G_1)-\zeta_3)\\
    &-\frac{3+x}{18(x-1)^2}\bigg(G_{0,0,0,0}-G_{0,0,0,1}-G_{0,0,1,0}+G_{0,0,1,1}+\frac{\pi^2}{6}\qty(2G_{0,0}+G_{0,1})-\zeta_3G_0+\frac{17\pi^4}{360}\Big)\Bigg),
\end{split}
\end{equation}
where we use the abbreviation $G_{\vec{a}}\equiv G_{\vec{a}}(x^{-1})$, and the dimensionless parameter $x$ is related to the Mandelstam variables $s,t$ or the scattering angle $\theta$ by
\begin{equation}
x \equiv 1+\frac{t}{s}=\frac{1+\cos \theta}{2}.
\end{equation}
In the same region the result for the non-planar double box integral $I_{1234}^{\rm ndb}$ reads
\begin{equation}\label{eq:IndbResult}
\begin{split}
    I_{1234}^{\rm ndb}=&(-s)^{1-2\epsilon}\Bigg(\frac{1}{288}\frac{1}{\epsilon^2}+\frac{37}{1728}\frac{1}{\epsilon}+\frac{-60-3607x+3625x^2-36x^3+18x^4}{51840(x-1)x}\\
    &+\frac{(-12+42x-47x^2+2x^3-63x^4+18x^5)}{25920(x-1)^2}\qty(G_1+i\pi)\\
    &-\frac{60-328x+727x^2-798x^3+399x^4}{25920(x-1)^2x^2}\qty(G_0-G_1)+\frac{\pi^2}{1728}\\
    &-\frac{30-173x+428x^2-600x^3+525x^4-270x^5+90x^6}{12960(x-1)^3x^3}\qty(G_{0,0}+i\pi G_0)\\
    &+\frac{30-113x+160x^2-107x^3+36x^4+6x^5}{12960(x-1)x^3}\qty(G_{0,1}+G_{1,0}+i\pi\qty(G_0+G_1)-\frac{\pi^2}{2})\\
    &+\frac{(x-1)^3(-30-7x+4x^2+9x^3+9x^4)}{12960x^3}\qty(G_{1,1}+i\pi G_1)-\frac{(3x-2)(3x-1)\zeta_3}{720(x-1)^2x^2}\\
    &+\frac{(2x-1)(2-5x+5x^2)}{2160(-1+x)^2x^2}\qty(G_{0,0,1}+G_{0,1,0}-2G_{1,0,0}+i\pi\qty(G_{0,0}+G_{0,1}-2G_{1,0})-\frac{\pi^2}{2}G_0+\frac{i\pi^3}{2})\\
    &+\frac{(x-1)^3(2+x)}{2160x^2}\qty(G_{1,0,1}+G_{1,1,0}-2G_{0,1,1}+i\pi\qty(G_{1,0}+G_{1,1}-2G_{0,1})-\frac{\pi^2}{2}G_1)\Bigg).
\end{split}
\end{equation}
The remaining integrals in \eqref{eq:tildeA2loop} can be obtained from \eqref{eq:IpdbResult} and \eqref{eq:IndbResult} by permuting the particle labels. For readers' convenience we also record these results in the ancillary file. Note that the expressions from permutations may live in different physical regions, and so before assembling them together one needs to analytically continued the Mandelstam variables to the same region following the standard $i\varepsilon$ prescription. For our computation we finally lands on the region $s,t<0$ and $u>0$, in which \eqref{eq:IpdbResult} and \eqref{eq:IndbResult} directly apply.

In order to perform the comparison in the bulk-point limit we need to analytically continue our ansatz for the correlator from Euclidean region to physical region as well. Following \cite{Drummond:2022dxw}, our prescription is to continue $z$ counter-clockwisely around 0 and $\zb$ clockwisely around $1$, and then set $z=\zb+2 \omega \zb \sqrt{1-\zb}$. The bulk-point limit $z\to\bar{z}$ is then reached by setting $\omega\to0$. As was already pointed out in Section \ref{sec:H3constraints}, when taking this limit at two loops $[\Delta^{(4)}]^2\mathcal{L}^{(3)}$ dominates over the modification terms, and the action of $\Delta^{(4)}$ reduces to a simple multiplication factor. Therefore it suffices to compare $\mathcal{A}^{\text{2-loop}}$ with the pre-correlator $\mathcal{L}^{(3)}$. The detailed connection between these two objects is
\begin{equation}\label{appeq:flatspacelimit}
    \lim_{\omega\to0}\left.(z-\zb)^5 \mathcal{L}^{(3)}\left(z^{\circlearrowleft_0}, \zb^{\circlearrowleft_1} \right)\right|_{z=\zb+2 \omega \zb \sqrt{1-\zb}}  
    = \left.  48\pi^2 (1-\zb)^2 \zb^6 s^{-2}  \tilde{\mathcal{A}}^{\text{2-loop}}(x)\right|_{x=1 / \zb},
\end{equation}
Note the above relation is already expressed in terms of the reduced amplitude $\tilde{\mathcal{A}}^{\text{2-loop}}$. Although the full amplitude $\mathcal{A}^{\text{2-loop}}$ contains an extra factor $stA^{\rm tree}_{1234}$ including polarization vectors, in practical computation we do not have to bother manipulating it. The reason is that comparison with the leading log data in \eqref{eq:leadinglogcompareH3} already fully determines contributions of some MPLs of highest transcendental weight, and matching them with the corresponding MPL contributions in $\tilde{\mathcal{A}}^{\text{2-loop}}$ easily determines the factors appearing in the above relation \eqref{appeq:flatspacelimit} (e.g., matching coefficients of $G_{0,0,0,0}$ or $G_{1,1,1,1}$ in the limit). Comparison between the remaining contributions then generates constraints for the undetermined variables in the ansatz.

\cite{Alday:2017vkk} proposed a simpler connection than \eqref{appeq:flatspacelimit}, the relation between discontinuities $\operatorname{dDisc}\mathcal{H}$ and $\operatorname{Disc} \mathcal{A}$, in the context of graviton scattering. Similar connection should also apply to the gluon scattering studied here. Ideally one would not expect much difference between the comparison at the level of the full correlator and that of the discontinuity, because in principle the correlator $\mathcal{H}$ can be reconstructed from its double discontinuity $\operatorname{dDisc}\mathcal{H}$ through Lorentz inversion \cite{Caron-Huot:2017vep}, which is the CFT counterpart of the dispersion relation relating $\mathcal{A}$ and its discontinuity $\operatorname{Disc} \mathcal{A}$ in flat space. However, at loop level these dispersion relations can be polluted by the presence of finite spin contributions to the correlator/amplitude, which imposes extra data in addition to the discontinuities. Therefore one expects that constraints from \eqref{appeq:flatspacelimit} are stronger.

\section{Recursion of twist-4 data at $\log^2 u$} \label{sec:recursion}

The reduced correlator $\mathcal{H}$ can be organized in terms of power expansions in $\log u$ in the small $u$ limit, where the power of $\log u$ goes up to $n+1$ at $n$ loops. The coefficient functions of these $\log u$ powers encode different combinations of the expansion coefficients of the CFT data with respect to $a_F$, which are schematically listed in Table \ref{table:data}. The goal of this appendix is to compute the  combination  $\langle a_{\mathbf{a}}^{(1)}(\gamma_{\mathbf{a}}^{(1)})^2+2 a_{\mathbf{a}}^{(0)} \gamma_{\mathbf{a}}^{(1)} \gamma_{\mathbf{a}}^{(2)}\rangle$ for twist-4 operators. This data contributes to the two-loop correlator in the $\log^2u$ coefficient, as explicitly shown in \eqref{eq:H3logu2twist4}, and we use them as one of the inputs in our bootstrap algorithm. As is clear from the expression, only tree-level and one-loop correlators are needed. Moreover, the twist-4 operators are free of operator mixing.  This fact makes it possible to extract their CFT data from just the $\mathcal{O}_2$ correlators alone. The angle brackets $\langle \ldots\rangle$ will also be dropped as they are no longer necessary in this case. 
\begin{table}[h]
\centering
\begin{tabular}{c|cccc}
        \toprule
        order & $a^{1}_F$&  $a^{2}_F$ &  $a^{3}_F$&\\
        \midrule
        $\;\log(u)^0\;$  & $\langle a_{\mathbf{a}}^{(1)}\rangle$ &$\langle a_{\mathbf{a}}^{(2)}\rangle$  &$\langle a_{\mathbf{a}}^{(3)}\rangle$ \\
        $\log(u)^1$ & $\langle a_{\mathbf{a}}^{(0)}\gamma_{\mathbf{a}}^{(1)}\rangle$ & $\langle a_{\mathbf{a}}^{(1)}\gamma_{\mathbf{a}}^{(1)}+a_{\mathbf{a}}^{(0)}\gamma^{(2)}\rangle$ &  $ \;\langle a_{\mathbf{a}}^{(2)}\gamma_{\mathbf{a}}^{(1)}+a_{\mathbf{a}}^{(1)}\gamma_{\mathbf{a}}^{(2)}+a_{\mathbf{a}}^{(0)}\gamma_{\mathbf{a}}^{(3)}\rangle\;$  \\
        $\log(u)^2$ &  & $\langle a_{\mathbf{a}}^{(0)}(\gamma_{\mathbf{a}}^{(1)})^2\rangle$ & $\;\langle a^{(1)}(\gamma_{\mathbf{a}}^{(1)})^2+2a_{\mathbf{a}}^{(0)}\gamma_{\mathbf{a}}^{(1)}\gamma_{\mathbf{a}}^{(2)}\rangle\;$  \\
         $\log(u)^3$  &  & & $\;\langle a_{\mathbf{a}}^{(0)}(\gamma_{\mathbf{a}}^{(1)})^3\rangle\;$  \\
        \bottomrule
    \end{tabular}
\caption{The OPE data encoded in the coefficient functions of different $\log u$ powers for correlators up to two loops.}
\label{table:data}
\end{table}

The most efficient way to extract the CFT data from explicit correlators is to use the Lorentzian inversion formula \cite{Caron-Huot:2017vep}. Applying it to the tree-level correlator $\mathcal{H}^{(1)}$ as well as the $a_F^{1}$ term in $\mathcal{G}_0$, we obtain the following data for the twist-4 operator with spin $\ell$
\begin{align}
    \label{agamma}a_{\mathbf{a}}^{(0)} \gamma_{\mathbf{a}}^{(1)}=&(-\mathtt{c}_t+(-1)^ \ell\mathtt{c}_u)_{\mathbf{a}}\ \frac{2 \Gamma ( \ell+3)^2 }{\Gamma (2  \ell+5)}\;,\\
     a_{\mathbf{a}}^{(1)}=&(-\mathtt{c}_t+(-1)^ \ell\mathtt{c}_u)_{\mathbf{a}}\ \frac{\Gamma ( \ell+3)^2}{\Gamma (2  \ell+5)} \left[2 \psi^{(0)}( \ell+3)- 2 \psi^{(0)}(2  \ell+5) + \frac{1}{2} \right]\;.
\end{align}
 Here for the color part $(\# )_{\mathbf{a}}$ is defined by  projectors as in \eqref{eq:csdef}. Similarly, from the one-loop correlator $\mathcal{H}^{(2)}$, we can extract the following combinations
\begin{align}
\label{agamma2} a_{\mathbf{a}}^{(0)}(\gamma_{\mathbf{a}}^{(1)})^2=&(\mathtt{d}_{st}+ (-1)^\ell \mathtt{d}_{su})_{\mathbf{a}}\ \frac{\Gamma (\ell+3)^2}{\Gamma (2 \ell+5)}\frac{8}{(\ell+1) (\ell+4)}\;,\\
        \begin{split}
 a_{\mathbf{a}}^{(1)}\gamma_{\mathbf{a}}^{(1)}+ a_{\mathbf{a}}^{(0)}  \gamma_{\mathbf{a}}^{(2)}
        =& (\mathtt{d}_{st}+ (-1)^\ell \mathtt{d}_{su})_{\mathbf{a}}\  \frac{\Gamma (\ell+3)^2}{\Gamma (2 \ell+5)} \left[ -\frac{4 \left(2\ell^3+23\ell^2+65\ell+32\right)}{\ell (\ell+1)^2 (\ell+4)^2 (\ell+5)} \right.    \\ 
        &\qquad + \left.  \frac{2 \left(12\psi ^{(0)}(\ell+3)-12\psi ^{(0)}(2 \ell+5)+11\right)}{3(\ell+1) (\ell+4)} \right]\\ 
        & -(1+(-1)^\ell) (\mathtt{d}_{tu})_{\mathbf{a}}\ \frac{\Gamma (\ell+3)^2}{\Gamma (2 \ell+5)}\frac{2 \left(5 \ell^2+25 \ell+24\right) }{3 \ell (\ell+1) (\ell+4) (\ell+5)}\;.
    \end{split}
\end{align}
By comparing (\ref{agamma}) and (\ref{agamma2}), we get
\begin{equation}
 \gamma_{\mathbf{a}}^{(1)}=(-\mathtt{c}_t+(-1)^\ell\mathtt{c}_u)_{\mathbf{a}}\ \frac{2}{(\ell+1) (\ell+4)}\;.
\end{equation}
We could also solve for $a_{\mathbf{a}}^{(0)}, a_{\mathbf{a}}^{(1)},\gamma_{\mathbf{a}}^{(2)}$. However, to compute the wanted data it is sufficient to consider the following combination
\begin{equation}
 a_{\mathbf{a}}^{(1)}(\gamma_{\mathbf{a}}^{(1)})^2+2 a_{\mathbf{a}}^{(0)} \gamma_{\mathbf{a}}^{(1)} \gamma_{\mathbf{a}}^{(2)}=2\left( a_{\mathbf{a}}^{(1)} \gamma_{\mathbf{a}}^{(1)}+a_{\mathbf{a}}^{(0)} \gamma_{\mathbf{a}}^{(2)}\right)\left(\gamma_{\mathbf{a}}^{(1)}\right)-\left( a_{\mathbf{a}}^{(1)}\right)\left( \gamma_{\mathbf{a}}^{(1)}\right)^2\;,
\end{equation}
and get
\begin{equation}
    \begin{split}
        {}&a_{\mathbf{a}}^{(1)}(\gamma_{\mathbf{a}}^{(1)})^2+2 a_{\mathbf{a}}^{(0)} \gamma_{\mathbf{a}}^{(1)} \gamma_{\mathbf{a}}^{(2)}\\
        =&- (1+(-1)^\ell) (\mathtt{f}_s)_{\mathbf{a}} \ \frac{\Gamma (\ell+3)^2}{\Gamma (2 \ell+5)}\frac{16 \left(5 \ell^2+25 \ell+24\right) }{3 \ell (\ell+1)^2 (\ell+4)^2 (\ell+5)}\\
        &+(\mathtt{e}_{s_1}+(-1)^\ell \mathtt{e}_{s_2})_{\mathbf{a}}\ \frac{\Gamma(\ell+3)^2}{\Gamma(2\ell+5)}\left[-\frac{32 \left(2\ell^3+23\ell^2+65\ell+32\right)}{\ell (\ell+1)^3 (\ell+4)^3 (\ell+5)}\right.\\
        &\qquad \quad + \left.  \frac{8 \left(12\psi ^{(0)}(\ell+3)-12\psi ^{(0)}(2 \ell+5)+19 \right)}{3(\ell+1)^2 (\ell+4)^2} \right]\;.
    \end{split}
\end{equation}
Note this result should only be trusted down to $\ell=2$. This is because it uses the one-loop data. The presence of the contact counterterm at one loop spoils the analyticity in spin at $\ell=0$.

\bibliography{refs} 
\bibliographystyle{utphys}
\end{document}